%% file: ColorDecompose.tex
\documentclass[a4paper,11pt]{article}

\usepackage{jheppub} 
\usepackage{epstopdf} 
\usepackage[T1]{fontenc} 
\usepackage{bbm}

\usepackage{graphics}
\usepackage{inputenc}
\usepackage{xspace}
\usepackage{amsmath}
\usepackage{amssymb}
\usepackage{url}
\usepackage{rotating}
\usepackage{enumerate}
\usepackage{graphicx}
\usepackage{def} 

\title{\boldmath Decomposing color structure into multiplet bases}

\author[]{Malin Sjodahl}
\author[]{ and Johan Thor\'en}

\affiliation[]{Department of Astronomy and Theoretical Physics, Lund
  University, S{\"o}lvegatan 14A, 223\,62 Lund, Sweden}

\emailAdd{malin.sjodahl@thep.lu.se}
\emailAdd{johan.thoren@thep.lu.se}

\abstract{We illustrate how QCD color structure elegantly can be 
decomposed into orthogonal multiplet bases corresponding to irreducible
representations of SU($\Nc$) with the aid of Wigner $3j$ and $6j$ coefficients.
We also show how to calculate the relevant $3j$ and $6j$ coefficients using multiplet 
bases and birdtrack techniques and argue that only a relatively small 
number of Wigner $3j$ and $6j$ coefficients are required.  
For up to six gluons plus quark-antiquark pairs we explicitly calculate 
all $6j$ coefficients required for up to NLO calculations.
}

\begin{document} 
\preprint{LU-TP 15-27, MCNET-15-18}
\maketitle
\flushbottom

\section{Introduction}
\label{sec:introduction}

One of the challenges associated with the high multiplicity 
of color charged particles at the LHC is the treatment of 
the non-Abelian color structure in QCD.
The traditional method for mastering this issue is to use non-orthogonal bases,
most notably trace bases \cite{Paton:1969je,Dittner:1972hm,Cvi76,Cvitanovic:1980bu,Mangano:1987xk,Mangano:1988kk,Nagy:2007ty,Sjodahl:2009wx,Alwall:2011uj,Platzer:2012np,Sjodahl:2012nk,Sjodahl:2014opa}, where the basis vectors are products of open and closed 
quark-lines, and color flow bases \cite{Maltoni:2002mq}, where the adjoint representation of the gluon has
been rewritten in terms of quark indices.

While these strategies come with several advantages --- the conceptual simplicity,
the simplicity of gluon emission, the simplicity of gluon exchange 
\cite{Sjodahl:2009wx}, the natural interpretation in terms of 
flow of color \cite{Maltoni:2002mq}
and the existence of recursion relations
for scattering amplitudes \cite{Parke:1986gb,Mangano:1987xk,Kleiss:1988ne,Berends:1987me,Cachazo:2004kj,Britto:2004ap,Bern:2008qj} ---
the non-orthogonality (and the overcompleteness) is a very severe drawback when it 
comes to squaring amplitudes for processes with many colored external legs.

In order to cure this, a general recipe for the construction of minimal and orthogonal
basis vectors has recently been proposed \cite{Keppeler:2012ih}.
For these bases the color space squaring of basis vectors scales only as the dimension of the
vector space, i.e., roughly as the factorial of the number
of gluons and $\qqbar$-pairs in the limit where the number of colors, $\Nc$,
goes to infinity, and roughly as an exponential for finite $\Nc$, 
rather than the square of the number of spanning vectors \cite{Keppeler:2012ih}.
These bases thus have the potential to speed up exact 
calculations in color space very significantly.
However, to realize this potential, it remains to argue that the decomposition 
of scattering amplitudes into multiplet bases can be accomplished efficiently.

In this paper we take important steps in this direction by showing how 
Feynman diagrams can be decomposed into multiplet bases in an efficient 
manner using Wigner $3j$ and $6j$ coefficients, which can be calculated
and stored once for all, see for example \cite{Cvi08, Bickerstaff6j}. 
Furthermore we argue that even for processes with
very many external legs we need only a manageable number of $3j$ and $6j$ 
coefficients and that, using the basis from \cite{Keppeler:2012ih},
these can be calculated.

This article is organized as follows: In \secref{sec:multiplet_bases}
we recapitulate the properties of multiplet bases 
(as constructed in \cite{Keppeler:2012ih}). Following this we illustrate how
the decomposition into these bases can be efficiently achieved 
using $3j$ and $6j$ coefficients in \secref{sec:decomposition}.
Then we show how to calculate all relevant $3j$ and $6j$ coefficients 
in \secref{sec:Wigner}, and argue that the number of such coefficients is 
very manageable. For leading order processes with up to four gluons plus 
$\qqbar$-pairs we explicitly show all needed $6j$ coefficients, whereas all $6j$
coefficients required for up to six gluons plus $\qqbar$-pairs for leading
and next to leading order processes are electronically
attached.
Finally we conclude and make an outlook in \secref{sec:conclusion}.

\section{Multiplet bases}
\label{sec:multiplet_bases}

Orthogonal bases corresponding to irreducible representations of 
subsets of particles may in principle be constructed
in many different ways. In the present paper we follow the construction
in \cite{Keppeler:2012ih}, and start by considering multiplet bases
for pure gluon processes.

\begin{figure}[tbp]
\centering 
\includegraphics[width=5 cm]{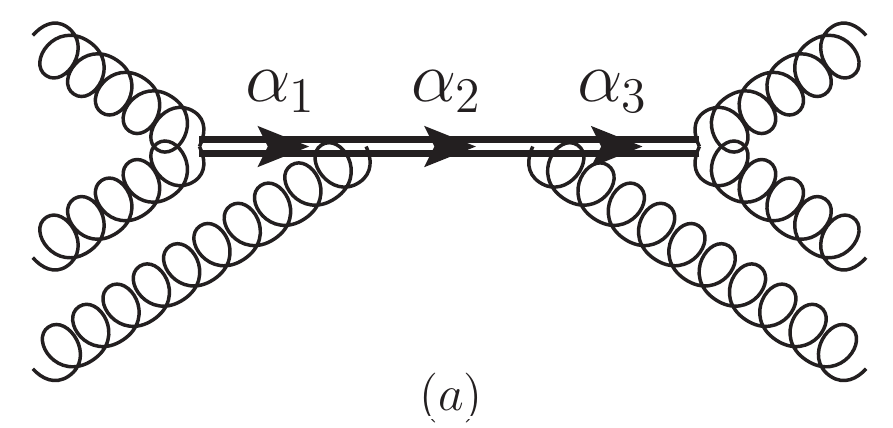}
\includegraphics[width=5 cm]{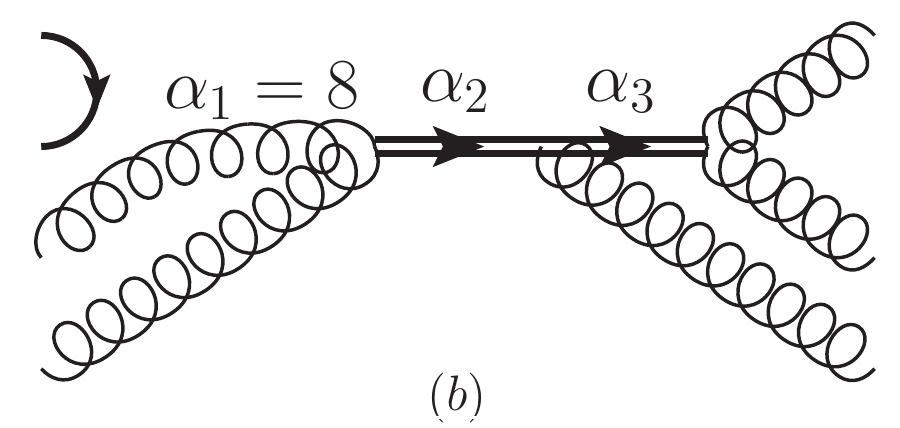}
\includegraphics[width=5 cm]{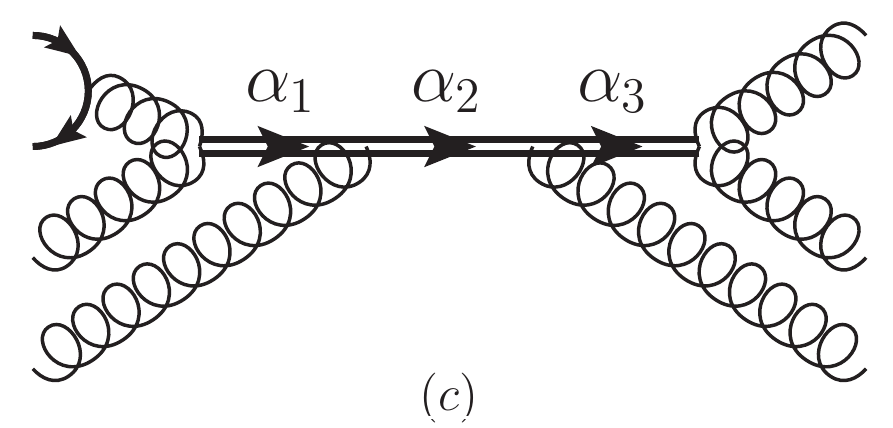}
\caption{\label{fig:6gVector} Example of a six-gluon multiplet basis vector (a),
and basis vectors for one $\qqbar$-pair and five gluons (b,c). 
In the six-gluon case the two first incoming gluons are forced to be in 
representations $\alpha_1$ etc.
}
\end{figure}

Multiplet bases are based on sub-grouping partons into states 
transforming under irreducible representations of SU($\Nc$).
More specifically, we may divide $\Ng$ gluons into 
$\lceil \frac{\Ng}{2}\rceil$ ``incoming'' gluons and 
$\lfloor \frac{\Ng}{2} \rfloor $ ``outgoing'' gluons, 
and then subgroup the gluons on the incoming and outgoing side,
and force the sets to transform under irreducible representations, 
as illustrated in \figref{fig:6gVector}(a), 
where double lines denote arbitrary representations.

For the treatment of quarks, we note that for each incoming quark
(outgoing antiquark) there is an outgoing quark (incoming antiquark).
The quarks and antiquarks may thus be grouped into pairs which together 
transform either as singlets or as octets, since 
$3\otimes \overline{3}=1 \oplus 8$\footnote{We use the terminology octets, 27-plets etc.\ 
even for $\Nc\ne 3$, although the dimensions of the representations clearly vary with $\Nc$.
The definitions of all representations relevant for this paper, for general $\Nc$, can be found for example in section 1.3 and 4 of \cite{Keppeler:2012ih}, where the Young diagrams of the representations and the smallest $\Nc$ for which the representations exist are given. The general $\Nc$ dimensions of all encountered representations can be found in the attached .m-file, WignerCoefficients.m. 
}.
Each $\qqbar$-pair transforming under the singlet representation corresponds to a 
$\delta^{q}{}_{\qbar}$-function in color space, whereas a $\qqbar$-pair forming
an octet can be replaced by a gluon index using the SU($\Nc$) generators
$(t^g)^q{}_{\qbar}$.
Knowing orthogonal bases for processes with an arbitrary number of gluons we
can therefore construct bases for $\Nq$ $\qqbar$-pairs by letting
$\Nq$ of the gluons split into a $\qqbar$-pair or be removed and replaced by
$\delta^{q}{}_{\qbar}$.

Being constructed to sort groups of partons into states transforming
under irreducible representations, the multiplet basis vectors have 
no direct relation to the order in perturbation theory; the bases 
are valid to all orders. Instead the bases can trivially be made
minimal for a given $\Nc$ by crossing out basis vectors corresponding
to representations which only appear for higher $\Nc$.
The dimension of the full vector space is thereby reduced from 
growing roughly as a factorial in the number of gluons plus 
$\qqbar$-pairs, to growing as an exponential \cite{Keppeler:2012ih}, cf. \tabref{tab:dimension}.
While this reduction in dimension may become significant for a large
number of colored partons, the main gain is clearly the orthogonality,
reducing the number of terms appearing in the squaring of an amplitude
from the square of the number of spanning vectors to
the number of spanning vectors.

\begin{table}[t]
  \begin{center}
    \begin{tabular}{|r r r | r  r r| r r r|}
      \hline\hline

      \multicolumn{3}{|c|}{$\Nq=0$} & \multicolumn{3}{c}{$\Nq=1$} & \multicolumn{3}{|c|}{$\Nq=2$}\\ 
      \hline
      $\Ng$ & $\Nc=3$ & $\Nc\to \infty$ &
      $\Ng$  & $\Nc=3$ &  $\Nc\to \infty$ &
      $\Ng$  & $\Nc=3$ &  $\Nc\to \infty$ 
      \\ [0.5ex] 
      \hline 
      4  & 8        & 9           & 3 & 10        & 11         & 2 &        13 &              14\\
      5  & 32       & 44          & 4 & 40        & 53         & 3 &        50 &              64\\ 
      6  & 145      & 265         & 5 & 177       & 309        & 4 &       217 &             362\\
      7  & 702      & 1 854       & 6 & 847       &  2 119     & 5 &     1 024 &           2 428\\
      8  & 3 598    & 14 833      & 7 & 4 300     &  16 687    & 6 &     5 147 &          18 806\\
      9  & 19 280   & 133 496     & 8 & 22 878    & 148 329    & 7 &    27 178 &         165 016\\
      10 & 107 160  & 1 334 961   & 9 & 126 440   & 1 468 457  & 8 &   149 318 &       1 616 786\\
      11 & 614 000  & 14 684 570  & 10& 721 160   & 16 019 531 & 9 &   847 600 &      17 487 988\\
      12 & 3 609 760& 176 214 841   & 11& 4 223 760 & 190 899 411  & 10& 4 944 920 & 206 918 942 \\
      \hline
    \end{tabular}
  \end{center}
  \caption{The dimension of the full vector space (all orders) for $\Nc=3$ and in the 
    $\Nc\to \infty$ limit. In the case of gluons only the dimension of the 
      relevant space can be further reduced by imposing charge conjugation invariance.
    \label{tab:dimension}}
\end{table}
\section{Decomposition}
\label{sec:decomposition}

Due to confinement we are only interested in color summed/averaged quantities, 
and it is not hard to argue that the action of summing over all external indices
corresponds to evaluating a scalar product. 
Letting $\Col_1$ and $\Col_2$ denote the color structures of two amplitudes 
the scalar product
is given by 
\begin{equation}
  \left\langle \Col_1 | \Col_2 \right\rangle
  =\sum_{a_1,\,a_2,\,...}\Col_1^{* a_1\,a_2...} \, \Col_2^{a_1\,a_2...} 
\label{eq:scalar_product}
\end{equation}
with $a_i=1,...,\Nc$ if parton $i$ is a quark or antiquark and $a_i=1,...,\Nc^2-1$ if parton $i$ is a gluon.

\begin{figure}[tbp]
\centering 
\includegraphics[width=11 cm]{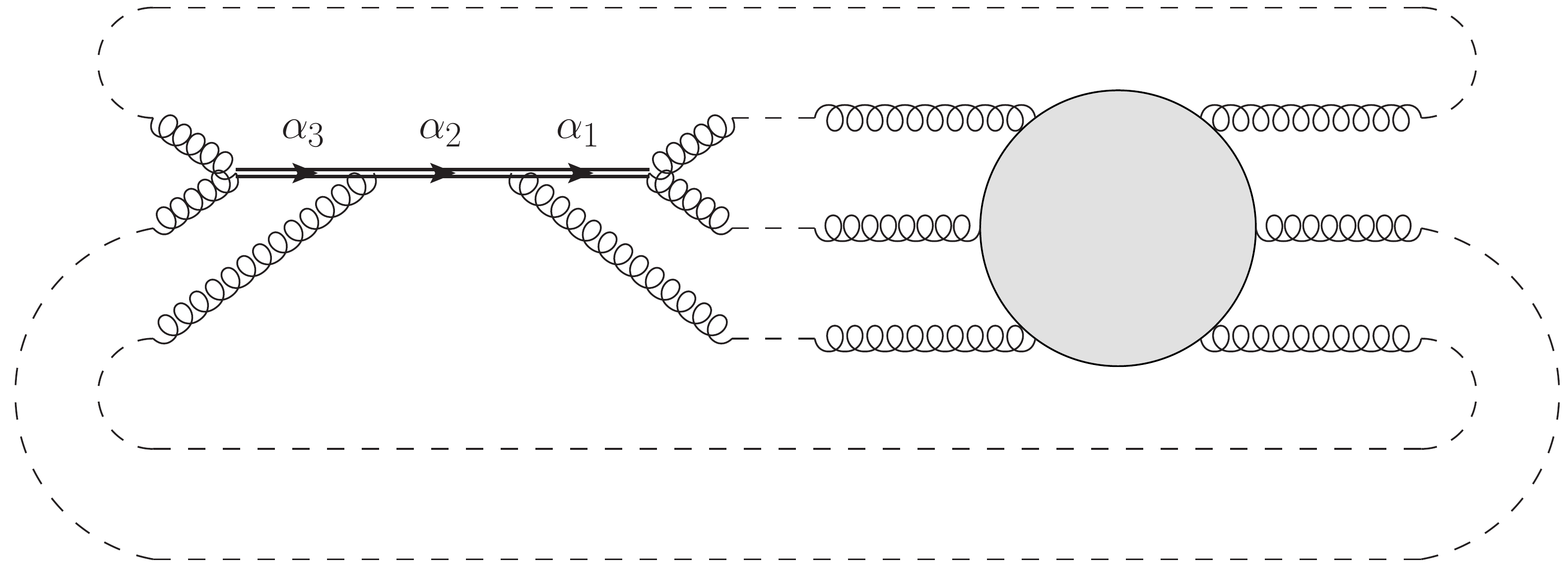}
\caption{\label{fig:scalar_product} Schematic example of scalar product between a vector in the six gluon multiplet basis and a general color structure, where the dashed lines indicate color contractions.}
\end{figure}

As scalar products correspond to color structures where all quark, antiquark
and gluon indices have been contracted, we may graphically represent 
them as fully contracted vacuum bubbles, as depicted in \figref{fig:scalar_product}.

One way of performing the color sum goes via removal of four-gluon
vertices in favor of triple-gluon vertices, rewriting of structure
constants in terms of traces, usage of the Fierz identity (completeness relation) to remove
gluon propagators, and finally counting of closed quark-loops, see for example
\cite{Sjodahl:2012nk,Sjodahl:2014opa}.
While this scalar product evaluation procedure scales much better 
than summing over explicit indices, the (potential) doubling in the number 
of terms with each structure constant and each Fierz identity, makes the 
evaluation of scalar products between Feynman diagrams and basis vectors 
relatively expensive for amplitudes with many (external and internal) gluons.
We will see below that using the birdtrack method we both reduce the number
of terms and avoid calculating scalar products with a large class of basis 
vectors for which the projection vanishes.

We thus suggest a strategy based on repeated usage 
of the completeness relation
\begin{eqnarray}\label{eq:CRDiagrammatic}
  \raisebox{-0.325\height}{
    \includegraphics[scale=0.45]{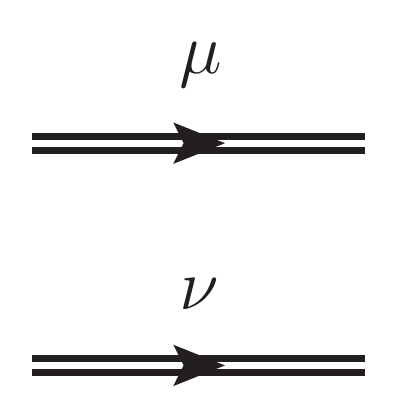}
  }
  =
    \sum_{\alpha\in\mu\otimes\nu}{
      \frac{d_\alpha}{\includegraphics[scale=0.3]{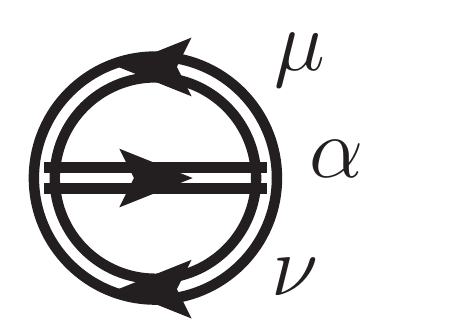}\hspace{-1.5mm}}
      \raisebox{-0.325\height}{
        \includegraphics[scale=0.45]{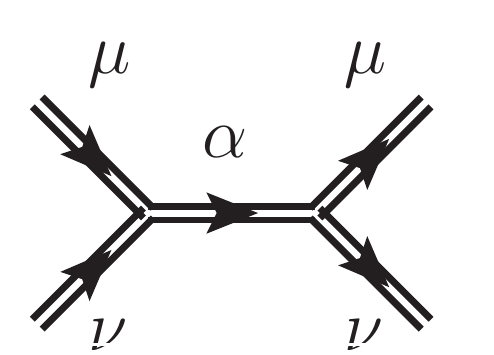}
      }
    }
\end{eqnarray}
and Schur's lemma
\begin{equation}\label{eq:SchursLemma}
\raisebox{-0.39\height}{
\includegraphics[scale=0.45]{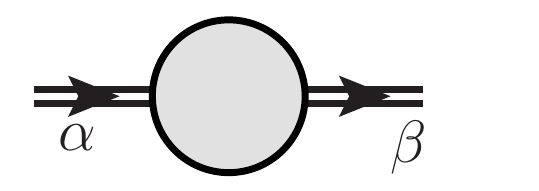}
}
\hspace{-5mm}
=
\frac{
\hspace{-2mm}
\raisebox{-0.4\height}{
\includegraphics[scale=0.45]{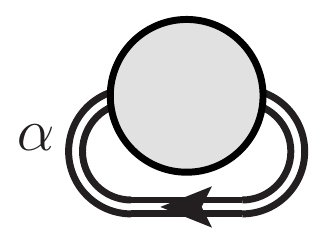}
}
\hspace{-2mm}
}{
d_\alpha
}
\delta^{\alpha}_{\;\beta}
\raisebox{-0.45\height}{
\includegraphics[scale=0.45]{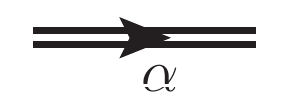}
}
\end{equation}
for performing color contractions for arbitrary representations,
see \cite{Cvi08} for a good introduction. Using the completeness relation, \eqref{eq:CRDiagrammatic}, and Schur's lemma, \eqref{eq:SchursLemma}, on a vertex correction gives
\begin{equation}\label{eq:VertexCorrection}
\raisebox{-0.44\height}{
	\includegraphics[scale=0.4]{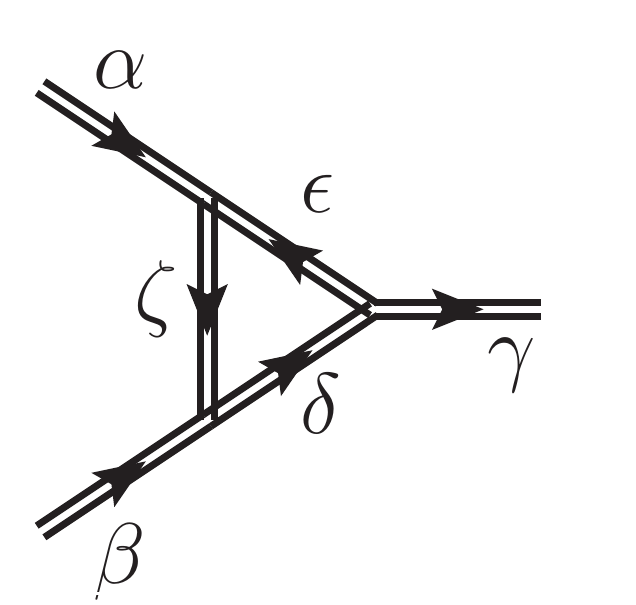}
}
\hspace{-2mm}
=
\sum_{a}
{
\frac{
	\raisebox{-0.45\height}{
		\includegraphics[scale=0.4]{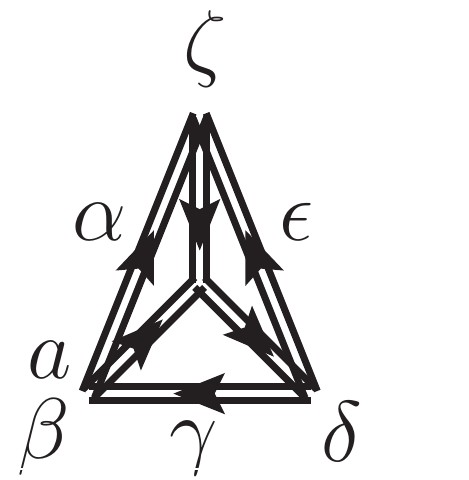}
	}
	\hspace{-3mm}
}{
	\raisebox{-0.45\height}{
		\includegraphics[scale=0.3]{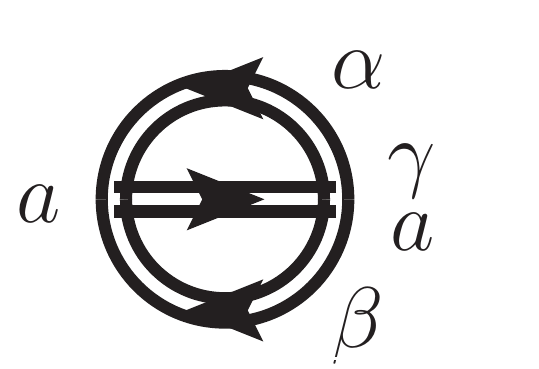}
	}
	\hspace{-3mm}
}
\raisebox{-0.43\height}{
	\includegraphics[scale=0.4]{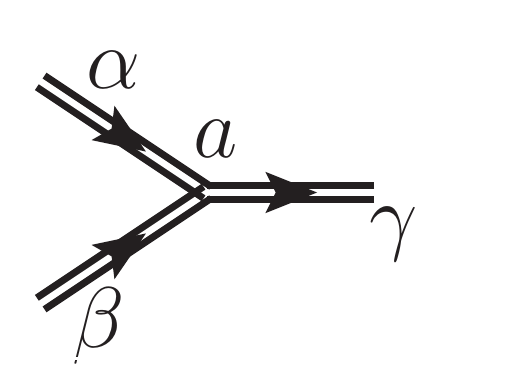}
}
}
\hspace{-3mm}.
\end{equation}
The sum in \eqref{eq:VertexCorrection} runs over all instances of 
the representation $\gamma$ in $\alpha\otimes\beta$, corresponding 
to different vertices $a$. 
For example, if $\alpha, \beta$ and $\gamma$ are all octets, the 
right hand side may  contain vertices with both the symmetric and antisymmetric
structure constant.
In general for QCD (any $\Nc$), if one of $\alpha$, $\beta$ and $\gamma$
is the octet representation, the sum runs over up to two ($\Nc-1$) vertices
if the other two representations are equal. If they are different the sum
only contains one term \cite{Keppeler:2012ih}.

The goal of the strategy is to express vacuum bubbles, 
for example the one in \figref{fig:scalar_product}, in terms of 
vacuum bubbles of the form in the numerator and denominator in
\eqref{eq:VertexCorrection}, known as Wigner $6j$ and $3j$ coefficients, 
respectively.
A vacuum bubble will contain loops of representations; as it is fully contracted every path through the bubble must close. For a loop of representations with $n$ vertices, completeness relations can be applied to relate the loop to a sum over loops with fewer vertices. For example, considering a loop with six vertices, one possibility of applying the completeness relation is  
\begin{equation}\label{eq:LoopContraction}
\raisebox{-0.45\height}{
	\includegraphics[scale=0.5]{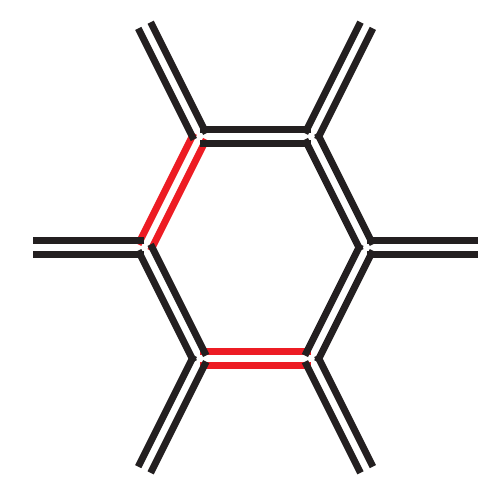}
}
=
\sum_{\alpha}{
\frac{
	d_\alpha
}{
	\hspace{-2mm}
	\raisebox{-0.45\height}{
		\includegraphics[scale=0.3]{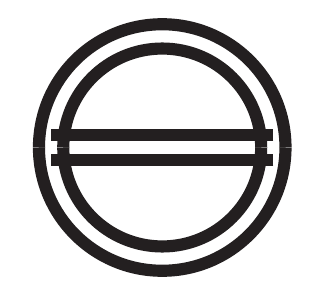}
	}
	\hspace{-2mm}
}
\raisebox{-0.45\height}{
	\includegraphics[scale=0.5]{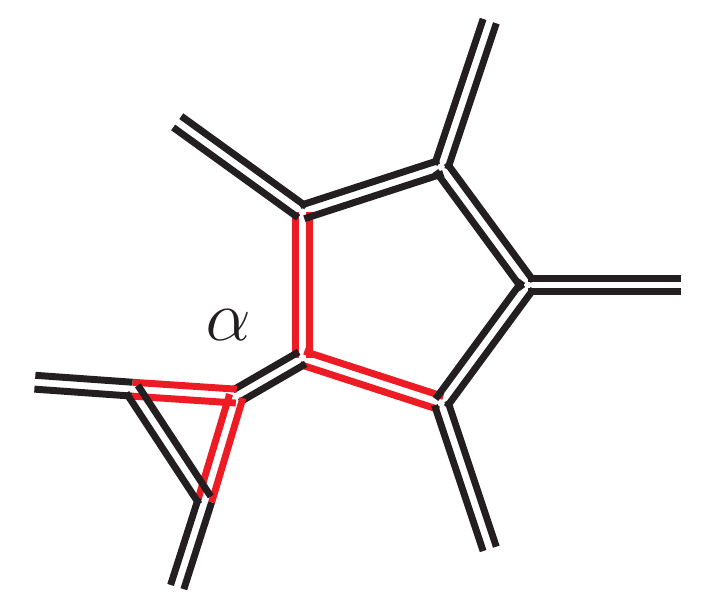}
}
}
=
\sum_{\alpha}{
\frac{
	d_\alpha
}{
	\hspace{-2mm}
	\raisebox{-0.45\height}{
		\includegraphics[scale=0.3]{Figures/Wig3jNoRepLabels}
	}
	\hspace{-2mm}
}
\frac{
	\hspace{-1mm}
	\raisebox{-0.2\height}{
		\includegraphics[scale=0.4]{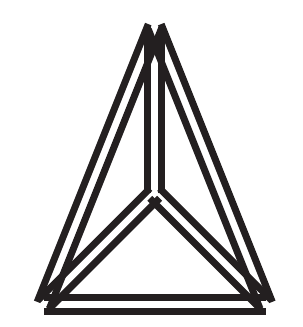}
	}
}{
	\hspace{-2mm}
	\raisebox{-0.45\height}{
		\includegraphics[scale=0.3]{Figures/Wig3jNoRepLabels}
	}
	\hspace{-2mm}
}
\raisebox{-0.45\height}{
	\includegraphics[scale=0.5]{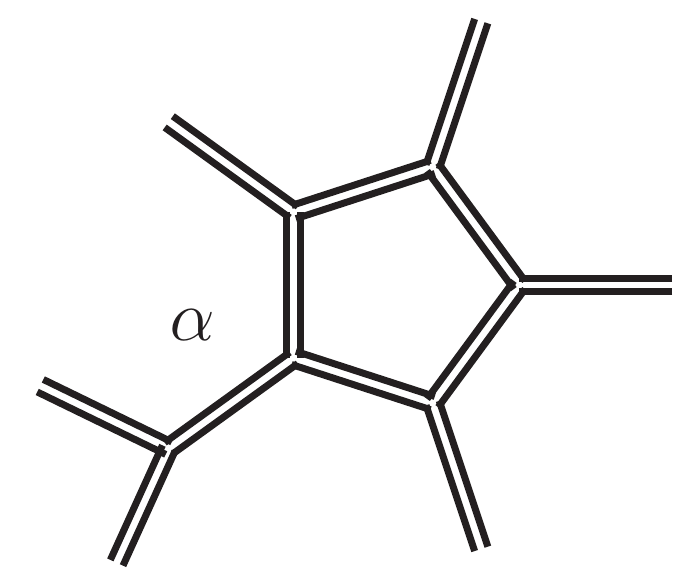}
}
},
\end{equation}
where we suppress representation and vertex indices.
This step of the procedure is independent of the number of vertices in the loop on the left hand side of \eqref{eq:LoopContraction}. Hence it can be applied repeatedly to an $n$-vertex loop until the expression is a sum over loops with three vertices. 
The remaining three-vertex loop, i.e., vertex correction, can then be removed using \eqref{eq:VertexCorrection}.
This step removes two vertices, since a three-vertex loop is replaced 
by a single vertex.
As this procedure is independent of the number of 
vertices in the loop, loops can be repeatedly contracted until
any vacuum bubble equals sums over 
Wigner $3j$ and $6j$ coefficients.

\subsection{An explicit example}
\label{sec:ExplicitExample}

As an example, let us denote gluons by plain lines without arrows 
and consider the color structure of the Feynman diagram 
\begin{equation}\label{eq:ExplicitExampleFeynmanDiagramqqbarTo5g}
\raisebox{-0.4\height}{
\includegraphics[scale=0.4]{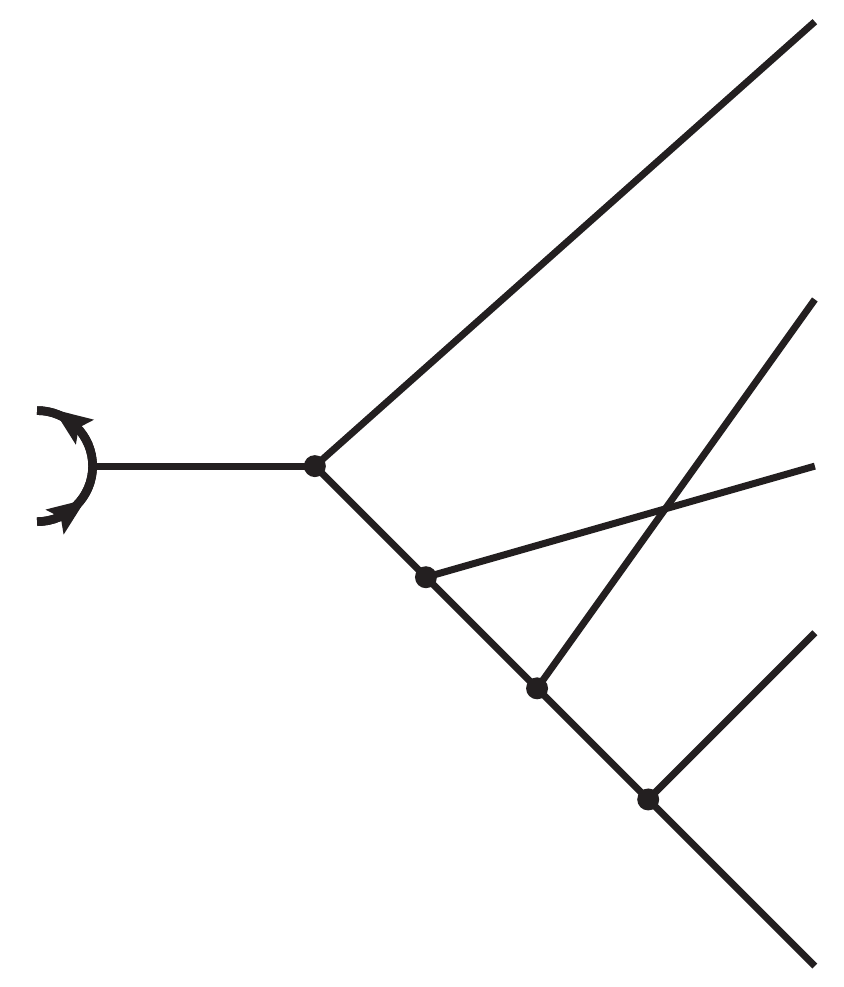}
},
\end{equation}
for $q\bar{q}\rightarrow{}5g$.
From a color structure perspective, any leg can be changed from the 
incoming side to the outgoing and vice versa. 
Hence \eqref{eq:ExplicitExampleFeynmanDiagramqqbarTo5g} 
may alternatively be drawn as
\begin{equation}\label{eq:ExplicitExampleFeynmanDiagram}
\raisebox{-0.4\height}{
\includegraphics[scale=0.4]{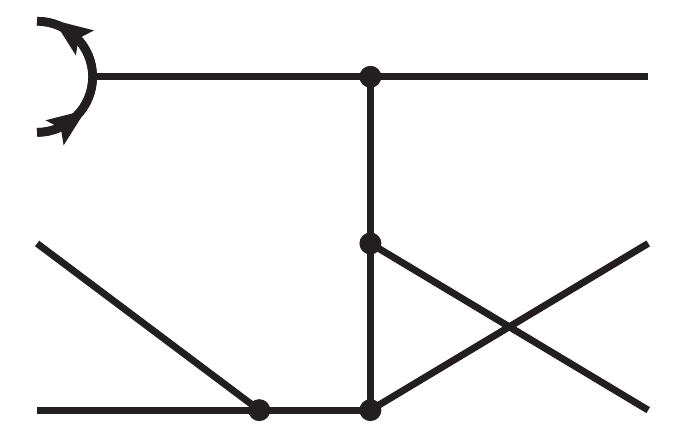}
}
.
\end{equation}
In \eqref{eq:ExplicitExampleFeynmanDiagram} the $q\bar{q}$-pair is in the adjoint representation, therefore it is orthogonal to all basis vectors with the $q\bar{q}$-pair in the singlet representation (cf. \figref{fig:6gVector}) by the tracelessness of the generators.

The vacuum bubble that has to be evaluated to express \eqref{eq:ExplicitExampleFeynmanDiagramqqbarTo5g} in a multiplet basis is
\begin{equation}\label{eq:ExampleScalarProduct}
\raisebox{-0.45\height}{
	\includegraphics[scale=0.4]{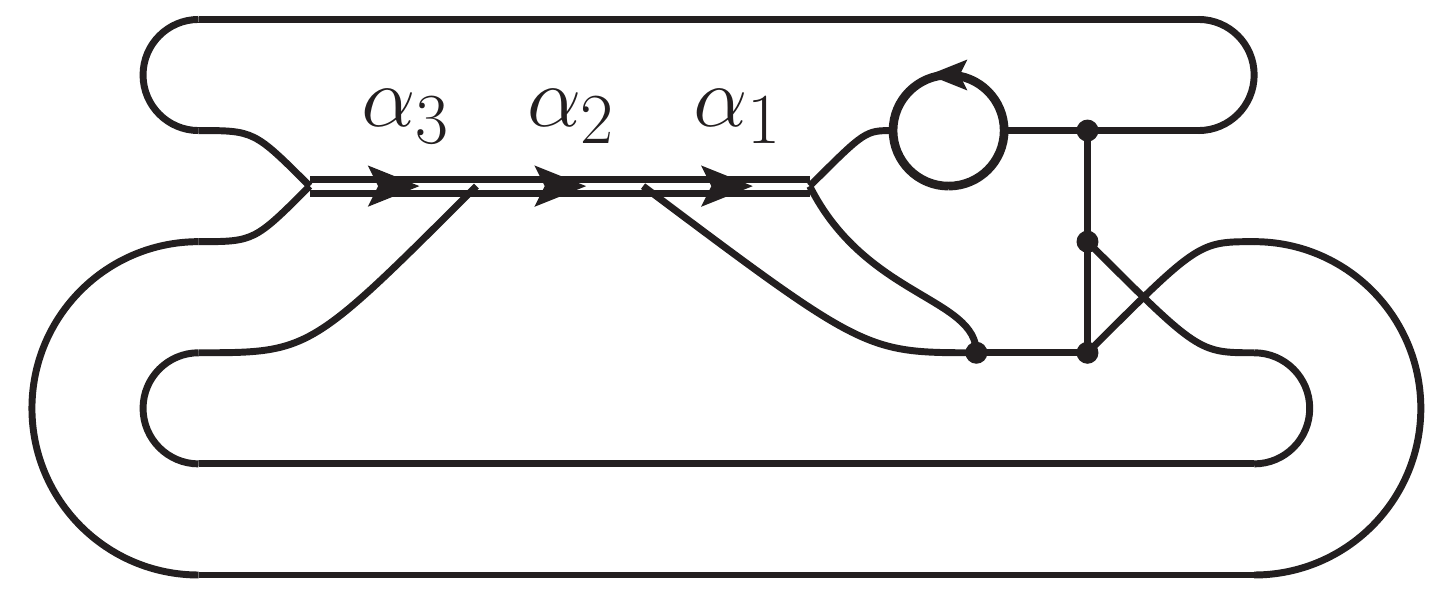}
}
=
\frac{
	\raisebox{-0.45\height}{
		\includegraphics[scale=0.4]{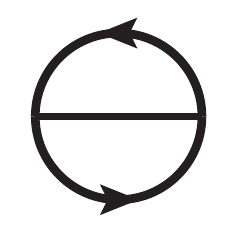}
	}
}{d_A}
\raisebox{-0.45\height}{
	\includegraphics[scale=0.4]{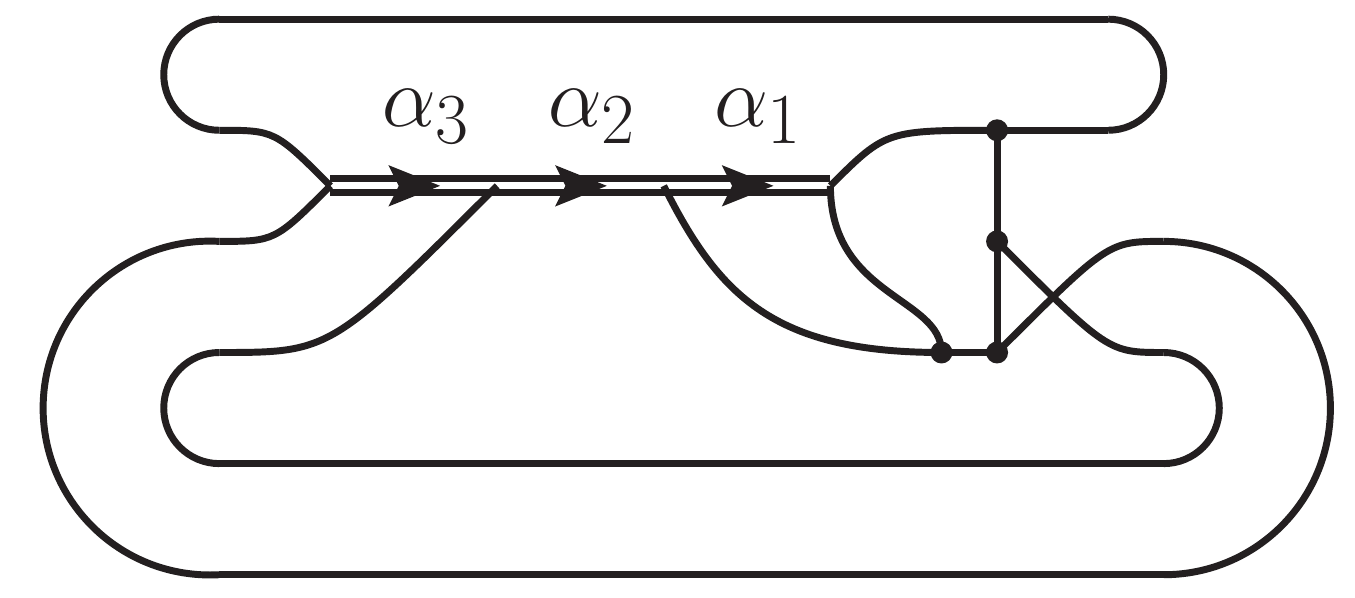}
}
,
\end{equation}
where the equality comes from applying Schur's lemma to remove the two-vertex loop with the quarks. The smallest loop on the right hand side of \eqref{eq:ExampleScalarProduct} is the vertex correction involving the representation $\alpha_1$, it can be removed using \eqref{eq:VertexCorrection}, resulting in
\begin{equation}\label{eq:ExampleFirstLoopRemoval}
\raisebox{-0.45\height}{
	\includegraphics[scale=0.4]{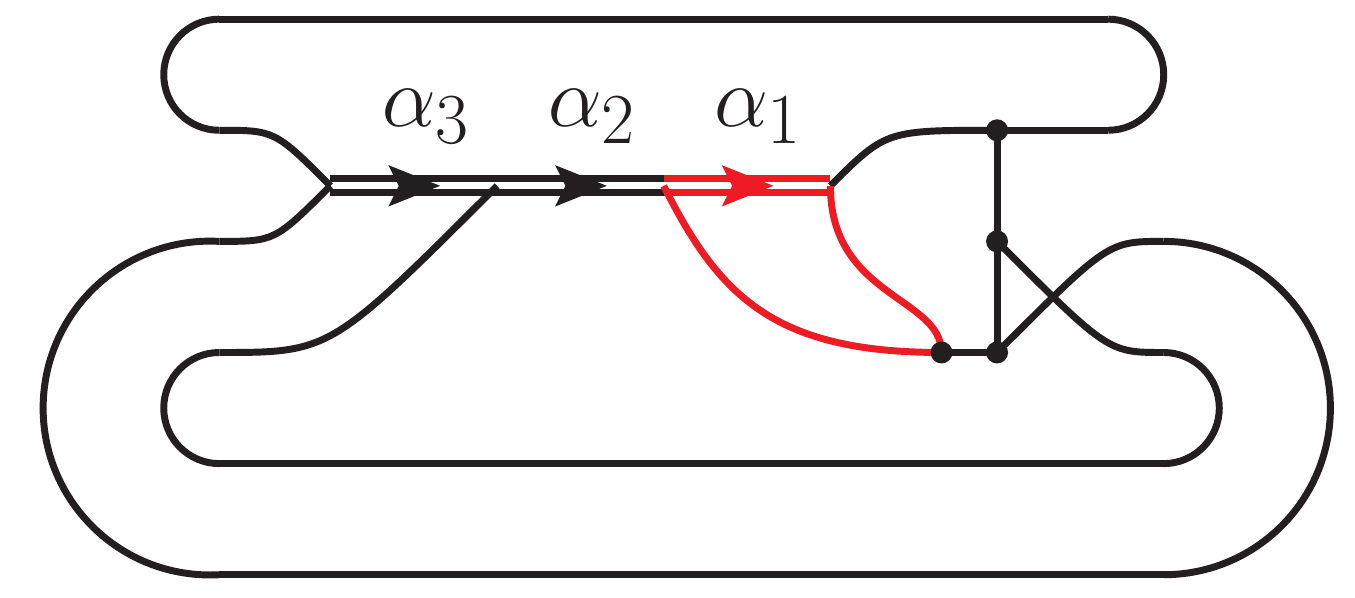}
}
=
\frac{
	\raisebox{-0.45\height}{
		\includegraphics[scale=0.4]{Figures/Example/WignerCoefficients/QuarkLoop/Wig3jQuarkLoop}
	}
}{d_A}
\sum_{a}
\frac{
	\hspace{0.5mm}
	\raisebox{-0.45\height}{
		\includegraphics[scale=0.4]{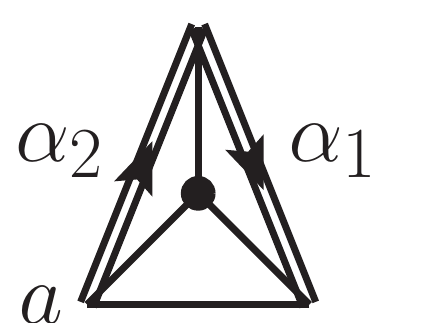}
	}
	\hspace{-3mm}
}{
	\hspace{2.5mm}
	\raisebox{-0.45\height}{
		\includegraphics[scale=0.4]{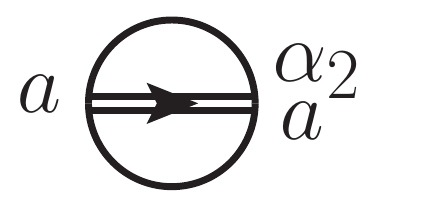}
	}
	\hspace{-3mm}
}
\raisebox{-0.45\height}{
	\includegraphics[scale=0.4]{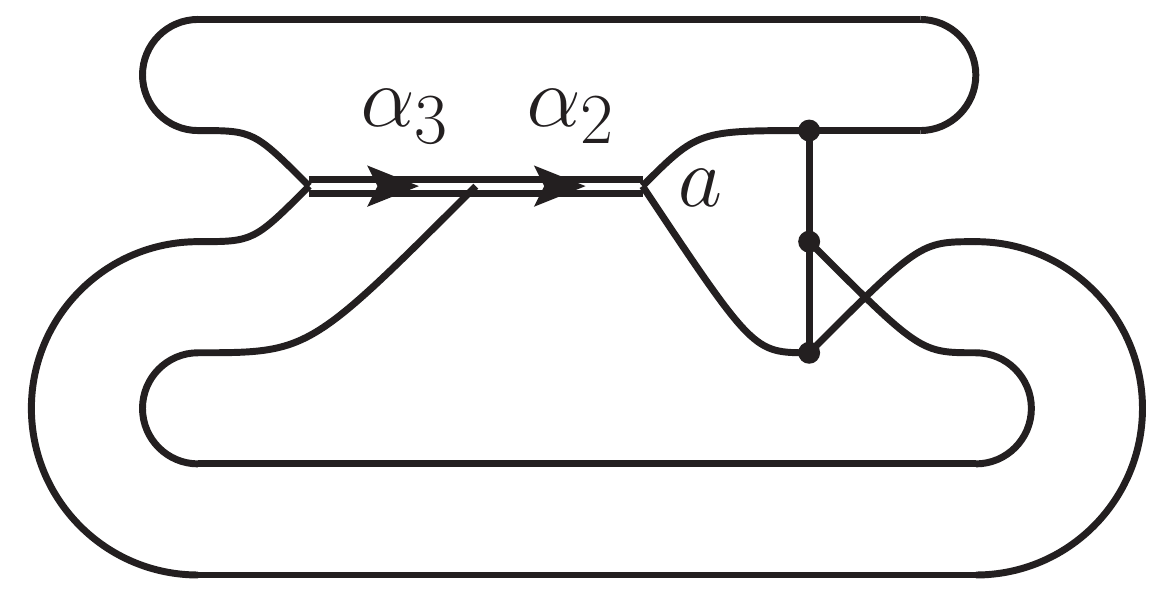}
}
.
\end{equation}
The shortest loop in the large vacuum bubble on the right hand side has four 
vertices and there are several possible choices of four-vertex loops, 
one is
\begin{equation}\label{eq:ExampleSecondLoopHighlighted}
\raisebox{-0.45\height}{
	\includegraphics[scale=0.4]{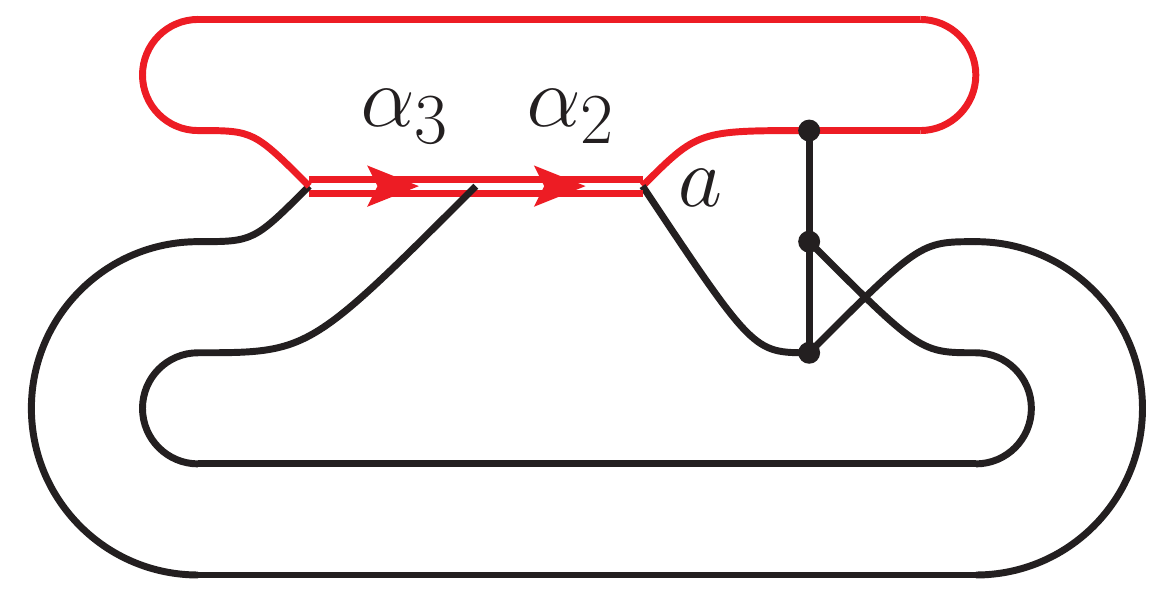}
}
.
\end{equation}
Contracting this loop gives
\begin{equation}\label{eq:ExampleSecondLoop}
\raisebox{-0.45\height}{
	\includegraphics[scale=0.4]{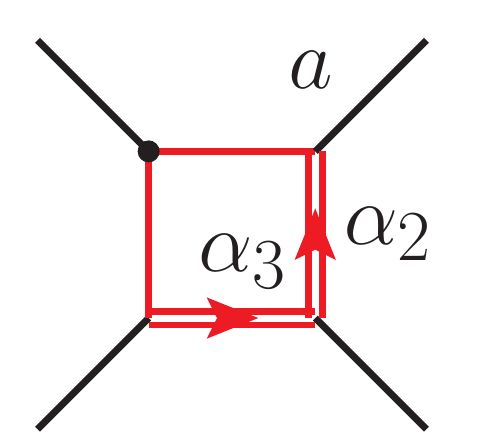}
}
\hspace{-2mm}
=
\sum_{\psi}{
\frac{d_\psi}{
	\hspace{-1mm}
	\raisebox{-0.45\height}{
		\includegraphics[scale=0.4]{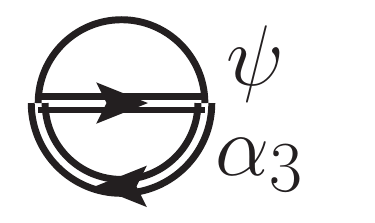}
	}
	\hspace{-3mm}
}
\raisebox{-0.45\height}{
	\includegraphics[scale=0.4]{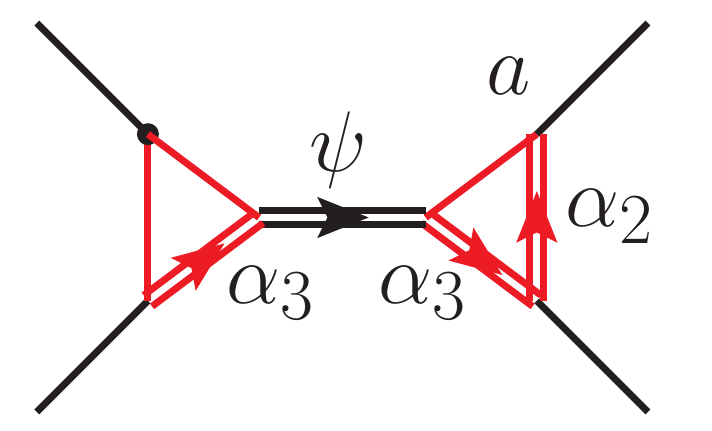}
}
}
\hspace{-2mm}
=
\sum_{\psi, b, c}{
\frac{d_\psi}{
	\hspace{-1mm}
	\raisebox{-0.45\height}{
		\includegraphics[scale=0.4]{Figures/Example/WignerCoefficients/SecondLoop/Wig3jCR}
	}
	\hspace{-3mm}
}
\frac{
	\raisebox{-0.2\height}{
		\includegraphics[scale=0.4]{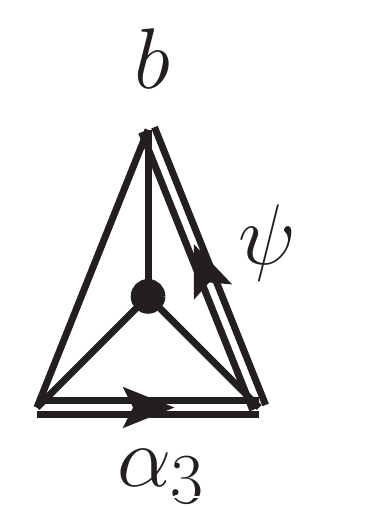}
	}
	\hspace{-4mm}
}{
	\hspace{1.5mm}
	\raisebox{-0.45\height}{
		\includegraphics[scale=0.4]{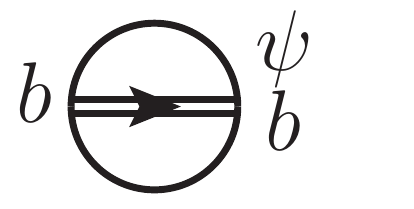}
	}
	\hspace{-4mm}
}	
\frac{
	\raisebox{-0.2\height}{
		\includegraphics[scale=0.4]{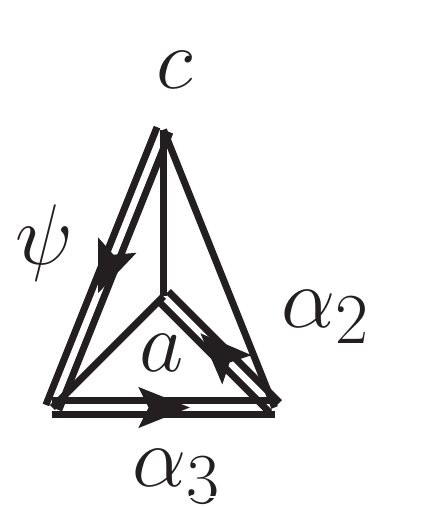}
	}
	\hspace{-4mm}
}{
	\hspace{1.5mm}
	\raisebox{-0.45\height}{
		\includegraphics[scale=0.4]{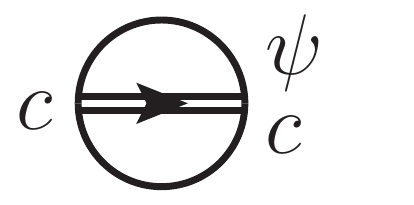}
	}
	\hspace{-4mm}
}
\raisebox{-0.45\height}{
	\includegraphics[scale=0.4]{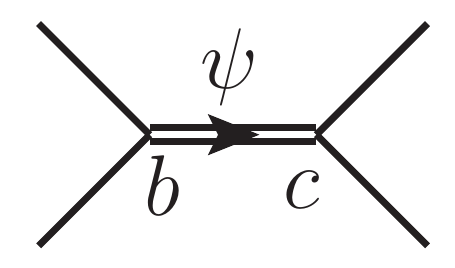}
}
}.
\end{equation}
Replacing the four-vertex loop in \eqref{eq:ExampleFirstLoopRemoval} with 
the right hand side of \eqref{eq:ExampleSecondLoop} results in
\begin{equation}\label{eq:ExampleSecondLoopRemoval}
\frac{
	\raisebox{-0.45\height}{
		\includegraphics[scale=0.4]{Figures/Example/WignerCoefficients/QuarkLoop/Wig3jQuarkLoop}
	}
}{d_A}
\sum_{a}
\frac{
	\hspace{0.5mm}
	\raisebox{-0.45\height}{
		\includegraphics[scale=0.4]{Figures/Example/WignerCoefficients/FirstLoop/Wig6j}
	}
	\hspace{-3mm}
}{
	\hspace{2.5mm}
	\raisebox{-0.45\height}{
		\includegraphics[scale=0.4]{Figures/Example/WignerCoefficients/FirstLoop/Wig3j}
	}
	\hspace{-3mm}
}
\sum_{\psi, b,c}{
\frac{d_\psi}{
	\hspace{0.5mm}
	\raisebox{-0.45\height}{
		\includegraphics[scale=0.4]{Figures/Example/WignerCoefficients/SecondLoop/Wig3jCR}
	}
	\hspace{-3mm}
}
\frac{
	\raisebox{-0.2\height}{
		\includegraphics[scale=0.4]{Figures/Example/WignerCoefficients/SecondLoop/Wig6jLeft}
	}
	\hspace{-4mm}
}{
	\hspace{1.5mm}
	\raisebox{-0.45\height}{
		\includegraphics[scale=0.4]{Figures/Example/WignerCoefficients/SecondLoop/Wig3jLeft}
	}
	\hspace{-4mm}
}	
\frac{
	\raisebox{-0.2\height}{
		\includegraphics[scale=0.4]{Figures/Example/WignerCoefficients/SecondLoop/Wig6jRight}
	}
	\hspace{-4mm}
}{
	\hspace{1.5mm}
	\raisebox{-0.45\height}{
		\includegraphics[scale=0.4]{Figures/Example/WignerCoefficients/SecondLoop/Wig3jRight}
	}
	\hspace{-4mm}
}
\raisebox{-0.45\height}{
	\includegraphics[scale=0.4]{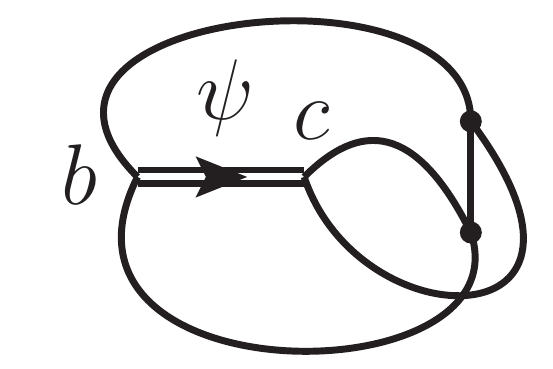}
}
}.
\end{equation}
The final expression is, by drawing the rightmost vacuum 
bubble in the same shape as the other $6j$ coefficients,
\begin{equation}\label{eq:ExampleFinal}
\raisebox{-0.45\height}{
	\includegraphics[scale=0.4]{Figures/Example/VacuumBubble/VacuumBubbleN10}
}
=
\frac{
	\raisebox{-0.45\height}{
		\includegraphics[scale=0.4]{Figures/Example/WignerCoefficients/QuarkLoop/Wig3jQuarkLoop}
	}
}{d_A}
\sum_{a,\psi,b,c}
\frac{
	\hspace{0.5mm}
	\raisebox{-0.45\height}{
		\includegraphics[scale=0.4]{Figures/Example/WignerCoefficients/FirstLoop/Wig6j}
	}
	\hspace{-3mm}
}{
	\hspace{2.5mm}
	\raisebox{-0.45\height}{
		\includegraphics[scale=0.4]{Figures/Example/WignerCoefficients/FirstLoop/Wig3j}
	}
	\hspace{-3mm}
}
{
\frac{d_\psi}{
	\hspace{-1mm}
	\raisebox{-0.45\height}{
		\includegraphics[scale=0.4]{Figures/Example/WignerCoefficients/SecondLoop/Wig3jCR}
	}
	\hspace{-3mm}
}
\frac{
	\raisebox{-0.2\height}{
		\includegraphics[scale=0.4]{Figures/Example/WignerCoefficients/SecondLoop/Wig6jLeft}
	}
	\hspace{-4mm}
}{
	\hspace{1.5mm}
	\raisebox{-0.45\height}{
		\includegraphics[scale=0.4]{Figures/Example/WignerCoefficients/SecondLoop/Wig3jLeft}
	}
	\hspace{-4mm}
}	
\frac{
	\raisebox{-0.2\height}{
		\includegraphics[scale=0.4]{Figures/Example/WignerCoefficients/SecondLoop/Wig6jRight}
	}
	\hspace{-4mm}
}{
	\hspace{1.5mm}
	\raisebox{-0.45\height}{
		\includegraphics[scale=0.4]{Figures/Example/WignerCoefficients/SecondLoop/Wig3jRight}
	}
	\hspace{-4mm}
}
\raisebox{-0.3\height}{
	\includegraphics[scale=0.4]{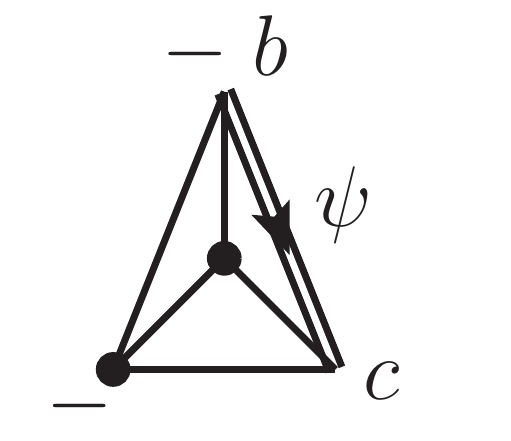}
}
}
\hspace{-3mm},
\end{equation}
where, in the last $6j$ coefficient, two vertices are drawn using 
Yutsis' notation \cite{YutsisNotation}
\begin{equation}\label{eq:YutsisVertexToNormalVertex}
\raisebox{-0.41\height}{
	\includegraphics[scale=0.45]{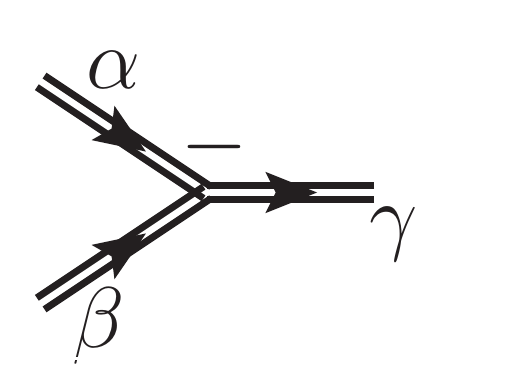}
}
\hspace{-3mm}
\equiv
\raisebox{-0.41\height}{
	\includegraphics[scale=0.45]{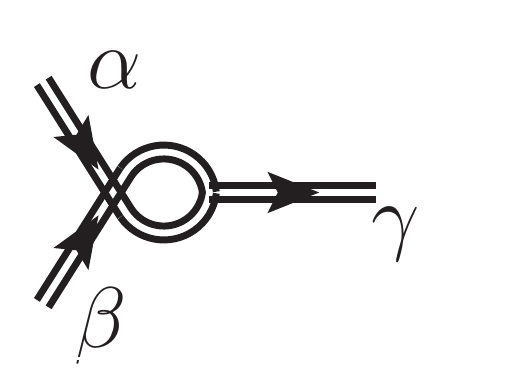}
}
\hspace{-3mm}
=
\sum_{a}
\frac{
\hspace{-1.5mm}
\raisebox{-0.45\height}{
	\includegraphics[scale=0.3]{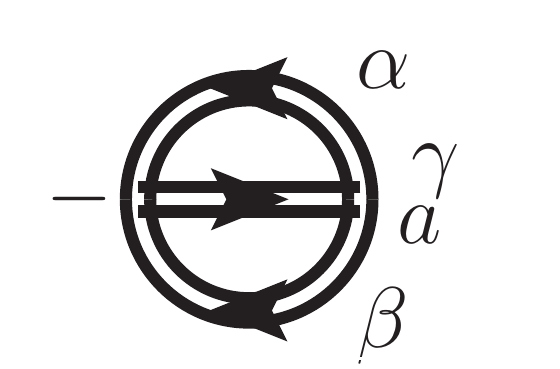}
}
\hspace{-3mm}
}{
\raisebox{-0.45\height}{
	\includegraphics[scale=0.3]{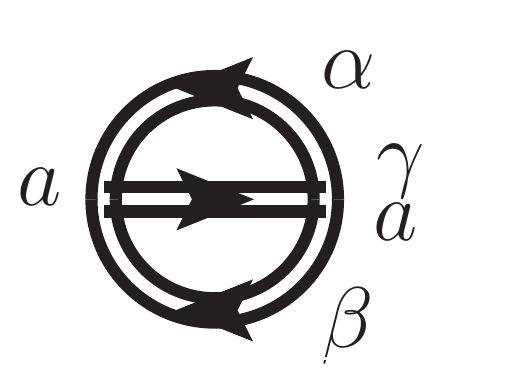}
}
\hspace{-3mm}
}
\raisebox{-0.43\height}{
	\includegraphics[scale=0.45]{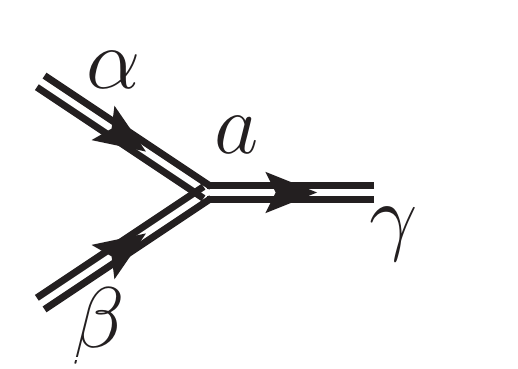}
}
\hspace{-3mm}.
\end{equation}
While the sum over vertices in \eqref{eq:YutsisVertexToNormalVertex}
in principle may contain more than one term, in all cases encountered here,
the vertices have been chosen such that only one term contributes, giving at most a minus sign.

From \eqref{eq:ExampleFinal} one can immediately determine that this color structure is orthogonal to many of the basis vectors. It is orthogonal to all of the basis vectors where the $q\bar{q}$-pair is in a singlet, as mentioned. It is also orthogonal to all basis vectors where $\alpha_2\not\in{}8\otimes{}8$, as seen from the 
vertex labeled $a$ on the right-hand side of \eqref{eq:ExampleFinal}.
Since the orthogonality is manifest from constraints on the representations,
the projection onto many basis vectors need not be calculated, which we 
expect to significantly speed up computations. 

The right hand side of \eqref{eq:ExampleFinal} can now be explicitly evaluated. 
As an example the vacuum bubble for the (unnormalized) basis vector with the 
representations $\alpha_1=27$, $\alpha_2=8$ and $\alpha_3=1$ is
\begin{equation}\label{eq:ExampleExplicitCalculation}
\raisebox{-0.45\height}{
	\includegraphics[scale=0.4]{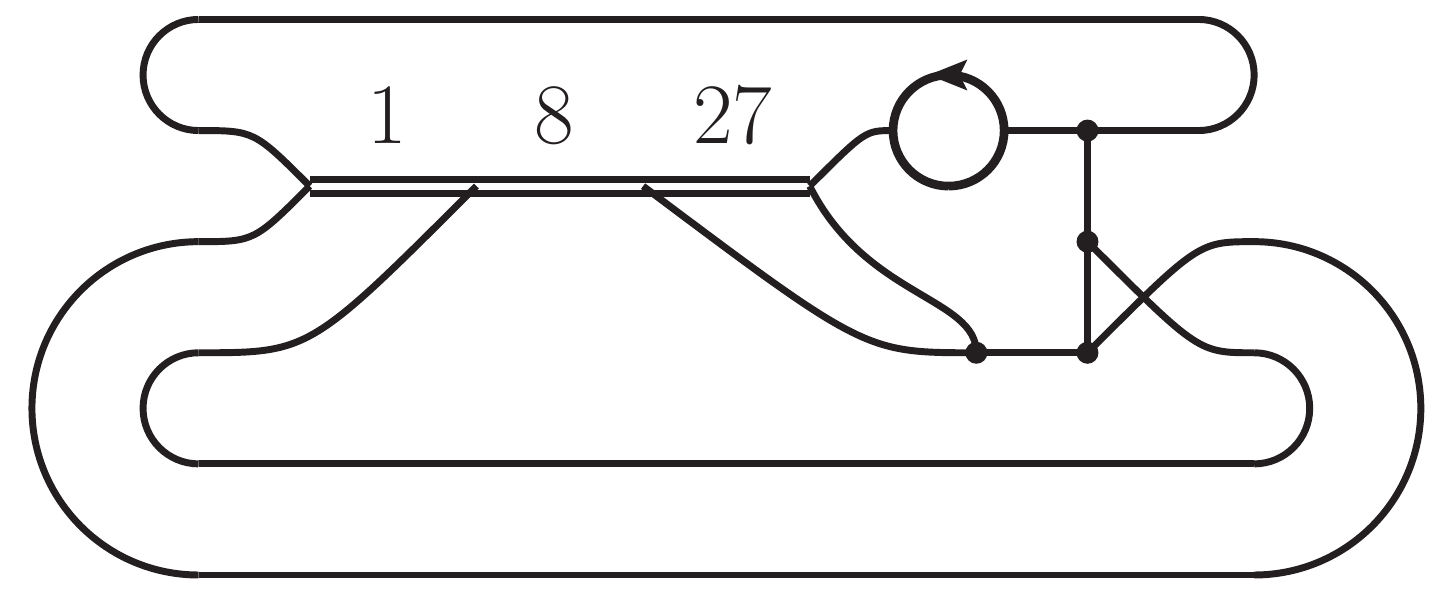}
}
=
-\frac{1}{2 \Nc (\Nc^2-1)^4}
,
\end{equation}
where the values of the Wigner $3j$ coefficients have been 
normalized to one, c.f., \eqref{eq:3jNormalization1} 
and \eqref{eq:3j888Normalization1}, and the $6j$ coefficients
are taken from \tabref{tab:4GluonWignerCoefficients}. 
Using the standard normalization of the antisymmetric
triple-gluon vertices in the original color structure,
we should multiply with a factor $(\sqrt{2 \Nc \TR (\Nc^2-1)})^4$,
where $\TR$ is defined by the generator normalization, 
$\Tr[t^at^b]=\TR \delta^{ab}$,
and typically chosen to be $1/2$ or 1.
Similarly, the $3j$ coefficient from the $\qqbar g$-vertices would in the 
standard normalization give a factor $\TR(\Nc^2-1)$, rather than 1.
If we want to project onto normalized basis vectors 
we also need to multiply with the square roots of the dimensions
of the representations $\alpha_1$, $\alpha_2$ and $\alpha_3$
and a factor $1/\sqrt{\TR}$ from the $\qqbar$-pair
giving an additional factor  
$\sqrt{1}\sqrt{(\Nc^2-1)}\sqrt{\Nc^2(\Nc^2+2\Nc-3)/4}/\sqrt{\TR}$. 
We note that the quadruple sum in \eqref{eq:ExampleFinal} typically 
only contains a few non-zero terms, making the above procedure for
scalar product evaluation by far superior to the method of 
rewriting all representations in the fundamental and anti-fundamental
representation, currently employed in for example 
\cite{Alwall:2011uj,Sjodahl:2012nk,Sjodahl:2014opa}.

We also remark that this procedure of decomposing color structure
is entirely general. 
In the following we focus on leading order 
(LO) and next to leading order (NLO) processes,
but there is no conceptual difference in treating color 
structure at higher orders.

An increasingly popular strategy for evaluating amplitudes
is to use recursion relations of various kinds rather than
plain Feynman diagrams. While the present paper addresses
generic (Feynman diagram) color structure decomposition, it
is also of interest to study how efficient multiplet basis
color decomposition can be achieved in recursive approaches.
This is done in detail in \cite{Du:2015apa}, for the case
of BCFW \cite{Britto:2004ap,Britto:2005fq} recursion of tree-level
MHV gluon amplitudes, where it is argued that the recursion
can be formulated and efficiently implemented directly in the
multiplet basis (cf. table~(1) in \cite{Du:2015apa}).
Using this recursion, the amplitudes can --- if desired ---
also be expressed in terms of color-ordered  amplitudes
(see for example eqs.~(2.15) and (B.5) in \cite{Du:2015apa}).

\section{Wigner $3j$ and $6j$ coefficients}
\label{sec:Wigner}

The procedure of \secref{sec:decomposition} expresses any vacuum bubble as a sum over factors of Wigner $3j$ and $6j$ coefficients.
The Wigner $3j$ coefficients are easily evaluated by contracting two vertices, 
\begin{equation}\label{eq:Wigner3jCalculation}
  \parbox{1.5cm}{\epsfig{file=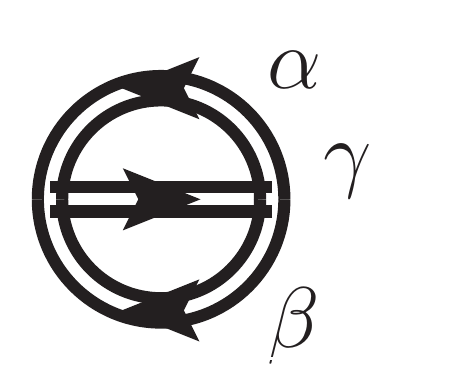,width=1.5cm}}.
\end{equation}
By construction the Wigner $3j$ coefficients of two different instances of the vertex between three representations must be zero in order for the completeness relation to hold.

In the initial vacuum bubble there are only vertices where at least one representation is the adjoint representation. As is argued in \appref{sec:RepresentationConstraints}, the loops for the contraction strategy of \secref{sec:decomposition} can be chosen such that the completeness relations are only applied where at least one of $\mu$ and $\nu$ in the completeness relation \eqref{eq:CRDiagrammatic} is the adjoint representation. This results in that every new vertex introduced from the completeness relation also contains at least one adjoint representation. There are only two different types of Wigner $6j$ coefficients where every vertex has at least one adjoint representation that cannot be simplified to Wigner $3j$ coefficients by applying Schur's lemma,
\begin{equation}\label{eq:Wigner6j3Representations}
\raisebox{-0.45\height}{
	\includegraphics[scale=0.4]{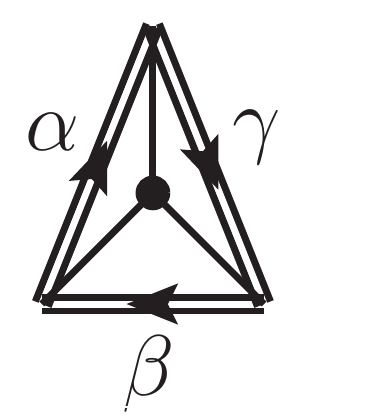}
}
\hspace{-4mm},\;
\raisebox{-0.45\height}{
	\includegraphics[scale=0.4]{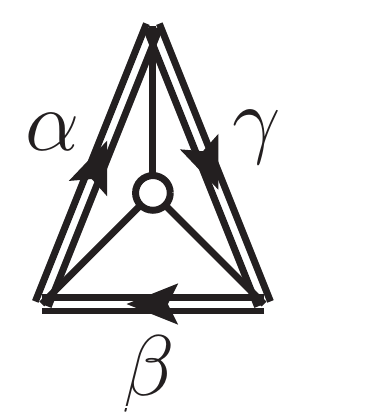}
}
\end{equation}
and
\begin{equation}\label{eq:Wigner6j4Representations}
\raisebox{-0.45\height}{
	\includegraphics[scale=0.4]{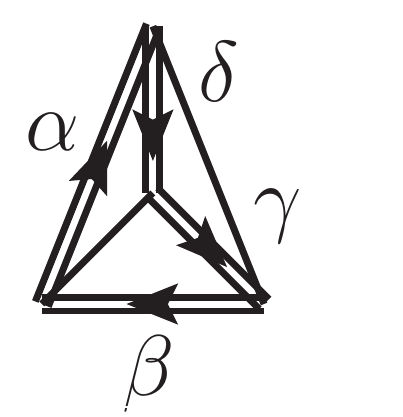}
}
\hspace{-3mm}.
\end{equation}
In \eqref{eq:Wigner6j3Representations} the 
black blob in the left $6j$ coefficient is the antisymmetric structure constant of $SU(\Nc)$,
$if^{abc}=\frac{1}{T_R}\left[\Tr{(t^{a}t^{b}t^{c})}-\Tr{(t^{c}t^{b}t^{a})}\right]$,
while the white blob is the symmetric structure constant,
$ d^{abc}=\frac{1}{T_R}\left[\Tr{(t^{a}t^{b}t^{c})}+\Tr{(t^{c}t^{b}t^{a})}\right]$. 

Internal quark-lines can be handled by applying the completeness relation and rewriting traces over three generators in terms of
$if^{abc}$ and $d^{abc}$.
A quark loop with an arbitrary number of gluons attached can be rewritten as 
\begin{align}\label{eq:InternalqLines}\nonumber
\raisebox{-0.45\height}{
	\includegraphics[scale=0.4]{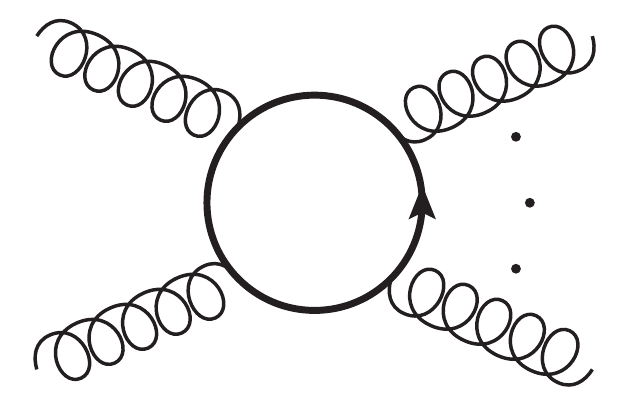}
}
&=
\frac{d_1}{
\raisebox{-0.45\height}{
	\includegraphics[scale=0.3]{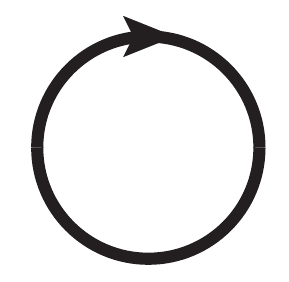}
}
}
\raisebox{-0.45\height}{
	\includegraphics[scale=0.4]{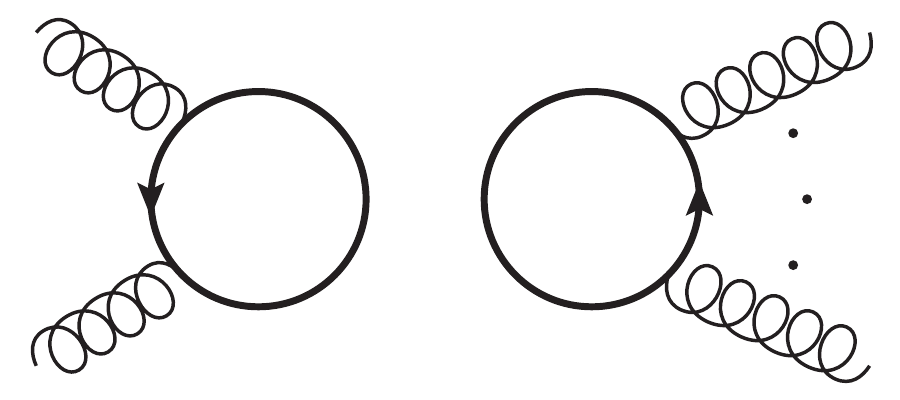}
}
+
\frac{d_A}{
\raisebox{-0.45\height}{
	\includegraphics[scale=0.3]{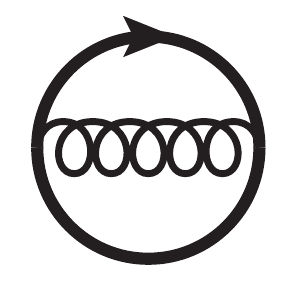}
}
}
\raisebox{-0.45\height}{
	\includegraphics[scale=0.4]{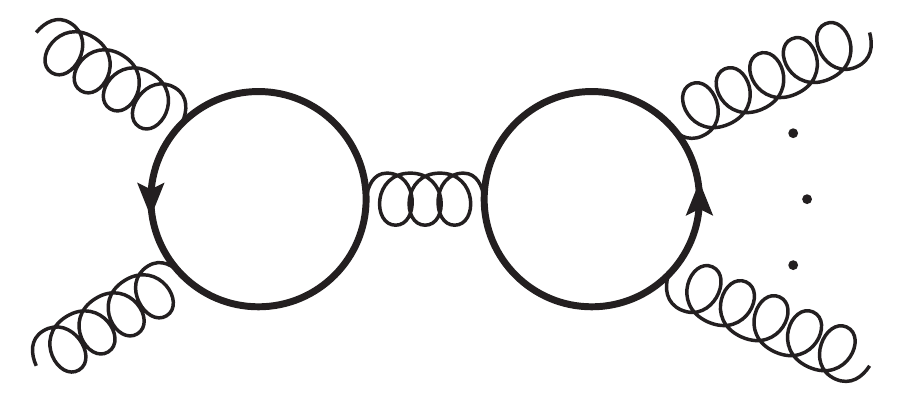}
}
\\
&=\nonumber
\frac{
	T_R
}{
	\hspace{-1mm}
	\raisebox{-0.45\height}{
		\includegraphics[scale=0.3]{Figures/InternalQuarks/Wig3j_Singlet}
	}
}
\raisebox{-0.45\height}{
	\includegraphics[scale=0.4]{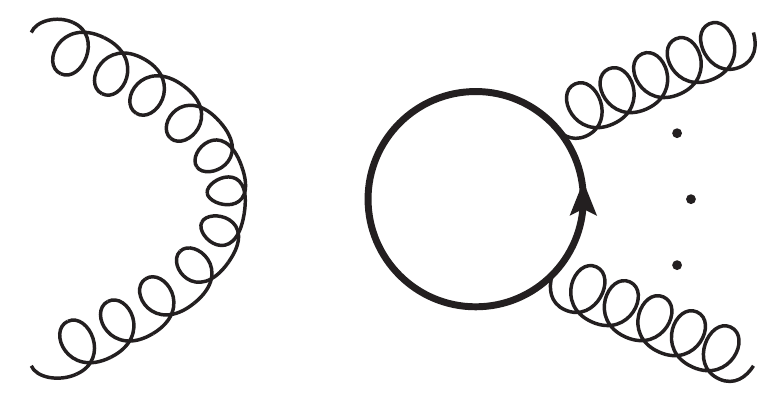}
}
\\
&+
\frac{
	1
	}{
	2 
}
\left(
\raisebox{-0.45\height}{
	\includegraphics[scale=0.4]{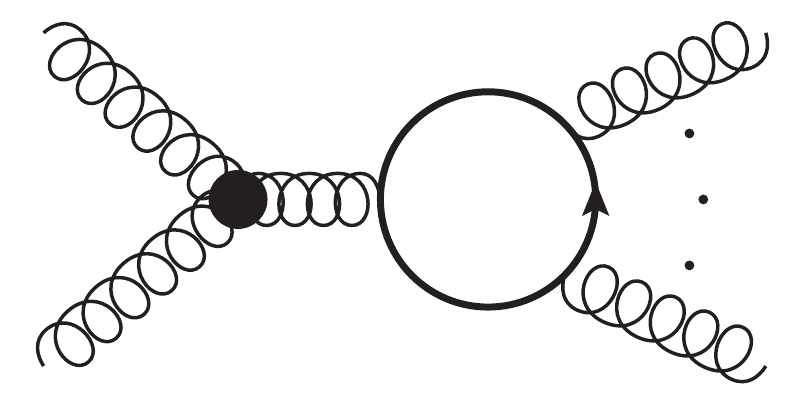}
}
+
\raisebox{-0.45\height}{
	\includegraphics[scale=0.4]{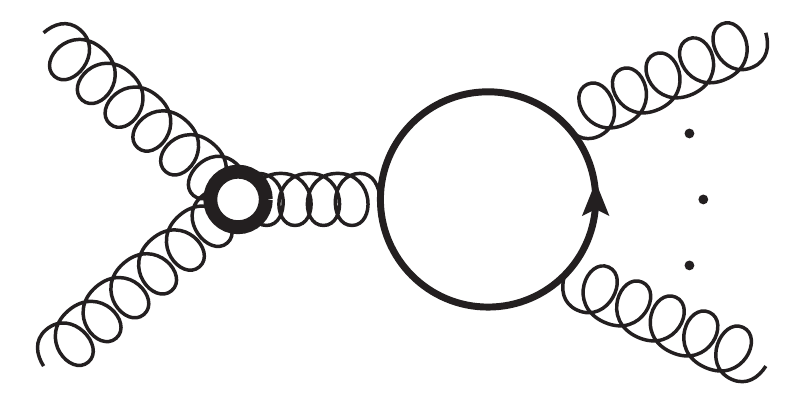}
}
\right)
,
\end{align}
where the generator normalization
\begin{equation}\label{eq:GeneratorNormalization}
T_R
=
\frac{
	\Tr{(t^{a}t^{a})}
}{
	d_A
}
=
\frac{
	\raisebox{-0.45\height}{
		\includegraphics[scale=0.3]{Figures/InternalQuarks/Wig3j_Octet}
	}
}{
	d_A
}
\end{equation}
has been used.
By applying this repeatedly, an arbitrarily large quark loop can be expressed only in terms of structure constants. This introduces $d^{abc}$ vertices, the last term in \eqref{eq:InternalqLines}, which in turn require the second coefficient in \eqref{eq:Wigner6j3Representations}.

\subsection{Evaluation of Wigner $6j$ coefficients}
\label{sec:WignerEvaluation}

Given a set of basis vectors from the basis construction described in \cite{Keppeler:2012ih}, the Wigner $6j$ coefficients can be calculated. The Wigner $6j$ coefficients 
of the form of \eqref{eq:Wigner6j4Representations}  can be evaluated 
by contracting two basis vectors which differ only by one
representation and where two gluons are crossed,
\begin{equation}\label{eq:Wigner6j4repCalculation}
\Tr{
\left[
\raisebox{-0.4\height}{
\includegraphics[scale=0.4]{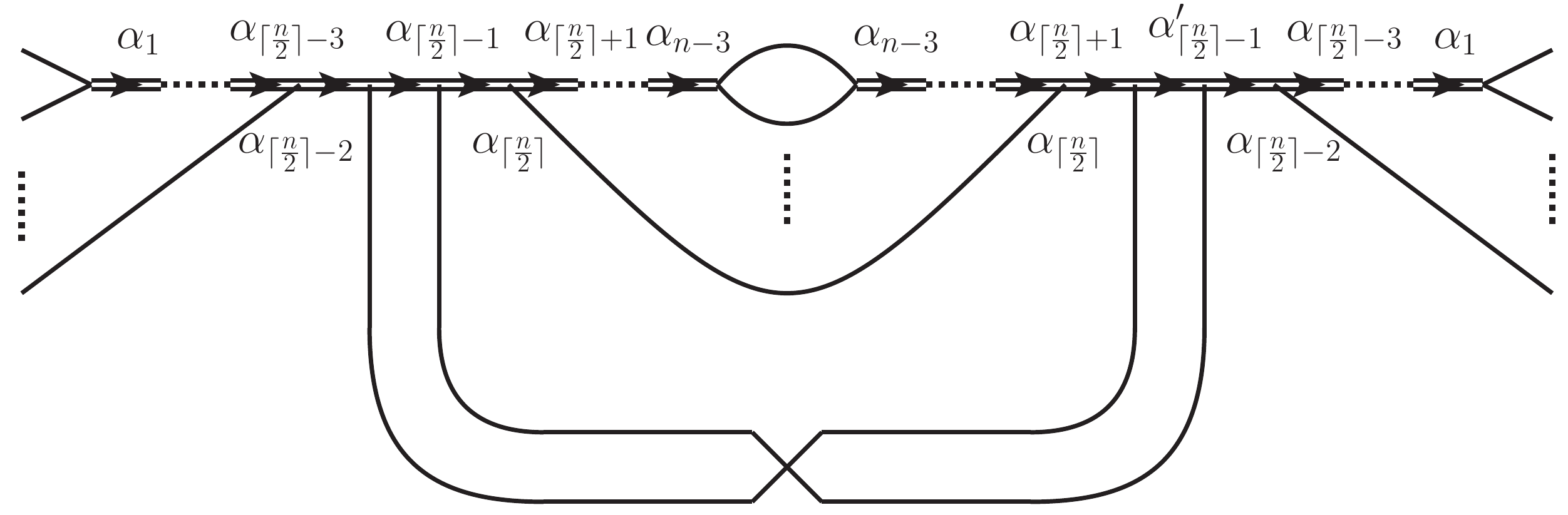}
}
\right].
}
\end{equation}
Here Schur's lemma can be applied repeatedly on the two-vertex loop in the middle, 
and on the two-vertex loop formed by the trace, giving factors of Wigner $3j$ 
coefficients over dimensions of representations. Hence 
\eqref{eq:Wigner6j4repCalculation} is equal to
\bea\label{eq:Wigner6j4repCalculationStep1}
& &
\prod_{i=1}^{\lceil\frac{n}{2}\rceil-2}{
	\frac{
		\raisebox{-0.45\height}{
		\includegraphics[scale=0.4]{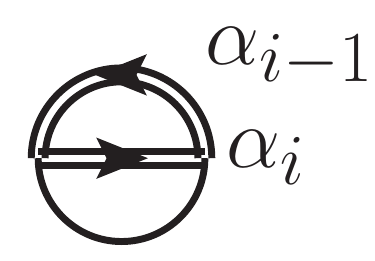}
		}
		\hspace{-3mm}
	}{
	d_{\alpha_i}
	}
}
\prod_{j=\lceil\frac{n}{2}\rceil}^{n-3}{
	\frac{
		\raisebox{-0.45\height}{
		\includegraphics[scale=0.4]{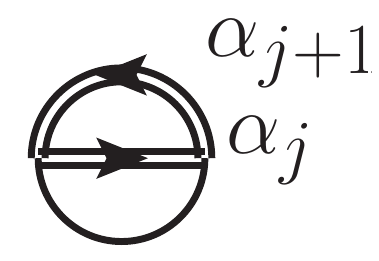}
		}
		\hspace{-3mm}
	}{
	d_{\alpha_j}
	}
}
\raisebox{-0.4\height}{
\includegraphics[scale=0.4]{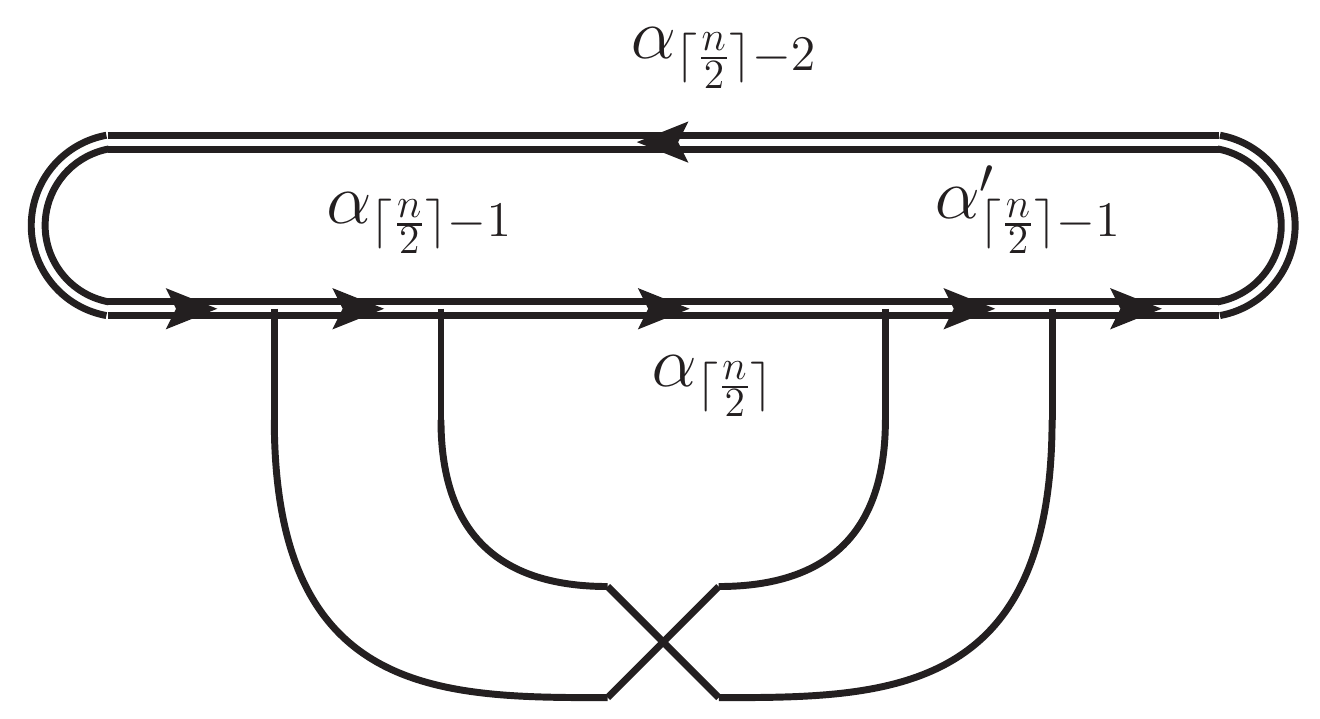}
}
\nn
&&=
\prod_{i=1}^{\lceil\frac{n}{2}\rceil-2}{
	\frac{
		\raisebox{-0.45\height}{
		\includegraphics[scale=0.4]{Figures/WignerEvaluation/WigProducts/Wig3j_LeftRight}
		}
		\hspace{-3mm}
	}{
	d_{\alpha_i}
	}
}
\prod_{j=\lceil\frac{n}{2}\rceil}^{n-3}{
	\frac{
		\raisebox{-0.45\height}{
		\includegraphics[scale=0.4]{Figures/WignerEvaluation/WigProducts/Wig3j_Middle}
		}
		\hspace{-3mm}
	}{
	d_{\alpha_j}
	}
}
\raisebox{-0.4\height}{
\includegraphics[scale=0.4]{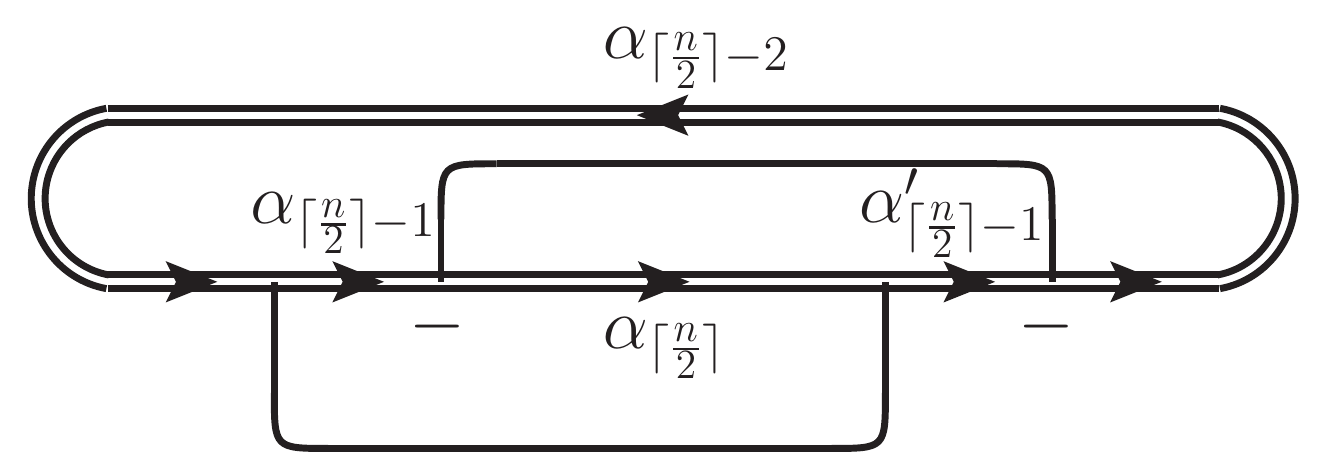}
}
\nn
&&=
\prod_{i=1}^{\lceil\frac{n}{2}\rceil-2}{
	\frac{
		\raisebox{-0.45\height}{
		\includegraphics[scale=0.4]{Figures/WignerEvaluation/WigProducts/Wig3j_LeftRight}
		}
		\hspace{-3mm}
	}{
	d_{\alpha_i}
	}
}
\prod_{j=\lceil\frac{n}{2}\rceil}^{n-3}{
	\frac{
		\raisebox{-0.45\height}{
		\includegraphics[scale=0.4]{Figures/WignerEvaluation/WigProducts/Wig3j_Middle}
		}
		\hspace{-3mm}
	}{
	d_{\alpha_j}
	}
}
\raisebox{-0.4\height}{
\includegraphics[scale=0.4]{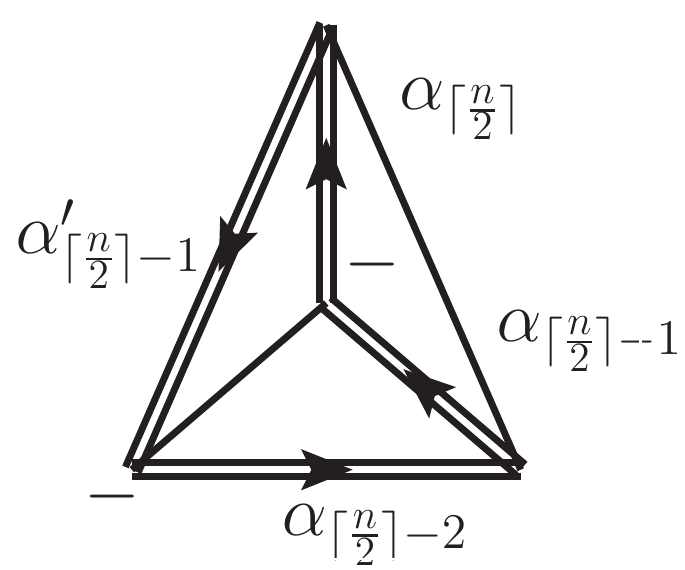}
},
\eea
where $\alpha_{0}$ and $\alpha_{n-2}$ are taken to be the octet 
representation. 
To note is that if the representations $\alpha_{\lceil\frac{n}{2}\rceil-1}$ and $\alpha_{\lceil\frac{n}{2}\rceil-1}'$ appear in basis vectors with fewer gluons, then these basis vectors can be used instead.

For the calculation of Wigner $6j$ coefficients of the form of \eqref{eq:Wigner6j3Representations},
for $n=\Ng+N_{q\bar{q}}$ partons, $n$-gluon basis vectors are 
contracted with $(n-1)$-gluon basis vectors. The difference 
of one external gluon between the two vectors is required since 
three gluons are to be contracted with one $if^{abc}$ or $d^{abc}$ vertex.
In the following a grey blob is used as a placeholder for the $if^{abc}$ and the $d^{abc}$ vertices. The vectors are contracted as
\bea\label{eq:Wigner6j3repCalculation}
& &
\Tr{
\left[
\raisebox{-0.4\height}{
\includegraphics[scale=0.4]{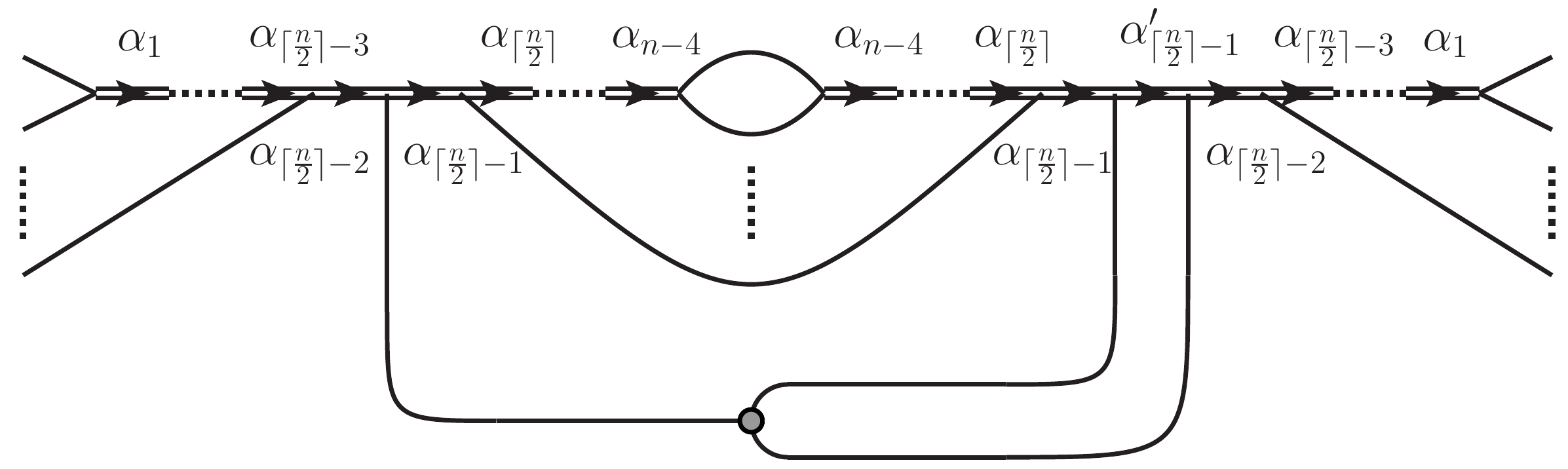}
}
\right]
}
\nn
&&=
\prod_{i=1}^{\lceil\frac{n}{2}\rceil-2}{
	\frac{
		\raisebox{-0.45\height}{
		\includegraphics[scale=0.4]{Figures/WignerEvaluation/WigProducts/Wig3j_LeftRight}
		}
		\hspace{-3mm}
	}{
	d_{\alpha_i}
	}
}
\prod_{j=\lceil\frac{n}{2}\rceil-1}^{n-4}{
	\frac{
		\raisebox{-0.45\height}{
		\includegraphics[scale=0.4]{Figures/WignerEvaluation/WigProducts/Wig3j_Middle}
		}
		\hspace{-3mm}
	}{
	d_{\alpha_j}
	}
}
\raisebox{-0.4\height}{
\includegraphics[scale=0.4]{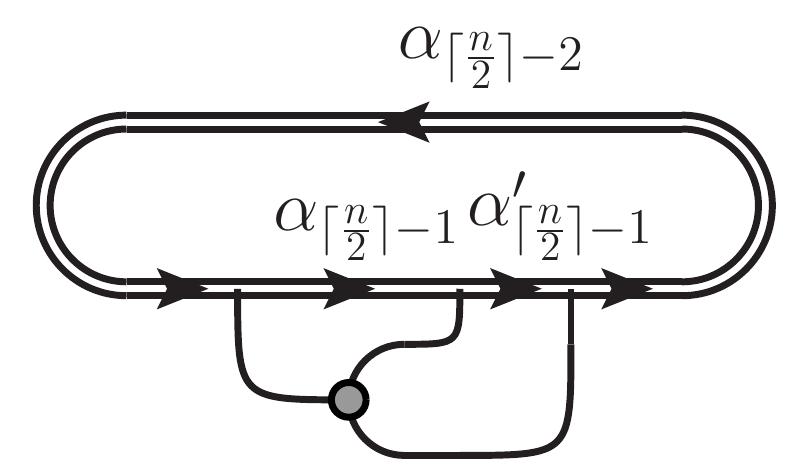}
}
\nn
&&=
\prod_{i=1}^{\lceil\frac{n}{2}\rceil-2}{
	\frac{
		\raisebox{-0.45\height}{
		\includegraphics[scale=0.4]{Figures/WignerEvaluation/WigProducts/Wig3j_LeftRight}
		}
		\hspace{-3mm}
	}{
	d_{\alpha_i}
	}
}
\prod_{j=\lceil\frac{n}{2}\rceil-1}^{n-4}{
	\frac{
		\raisebox{-0.45\height}{
		\includegraphics[scale=0.4]{Figures/WignerEvaluation/WigProducts/Wig3j_Middle}
		}
		\hspace{-3mm}
	}{
	d_{\alpha_j}
	}
}
\raisebox{-0.4\height}{
\includegraphics[scale=0.4]{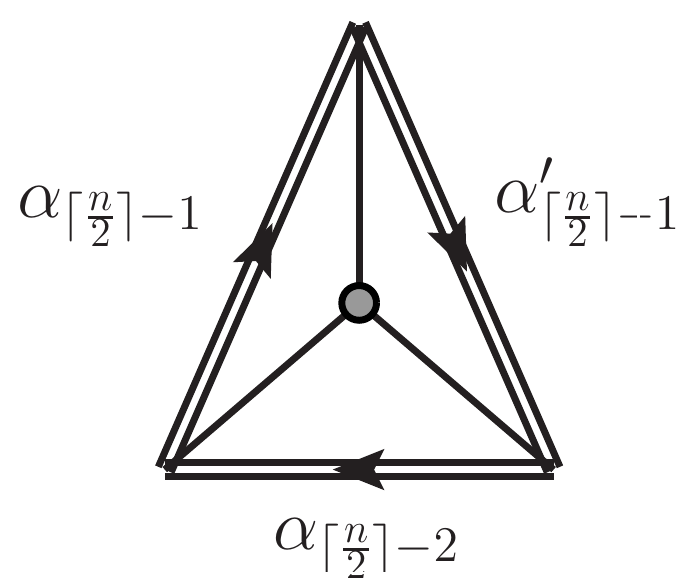}
}
,
\eea
where $\alpha_{0}$ and $\alpha_{n-3}$ are the octet representation.
That it is 
sufficient with an $(n-1)$-gluon and an $n$-gluon basis vectors 
is due to constraints that can be placed on the required 
Wigner coefficients, derived in \appref{sec:RepresentationConstraints}.
If the Feynman diagram to be decomposed contains quarks, the grey blob is 
either $if^{abc}$ or $d^{abc}$, but if it only contains gluons the Wigner coefficients with $d^{abc}$ are not required.

One might worry that the value of the Wigner coefficients of the 
higher representations could depend on the construction history of the 
vertices, such that, for example, the Wigner $6j$ coefficients 
containing a vertex with a 35-plet with the construction history 
$\alpha=(10,35)$ would differ from those with vertices constructed from 
$\alpha=(27,35)$. In the case of unique vertices, i.e., vertices 
between three representations which can only be combined to a vertex
in one way, Schur's lemma and isomorphism guarantees that this can not
happen (as long as the vertices have the same normalization and matching
sign conventions).
In the case of vertices that appear in several instances,
for example the two vertices involving the representations 35, 35 and 8,
one can prove that these sets of vertices can be chosen to give 
identical $6j$ coefficients. In \appref{sec:Uniqueness6j} we give the 
birdtrack proof of both of these statements.

\subsection{Required Wigner coefficients for LO and NLO color structures}
The Wigner coefficients of \eqref{eq:Wigner6j3Representations} and 
\eqref{eq:Wigner6j4Representations} have many symmetries which can be 
used to reduce the number of coefficients that has to be calculated. 
The symmetries of the relevant Wigner $6j$ coefficients,
proven in \appref{sec:Symmetries6js},
are:
\begin{enumerate}[(i)]
\item Rotation symmetry
\bea\label{eq:RotationalSymmetry}
\raisebox{-0.45\height}{
	\includegraphics[scale=0.45]{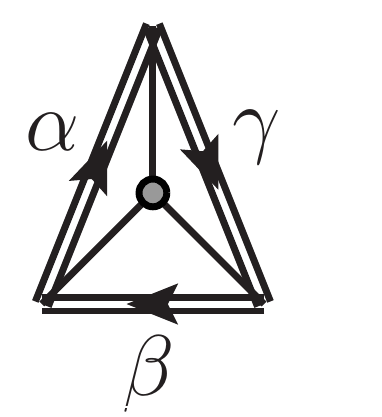}
}
\hspace{-3mm}
=
\raisebox{-0.38\height}{
	\includegraphics[scale=0.45]{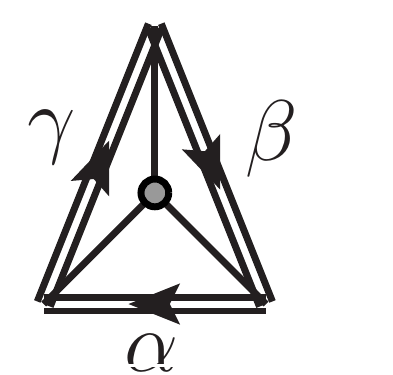}
}
\hspace{-5mm},& \;\;\;\;
\raisebox{-0.45\height}{
	\includegraphics[scale=0.45]{Figures/Wigner6js/Wig6j4Reps}
}
\hspace{-3mm}
=
\hspace{-3mm}
\raisebox{-0.33\height}{
	\includegraphics[scale=0.45]{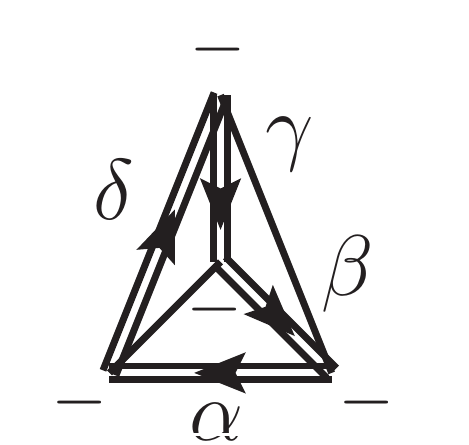}
}
.
\eea
\item Conjugation symmetry,
\bea\label{eq:ConjugationSymmetry}
\raisebox{-0.45\height}{
	\includegraphics[scale=0.45]{Figures/Wigner6js/Wig6j3RepsBoth}
}
\hspace{-3mm}
=
\raisebox{-0.38\height}{
	\includegraphics[scale=0.45]{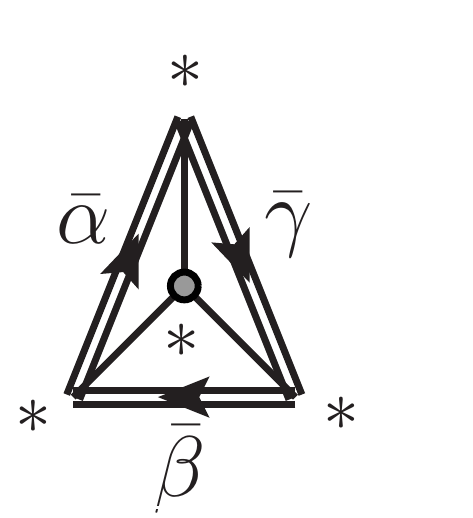}
}
\hspace{-5mm},& \;\;\;\;
\raisebox{-0.45\height}{
	\includegraphics[scale=0.45]{Figures/Wigner6js/Wig6j4Reps}
}
\hspace{-3mm}
=
\raisebox{-0.38\height}{
	\includegraphics[scale=0.45]{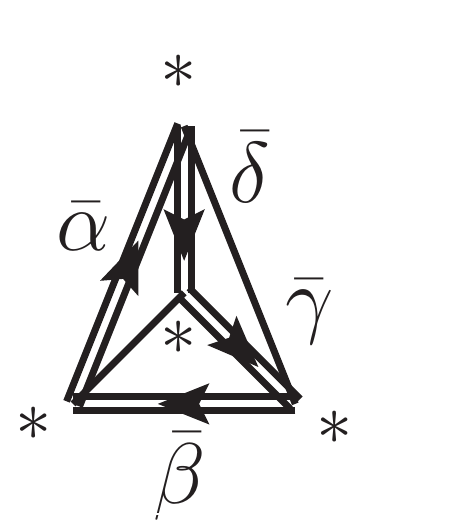}
}
\hspace{-3mm},
\eea
where the $*$ by the vertices is there to indicate that each vertex is to be 
understood as the conjugated version of the vertex on the left hand
side. 
Using the completeness relation we note that the conjugated 
vertex may in principle be a linear combination of various vertices,
\begin{equation}
\left(\raisebox{-0.41\height}{
	\includegraphics[scale=0.45]{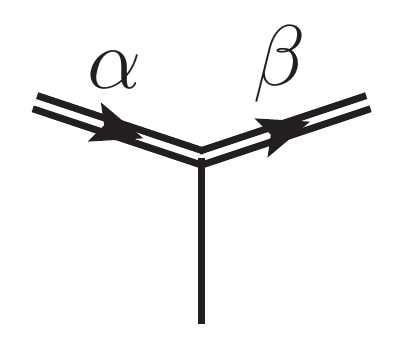}
}\right)^{*}
\hspace{-3mm}
\equiv
\raisebox{-0.41\height}{
	\includegraphics[scale=0.45]{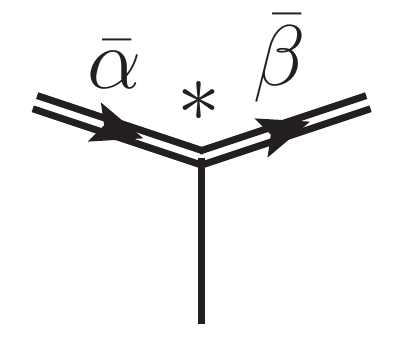}
}
\hspace{-3mm}
=
\sum_{a}
\frac{
\hspace{-1.5mm}
\raisebox{-0.45\height}{
	\includegraphics[scale=0.3]{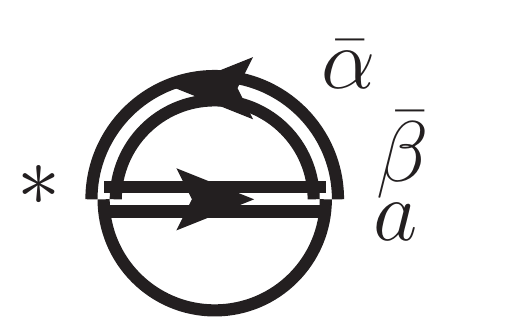}
}
\hspace{-3mm}
}{
\raisebox{-0.45\height}{
	\includegraphics[scale=0.3]{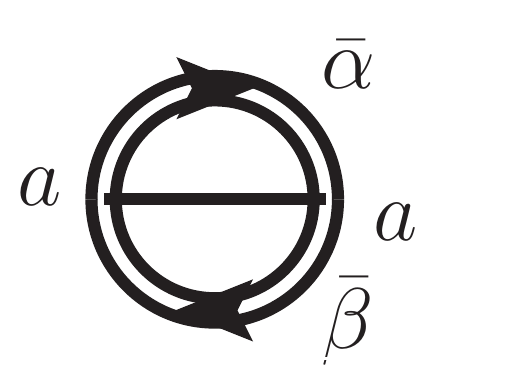}
}
\hspace{-3mm}
}
\raisebox{-0.43\height}{
	\includegraphics[scale=0.45]{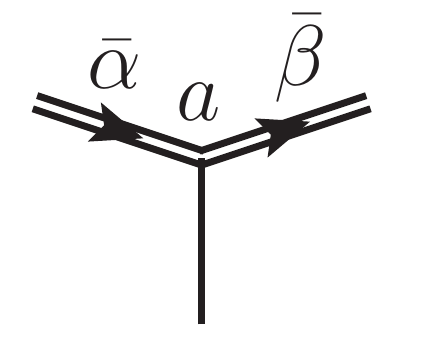}
}
\hspace{-3mm}.
\end{equation}
Note that even for real representations, conjugating internal lines
in the vertex may change it.
All vertices encountered in the present paper have been
chosen such that the sum above only contains one term, meaning
that conjugating  gives at most a minus sign, as for the triple-gluon
vertex. In general this also holds whenever $\alpha \ne \beta$,
since in $\alpha \otimes 8$, there is at most one instance of the
representation $\beta$ \cite{Keppeler:2012ih}.

\item Reversion symmetry
\bea\label{eq:MirrorSymmetry}
\raisebox{-0.45\height}{
	\includegraphics[scale=0.5]{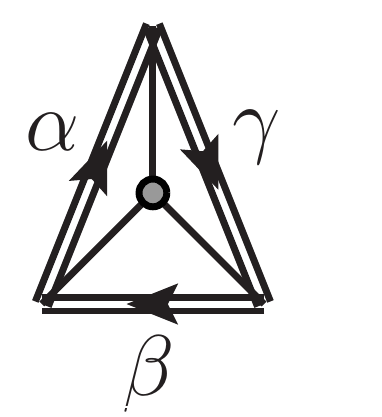}
}
\hspace{-3mm}
=
\raisebox{-0.4\height}{
	\includegraphics[scale=0.5]{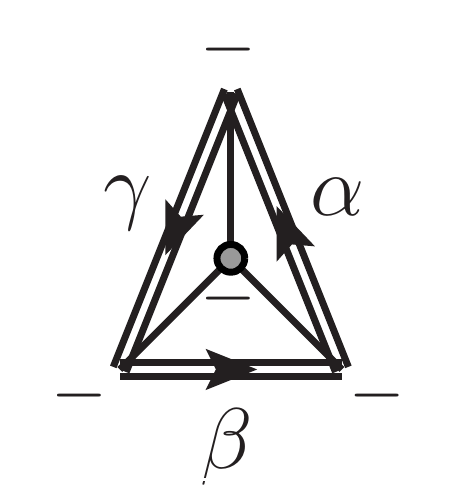}
}
\hspace{-4mm},& \;\;\;\;
\raisebox{-0.45\height}{
	\includegraphics[scale=0.5]{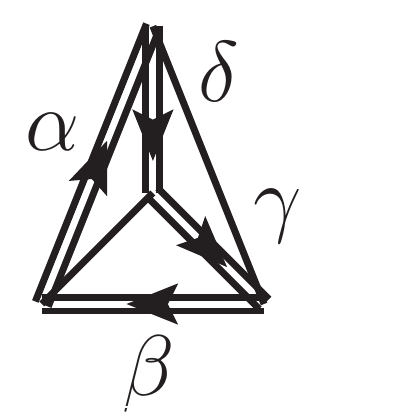}
}
\hspace{-3mm}
=
\raisebox{-0.4\height}{
	\includegraphics[scale=0.5]{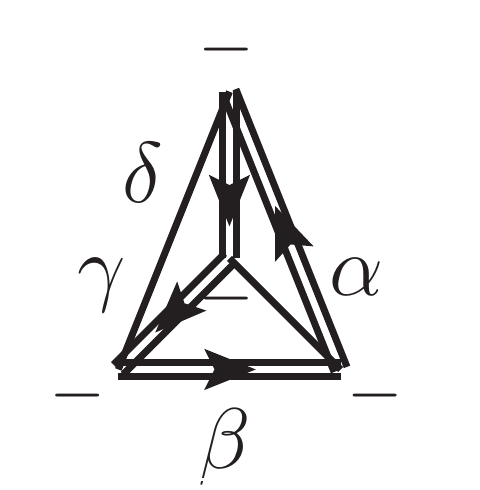}
}.
\eea
\end{enumerate}

With the symmetries stated in eqs.~(\ref{eq:RotationalSymmetry}-\ref{eq:MirrorSymmetry}), 
along with the constraints on the representations $\alpha$, $\beta$, 
$\gamma$ and $\delta$ proven in \appref{sec:RepresentationConstraints}, 
an upper limit for the required number of Wigner coefficients has been 
calculated, and is shown in \tabref{tab:N6j}. 
We note that, in comparison to the number of basis vectors (see \tabref{tab:dimension}), the number
of needed $6j$ coefficients --- which can be calculated once and for all ---
grows very slowly with the number of partons. It should nevertheless 
be remarked that the numbers in \tabref{tab:N6j}  necessarily are upper 
estimates
since 
the $6j$ coefficients depend on the choice of generalized vertices.
For the case when a representation $\gamma$ appears more than once
in the tensor product of two representations $\alpha \otimes \beta$,
such that there are several vertices connecting the representations
$\alpha$, $\beta$ and $\gamma$, there is in principle an infinite 
number of ways of defining the corresponding vertices.
Since the $6j$ coefficients depend on the vertices, this is reflected 
in the values of the $6j$ coefficients. In particular, a clever choice
of vertices may lead to the vanishing of some $6j$ coefficients which 
would not vanish for another choice. 
\begin{table}[tbp]
\centering
\begin{tabular}{ |l|r|r|r|r|r|r|r|r|r| }
\hline
$n=\Ng+N_{q\bar{q}}$ 	& 4 	& 5 	& 6 		& 7 		& 8 		& 9 		& 10 		& 11 		& 12	\\
\hline
LO gluons $\Nc = 3$ 	    & 21 & 39& 106& 152& 254& 318& 452& 536& 705\\
\hline
NLO $\Nc = 3$ 	    &29 &55& 120& 176& 272& 350& 476& 576& 733\\
\hline
LO gluons $\Nc \to \infty$ 	    &28& 68& 313& 636& 1 777& 3 095& 7 289	& 12 009 & 25 487 \\
\hline
NLO $\Nc \to \infty$     & 44& 108& 389& 808& 2 023& 3 693& 8 077	& 13 783 & 27 613 \\
\hline
\end{tabular}
\caption{\label{tab:N6j} Upper limits for the needed number of $6j$ coefficients
for tree-level gluon amplitudes and NLO color structures with up to $n$ gluons plus $\qqbar$-pairs. 
These numbers are calculated using the symmetries in 
eqs.~(\ref{eq:RotationalSymmetry}-\ref{eq:MirrorSymmetry})
and the representation constraints from \appref{sec:RepresentationConstraints}.
The actual numbers of non-vanishing $6j$ coefficients depend on the 
choice of (generalized) vertices. 
}
\end{table}

Before actually evaluating any $6j$ coefficient we need to decide on
the generalized vertices connecting three general representations.
We choose our convention such that all vertices are normalized to
one in the sense that the corresponding $3j$ coefficient is one,
\begin{equation}
  \parbox{1.5cm}{\epsfig{file=Figures/3jNormalization/Wig3j,width=1.5cm}}=1
  \;\mbox{ (for values of } 6j  \mbox{ coefficients)}.
  \label{eq:3jNormalization1}
\end{equation}
Especially we remark that this normalization is applied to the 
antisymmetric triple-gluon vertex

\begin{equation}
\parbox{1cm}{\epsfig{file=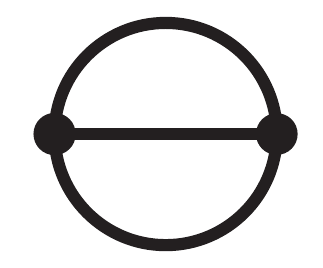,width=1cm}}=1
  \;\mbox{ (for values of } 6j  \mbox{ coefficients)}.
  \label{eq:3j888Normalization1}
\end{equation}
giving the vertex 
\begin{equation}
\parbox{2cm}{\epsfig{file=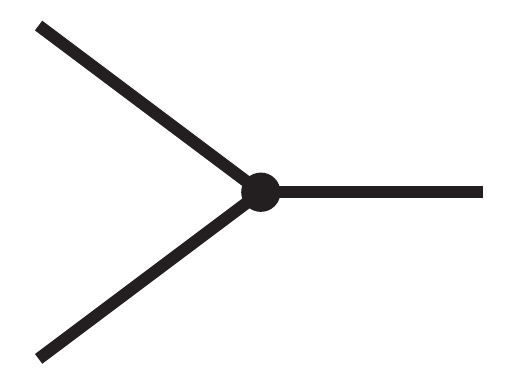,width=2cm}}
=
\frac{1}{\sqrt{2 \Nc \TR (\Nc^2-1)}}
\parbox{2cm}{\epsfig{file=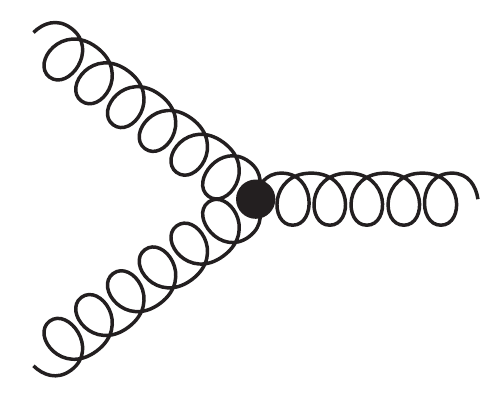,width=2cm}}
\;\mbox{ (for values of } 6j  \mbox{ coefficients)}
.
\end{equation}

\include{./Tables/ExplicitWigTable}

As an example, the non-vanishing Wigner $6j$ coefficients required for
QCD color structures appearing in calculations up to NLO with up to four external gluons 
plus $\q\qbar$-pairs are shown in 
\tabref{tab:4GluonWignerCoefficients}. Vanishing $6j$ coefficients,
$6j$ coefficients related by symmetries,
and $6j$ coefficients only required for $\Nc \geq4$ have been omitted. 
The sign conventions for the vertices involving two different representations
of $10$, $\overline{10}$, $27$ and $0$
and one octet are given by normalizing the vertices
\begin{eqnarray}
  \label{eq:a_vertex}
  \parbox{2cm}{\epsfig{file=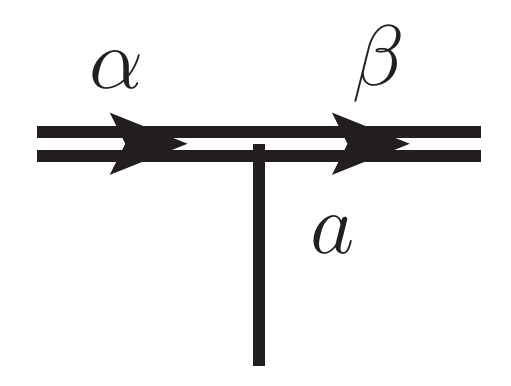,width=2cm}}\propto
  \parbox{3cm}{\epsfig{file=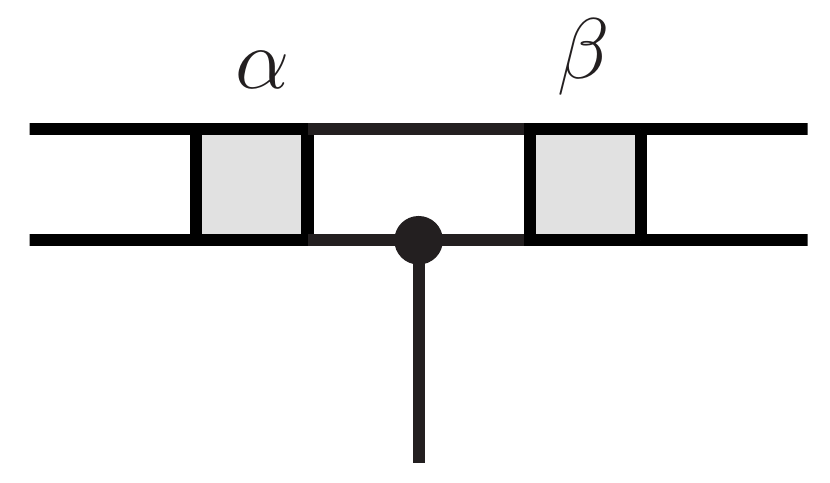,width=3cm}}
\end{eqnarray}
s.t. the normalization constant is positive (and fixed by requiring 
the $3j$ coefficient to be one).

For the two vertices involving 27, 8, and 27 one vertex is chosen
antisymmetric (a) under exchange of vertex order, and is constructed
as in \eqref{eq:a_vertex}, whereas the other vertex is chosen
symmetric (s), and proportional to 
\begin{eqnarray}
  \label{eq:s_vertex}
  \parbox{2cm}{\epsfig{file=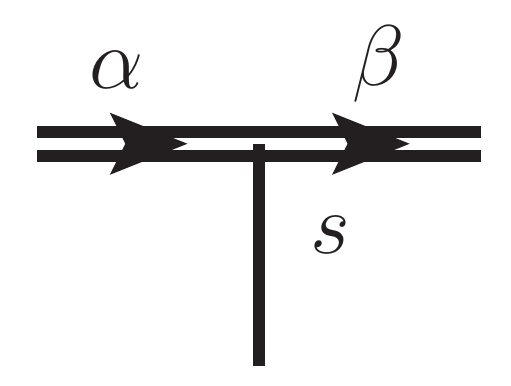,width=2cm}}\propto
  \parbox{3cm}{\epsfig{file=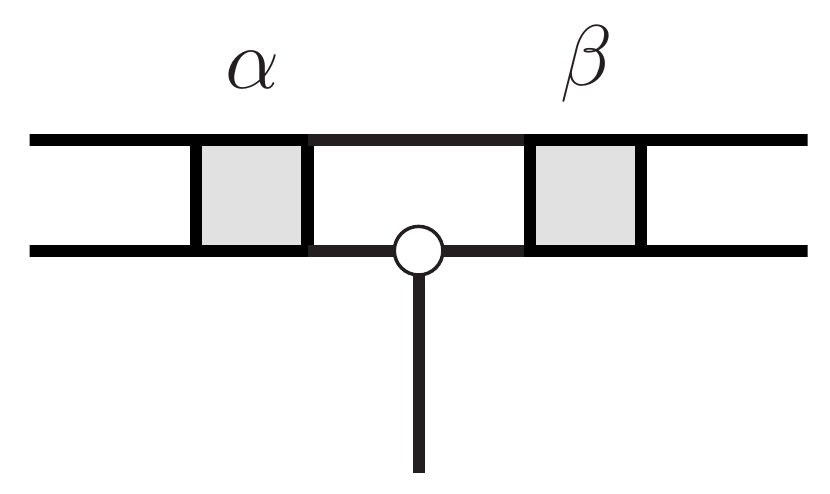,width=3cm}}
\end{eqnarray}
with positive proportionality constant.
For higher $\Nc$, the vertices between 0, 8, and 0,
are defined similarly.
For the decuplet vertices involving 10, 8 and 10, one vertex is chosen
as in \eqref{eq:a_vertex}, whereas the remaining
vertex --- which only appears for $\Nc\geq 4$ --- is given by the 
orthogonal linear combination of  
\eqref{eq:a_vertex} and \eqref{eq:s_vertex} with 
positive constant in front of \eqref{eq:s_vertex}
and negative in front of  \eqref{eq:a_vertex}.
The corresponding vertices with anti-decuplets, involving 
$\overline{10}$, 8 and $\overline{10}$, are defined as 
the conjugate of the decuplet versions, which implies a minus
sign in front of \eqref{eq:a_vertex}, for both cases.
These signs are also encoded in the electronically appended
five-gluon basis.

The sign conventions of the other vertices, which appear twice 
in the 6j coefficients in \tabref{tab:4GluonWignerCoefficients},
do not change the sign of the coefficients. 

Going beyond four gluons plus $\qqbar$-pairs, the full set of 
$6j$ coefficients required for up to NLO calculations
with up to six gluons plus $\qqbar$-pairs are attached in a
human and Mathematica readable .m-file.
Here the relevant sign conventions are
defined by the appended six- and seven-gluon basis vectors.
Since we view the basis vectors as constructed from
vertices, the usage of the birdtrack method imposes 
sign correlations between the basis vectors. For this 
reason --- and since we require that conjugating a vertex should 
give the vertex with conjugated representations when possible  
--- some six-gluon basis vectors have changed sign
w.r.t. the basis published along with \cite{Keppeler:2012ih}.

\section{Conclusion and outlook}
\label{sec:conclusion}

We have demonstrated how QCD color structure elegantly can be decomposed 
into multiplet bases with the aid of Wigner $3j$ and $6j$ coefficients, 
which we have shown how to calculate using the multiplet bases 
from \cite{Keppeler:2012ih}.

We have also argued that only a relatively small set of such coefficients
are needed, and that the number of required coefficients is severely
reduced for $\Nc=3$ compared to the limit $\Nc\to\infty$. 
For leading and next to leading order processes and up to six gluons 
plus $\qqbar$-pairs we have explicitly evaluated all 
necessary coefficients.

We remark that although the discussion in the present paper
has focused on the decomposition of color structure associated
with Feynman diagrams, the same principle can be applied in
other contexts, such as the color structure of gluon emission for 
parton showers, the
calculation of soft anomalous dimension matrices, or recursive 
approaches to scattering amplitudes \cite{Du:2015apa}.

\acknowledgments
We thank Johan Gr\"onqvist and Stefan Keppeler for useful
comments.
This work was supported by the Swedish Research Council 
(contract number 621-2012-27-44 and  621-2013-4287) and 
in part by the MCnetITN FP7 Marie Curie Initial Training
Network, contract PITN-GA-2012-315877.

\appendix

\section{Representations in $6j$ coefficients}
\label{sec:RepresentationConstraints}
This appendix contains a proof that Wigner coefficients of the form of \eqref{eq:Wigner6j3Representations} and \eqref{eq:Wigner6j4Representations} are sufficient for decomposing QCD color structures  by the method of \secref{sec:decomposition}. 
For the proof, specific choices of loops in the contraction procedure 
are examined.
It is shown that such loops can always be found in the original vacuum bubble, and in every subsequent step.
The last part of this appendix puts constraints on the representations occurring in the Wigner coefficients, for both tree-level and NLO color structures.
We stress that the contraction strategy of this 
appendix should be viewed as a proof of the constraints that can be put on the Wigner coefficients, and not as a suggestion for implementation of the method.
In actual examples, one can often, as in \secref{sec:ExplicitExample}, choose contractions such that most loops only involve two or three vertices.

For this appendix it will be important to differentiate between the adjoint representations appearing in the color structure to be decomposed, and the arbitrary representations in the basis vector or from completeness relations (which may well be adjoint representations). In this appendix any mention of the adjoint representation will exclusively refer to the adjoint representations from the initial color structure, i.e., they are adjoint representations for all of the basis vectors.
 The labeled representations, e.g., $\alpha_1$, $\alpha_2$ and $\alpha_3$ in \figref{fig:6gVector}, are referred to as the arbitrary representations.

\subsection{Gluon-only color structures}
\label{app:GluonOnlyCase}
To completely contract vacuum bubbles with both external quarks 
and gluons, three different types of loops will be required. In the following, 
tree-level and NLO gluon-only color structures are handled first. 
The additional type of loops required for an arbitrary number of quarks 
is dealt with afterwards. The vacuum bubble for LO gluon-only color structures
 will always contain at least two loops of the form 
\begin{equation}\label{eq:Loop1}
\raisebox{-0.4\height}{
	\includegraphics[scale=0.45]{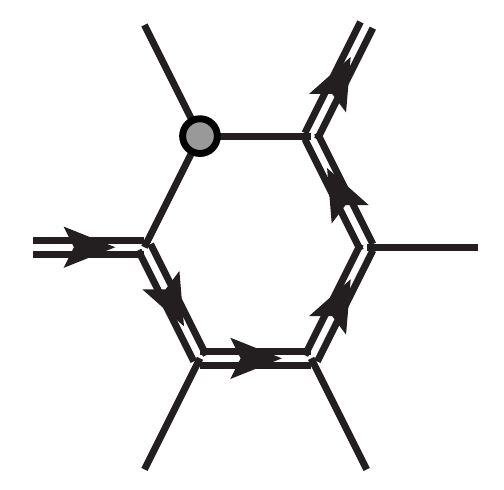}
}
,
\end{equation}
where the characterizing feature is that there is only one 
vertex from the initial color structure (the gray blob,
in the gluon-only case this is always $if^{abc}$). The total number of vertices in the loop does not matter for the principle. 
To reduce such a loop, the completeness relation, \eqref{eq:CRDiagrammatic}, and the contraction of a vertex correction, \eqref{eq:VertexCorrection}, can be applied to the two red representations in
\begin{equation}\label{eq:CRApplicationLoopType1}
  \raisebox{-0.4\height}{
    \includegraphics[scale=0.45]{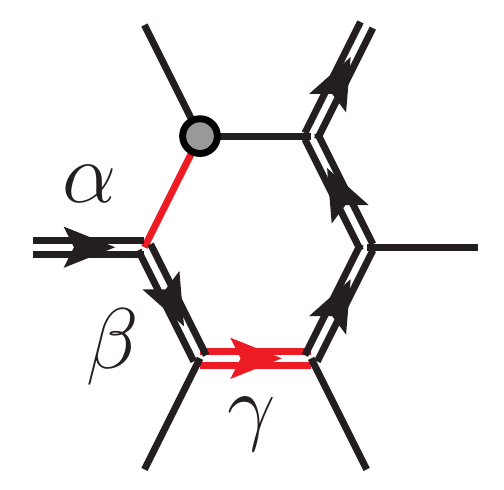}
  }
  =
  \sum_{\psi}{
    \frac{d_{\psi}}{
      \hspace{0.5mm}
      \raisebox{-0.45\height}{
	\includegraphics[scale=0.4]{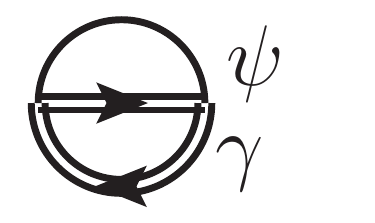}
      }
      \hspace{-3mm}
    }
    \raisebox{-0.4\height}{
      \includegraphics[scale=0.45]{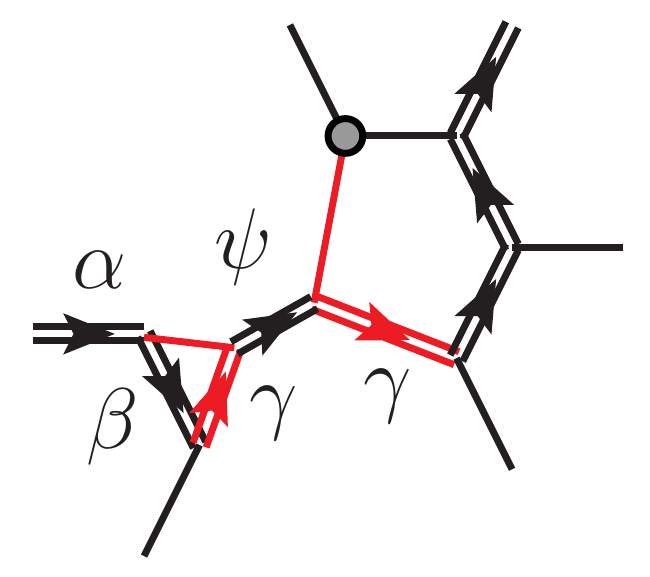}
    }
  }
=
\sum_{\psi,a}{
		\frac{d_{\psi}}{
		\hspace{0.5mm}
		\raisebox{-0.45\height}{
			\includegraphics[scale=0.4]{Figures/Representations/Wig3jCR}
		}
		\hspace{-3mm}
	}
	\frac{
		\hspace{-1mm}
		\raisebox{-0.1\height}{
			\includegraphics[scale=0.4]{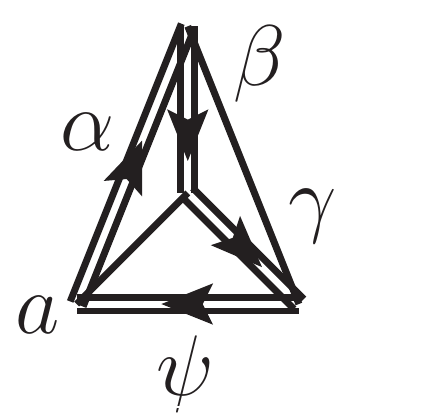}
		}
		\hspace{-5mm}
	}{
		\hspace{0.5mm}
		\raisebox{-0.45\height}{
			\includegraphics[scale=0.4]{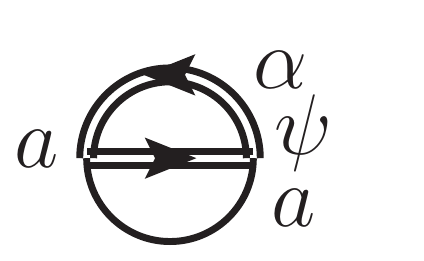}
		}
		\hspace{-4mm}
	}
\raisebox{-0.4\height}{
	\includegraphics[scale=0.45]{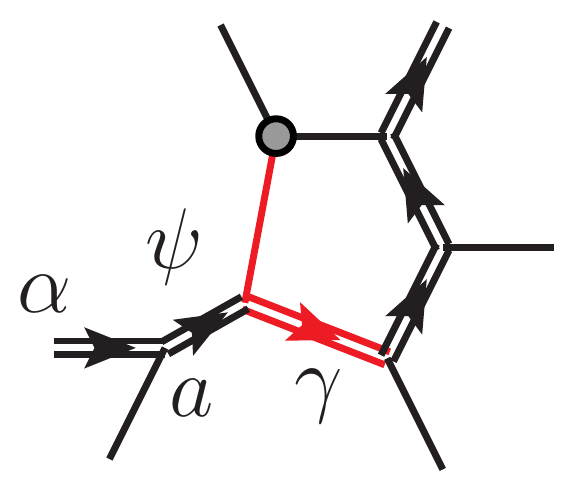}
}
}
.
\end{equation}
Repeating this procedure will result in a vertex correction 
containing the gray blob (for the loop in the above example, this 
step would need to be repeated two more times). The vertex correction with the gray blob gives
\begin{equation}\label{eq:CRApplicationLoopType1LastCR}
\raisebox{-0.49\height}{
	\includegraphics[scale=0.4]{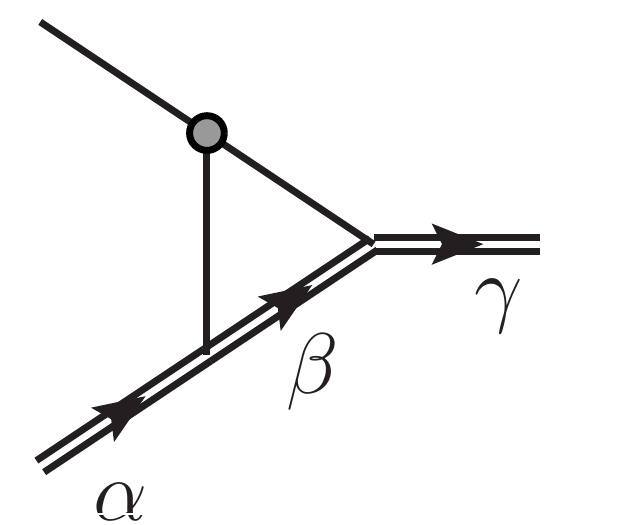}
}
\hspace{-3mm}
=
\sum_{a}{
\frac{
		\hspace{-0.5mm}
		\raisebox{-0.1\height}{
			\includegraphics[scale=0.4]{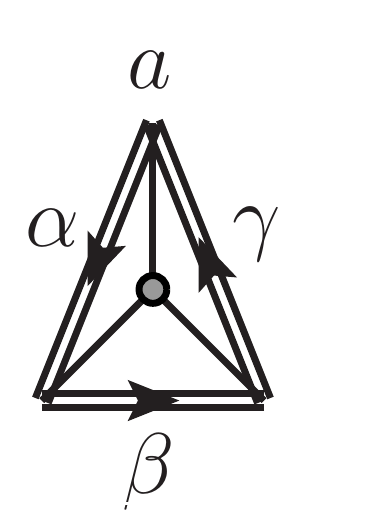}
		}
		\hspace{-4mm}
	}{
		\hspace{-0.5mm}
		\raisebox{-0.45\height}{
			\includegraphics[scale=0.4]{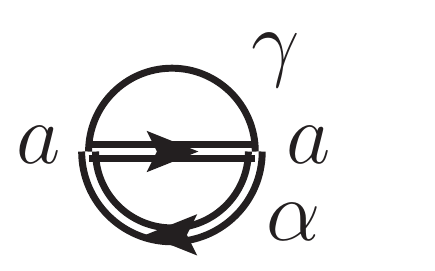}
		}
		\hspace{-4mm}
}
\raisebox{-0.5\height}{
	\includegraphics[scale=0.4]{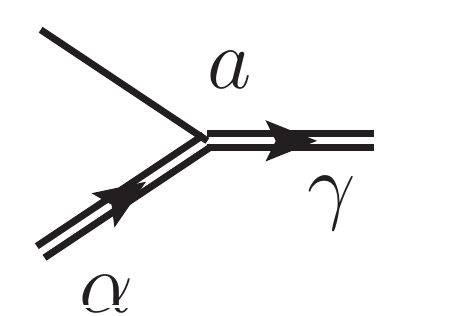}
}
}
\hspace{-3mm}
,
\end{equation}
for some representations $\alpha$, $\beta$ and $\gamma$.
This last step removes two vertices, one gray blob, i.e., a vertex from the initial color structure and one vertex between arbitrary representations. 
Since every loop that is contracted removes one vertex of each kind, 
the resulting vacuum bubble is topologically equivalent to a vacuum bubble 
for a tree-level color structure with $n-1$ external gluons. 
After a loop of the form of \eqref{eq:Loop1} has been contracted, there must thus exist at least two loops of the type in \eqref{eq:Loop1} in the resulting color structure by the above argument. Hence any LO gluon-only color structure can be completely contracted by repeatedly contracting loops of the form of \eqref{eq:Loop1}.

Only choosing loops of the form of \eqref{eq:Loop1} 
is sufficient for tree-level gluon-only color structures. For higher orders, 
it is not always possible to choose loops of this form. At NLO this happens
for diagrams where all external gluons are attached to the loop, such as
\begin{equation}\label{eq:NLOAllLoop}
  \raisebox{-0.4\height}{
    \includegraphics[scale=0.4]{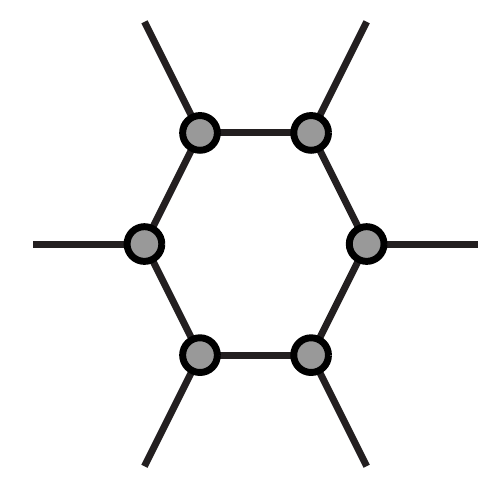}
  }.
\end{equation}
For all other gluon-only NLO color structures there is at least one $if^{abc}$
with two uncontracted indices, meaning that a loop of the from
in \eqref{eq:Loop1} can be found, such that it is possible to contract
loops as in \eqref{eq:CRApplicationLoopType1}.
For color structures of the form of \eqref{eq:NLOAllLoop}, there always exists loops of the form
\begin{equation}\label{eq:Loop3}
\raisebox{-0.4\height}{
	\includegraphics[scale=0.5]{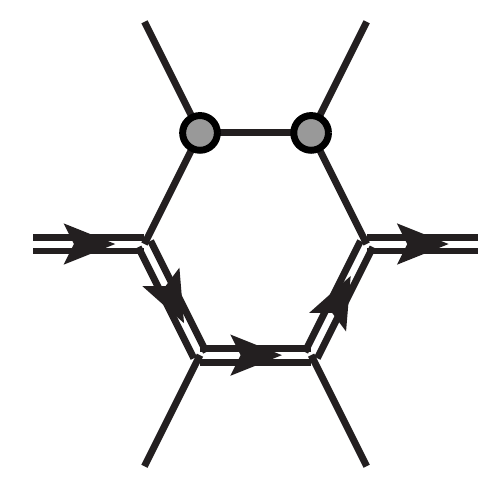}
}.
\end{equation}
Similarly to the loop in \eqref{eq:Loop1}, the steps detailed in 
\eqref{eq:CRApplicationLoopType1} remain valid. However, at the end,
instead of contracting a loop of the form in \eqref{eq:CRApplicationLoopType1LastCR},
a loop with four vertices is encountered,
\bea\label{eq:CRApplicationLoopType3}
\raisebox{-0.5\height}{
	\includegraphics[scale=0.4]{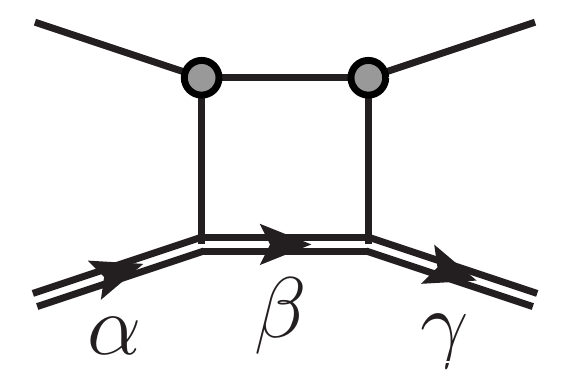}
}
\hspace{-3mm}
&=&
\sum_{\psi}{
\frac{d_\psi}{
	\hspace{-0.5mm}
	\raisebox{-0.45\height}{
		\includegraphics[scale=0.4]{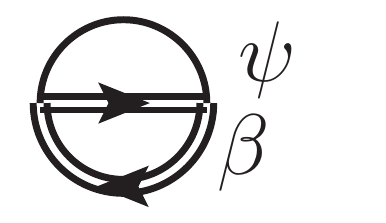}
	}
	\hspace{-4mm}
}
\raisebox{-0.5\height}{
	\includegraphics[scale=0.4]{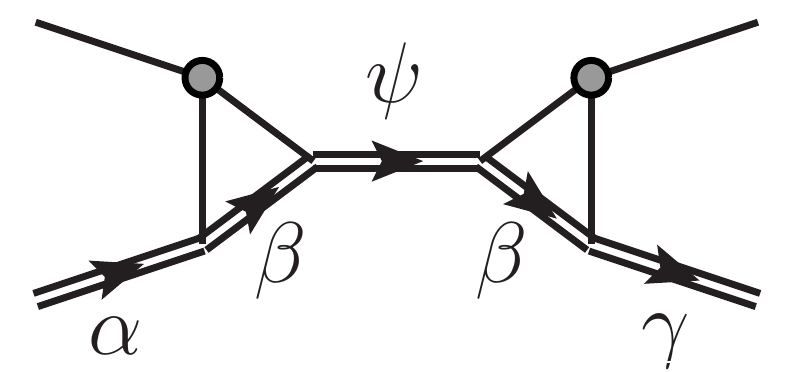}
}
}
\hspace{-3mm}
\nn
&=&
\sum_{\psi,a,b}{
\frac{d_\psi}{
	\hspace{-0.5mm}
	\raisebox{-0.45\height}{
		\includegraphics[scale=0.4]{Figures/Representations/NLO/Wig3j_CR}
	}
	\hspace{-4mm}
}
\frac{
		\hspace{-0.5mm}
		\raisebox{-0.1\height}{
			\includegraphics[scale=0.4]{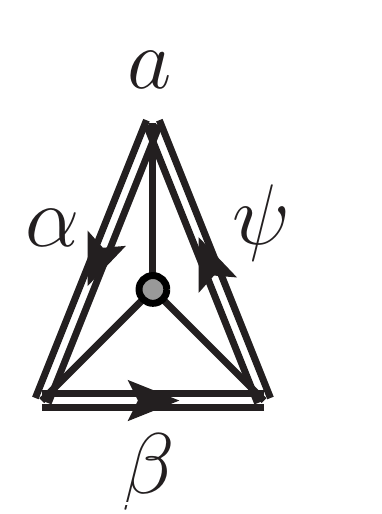}
		}
		\hspace{-4mm}
	}{
		\hspace{-1mm}
		\raisebox{-0.45\height}{
			\includegraphics[scale=0.4]{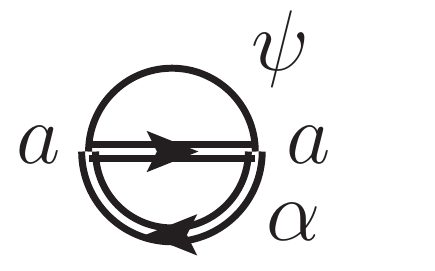}
		}
		\hspace{-4mm}
}
\frac{
		\hspace{-0.5mm}
		\raisebox{-0.1\height}{
			\includegraphics[scale=0.4]{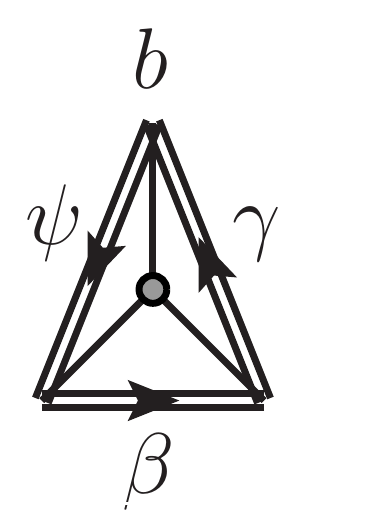}
		}
		\hspace{-4mm}
	}{
		\hspace{-1mm}
		\raisebox{-0.45\height}{
			\includegraphics[scale=0.4]{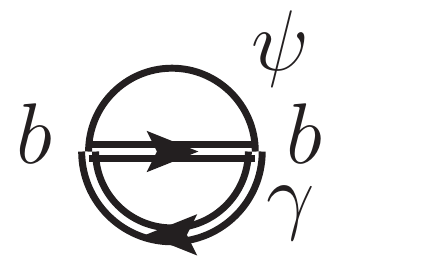}
		}
		\hspace{-4mm}
}
\raisebox{-0.49\height}{
	\includegraphics[scale=0.4]{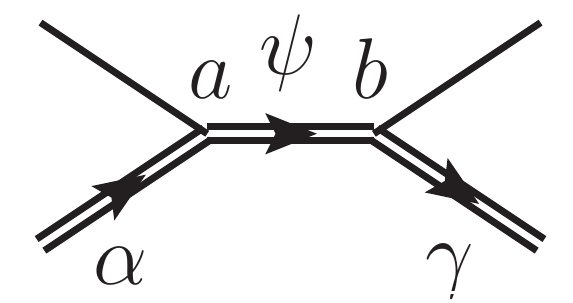}
}
}
\hspace{-1.5mm}
.
\eea
Handling a loop like this removes two vertices from the color structure, 
and none from the chain of arbitrary representations.
Since an NLO 
color structure has two vertices more than a tree-level color structure, 
the topology after the contraction (remembering that the two triple-gluon 
vertices of \eqref{eq:Loop3} are in the loop of the NLO color structure) 
is equivalent to that of a tree-level color structure.
Note that a loop 
of the type in \eqref{eq:Loop3}, need not be the first loop to be contracted 
(most NLO diagrams contain no loop of the form in \eqref{eq:NLOAllLoop}), 
but might at some step be encountered, and necessary to contract to 
continue the contraction of the vacuum bubble. In fact, all loop contractions apart from one, where \eqref{eq:Loop3} is inserted to break the NLO loop, can be of loops of the form in \eqref{eq:Loop1}.
In this way, we can thus contract any gluon-only NLO diagram.

Color structures of arbitrary order in perturbation theory can be decomposed by contracting similar loops to \eqref{eq:Loop1} and \eqref{eq:Loop3}, in general with three or more vertices from the color structure, instead of one or two, respectively. These have been avoided above in order to be able to put constraints on the representations appearing in the Wigner coefficients.

When dealing with internal quark loops,
the terms coming from the first term on the right hand side of 
\eqref{eq:InternalqLines}, appearing when removing the internal quark loop, 
will give rise to loops of the form
\begin{equation}\label{eq:Loop2}
\raisebox{-0.4\height}{
	\includegraphics[scale=0.5]{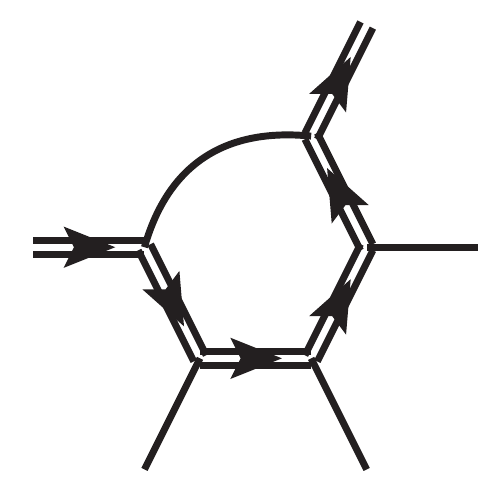}
}
\hspace{-1mm}.
\end{equation}
Also for this loop,
the first step for the first loop type, \eqref{eq:CRApplicationLoopType1}, 
can be performed analogously. The difference again occurs in the last 
step, where a vertex correction of a different type remains of the loop,
\begin{equation}
  \label{eq:CRApplicationLoopType2LastCR}
    \raisebox{-0.39\height}{
      \includegraphics[scale=0.4]{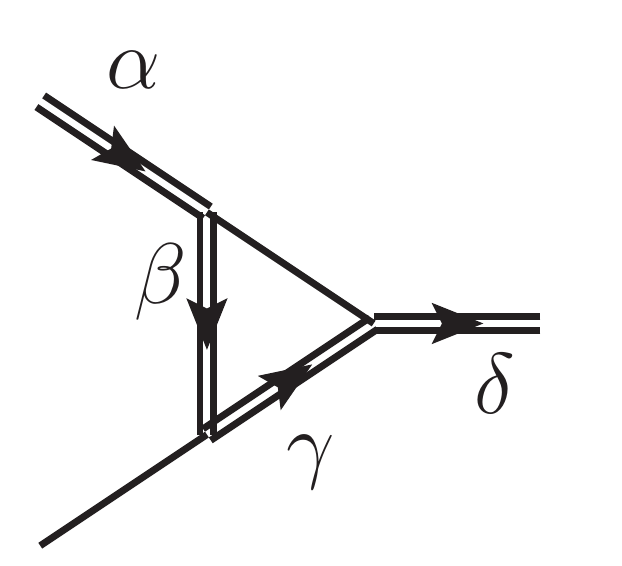}
    }
    \hspace{-3mm}
    =
    \sum_{a}{
      \frac{
	\hspace{-0.5mm}
	\raisebox{-0.1\height}{
	  \includegraphics[scale=0.4]{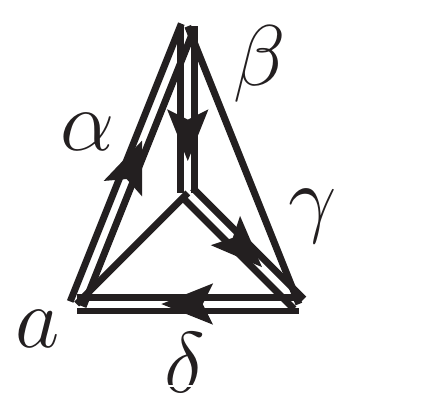}
	}
	\hspace{-4mm}
      }{
	\hspace{-1mm}
	\raisebox{-0.45\height}{
	  \includegraphics[scale=0.4]{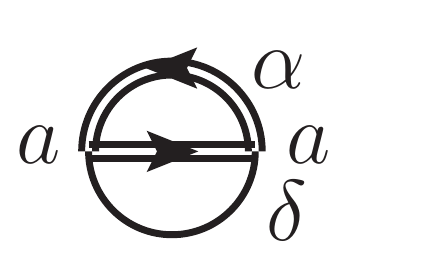}
	}
	\hspace{-4mm}
      }
      \raisebox{-0.33\height}{
	\includegraphics[scale=0.4]{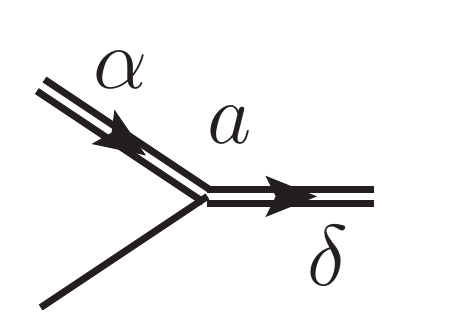}
      }
    }
\hspace{-2.5mm}
.
\end{equation}
In this step two vertices from the arbitrary representations are removed, 
and none from the initial color structure which, however, already, for these ``singlet'' terms, contain two vertices less. Thus contracting loops of this type allows for gluon-only color structures with an internal quark loop to be completely contracted as well.

\subsection{Color structures with external quarks}
For color structures with quark-antiquark pairs, the color structure after using \eqref{eq:InternalqLines} is not guaranteed to have any loops of the form of \eqref{eq:Loop1} or \eqref{eq:Loop2}. 
As will be argued below, the consequence of this is that loops of the type in \eqref{eq:Loop3} are required also at LO.
 When the quarks are combined into octet or singlet representations in the basis vectors to match the basis vectors in \figref{fig:6gVector} (b) and (c), they will form traces over some number of generators. These quark traces can be simplified by using completeness relations and removing loops. As we will see, after the quarks have been removed, an amplitude containing only gluons and $if^{abc}$ and $d^{abc}$ vertices remains. This gluon-only color structure will be of the same order as the starting color structure with quarks (or lower), hence it can be contracted as described in \secref{app:GluonOnlyCase} for LO and NLO color structures.

The color structure of a general QCD amplitude with $N_g$ gluons and $N_{q\bar{q}}$ quark-antiquark pairs will be of the form
\begin{equation}\label{eq:GeneralQCDAmplitude}
	| \Col \rangle
	=
	\raisebox{-0.45\height}{
		\includegraphics[scale=0.4]{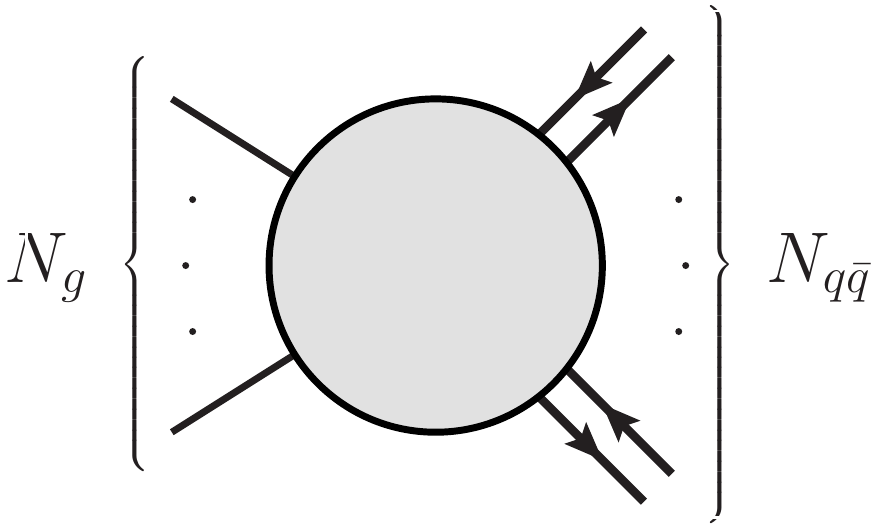}
	}
,
\end{equation}
where the gray blob can be of arbitrary order in perturbation theory.
Considering one of the external quarks, with some label $i$, the amplitude can be written as
\begin{equation}\label{eq:GeneralQCDAmplitudeQuarki}
	| \Col \rangle
	=
	\raisebox{-0.42\height}{
		\includegraphics[scale=0.4]{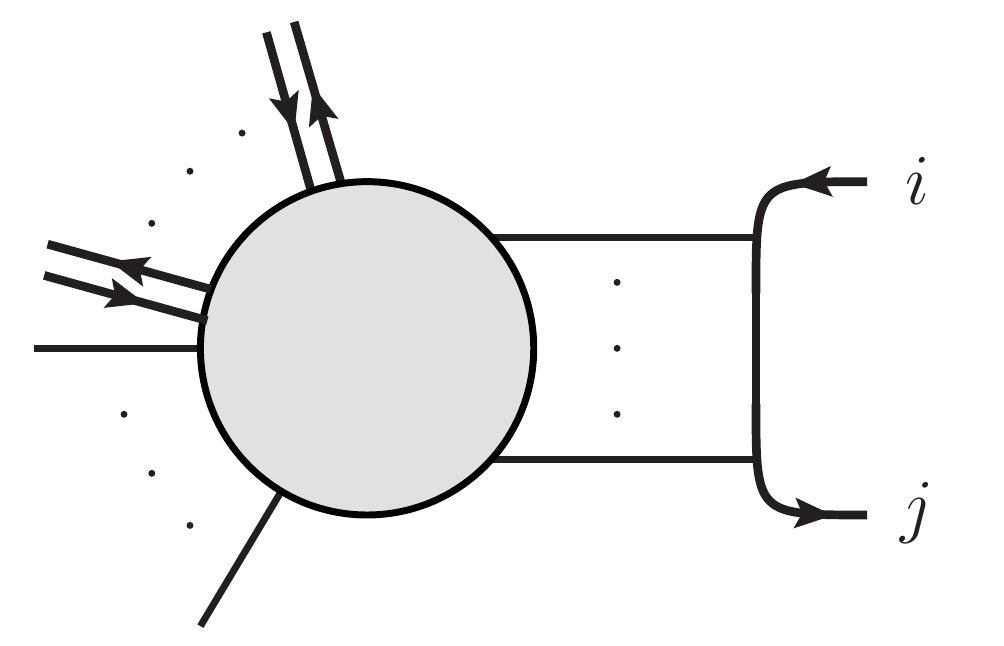}
	}
,
\end{equation}
where $j$ is one of the external antiquarks and the dots to the right represent that there can be any number of gluons connecting to the rest of the color structure. 

To manipulate \eqref{eq:GeneralQCDAmplitudeQuarki}, a completeness relation, \eqref{eq:CRDiagrammatic}, can be applied to the open quark-line with $i$ and $j$ (analogously to \eqref{eq:InternalqLines}) giving 
\begin{equation}\label{eq:QuarkLineSimplification}
	\raisebox{-0.45\height}{
		\includegraphics[scale=0.4]{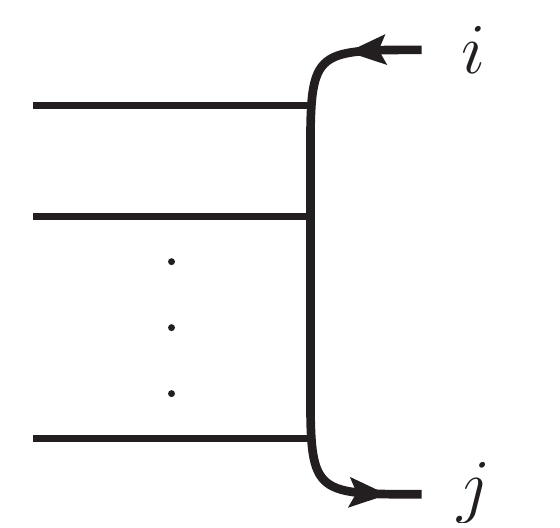}
	}
	=
	\frac{
		T_R
	}{
		\hspace{-1mm}
		\raisebox{-0.45\height}{
			\includegraphics[scale=0.3]{Figures/InternalQuarks/Wig3j_Singlet}
		}
	}
	\raisebox{-0.45\height}{
		\includegraphics[scale=0.4]{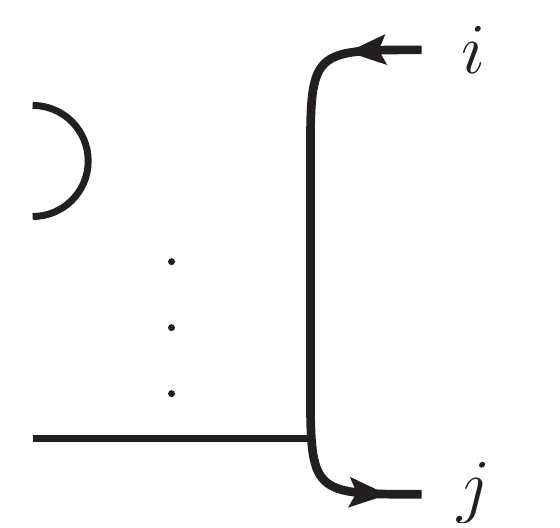}
	}
	+
	\frac{1}{2}
	\raisebox{-0.45\height}{
		\includegraphics[scale=0.4]{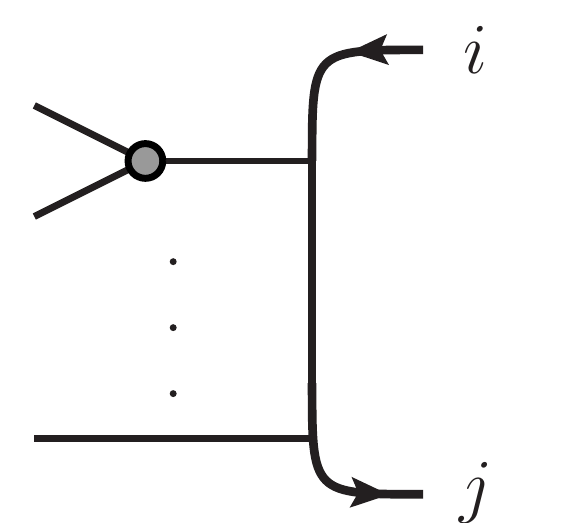}
	}
,
\end{equation}
where the big blob has been suppressed and the small gray blob represents $if^{abc}+d^{abc}$. Repeated application of \eqref{eq:QuarkLineSimplification} allows for the color structure to be written as a sum over different contributions where quark $i$ and antiquark $j$ are in either a singlet or an octet, i.e. \eqref{eq:GeneralQCDAmplitudeQuarki} can be written as
\begin{equation}\label{eq:GeneralQCDAmplitudeQuarkiSingletOctet}
	| \Col \rangle
	=
	\sum{
		\raisebox{-0.45\height}{
			\includegraphics[scale=0.4]{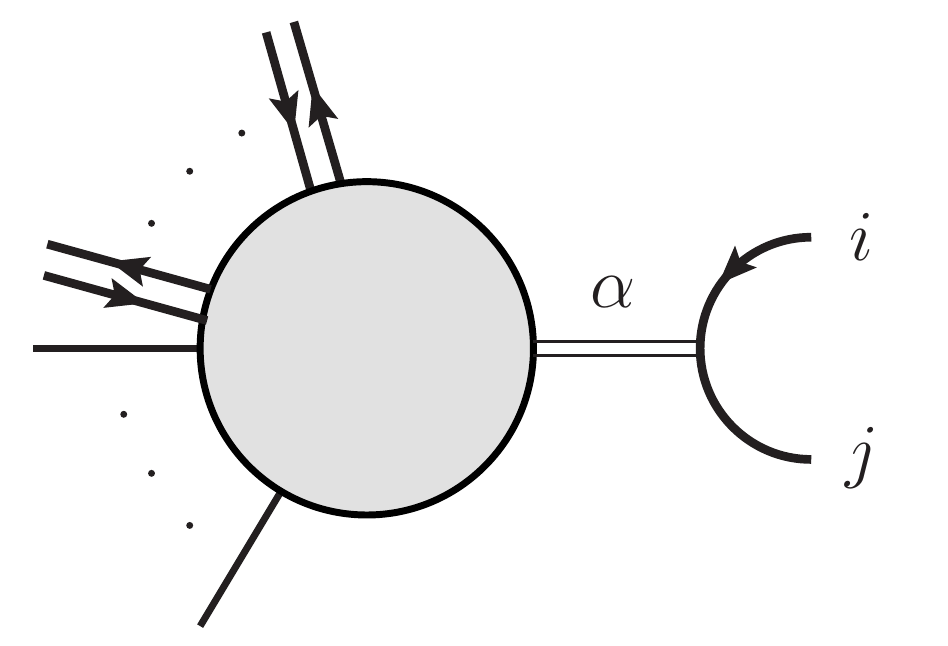}
		}
	}
,
\end{equation}
where the thin double line $\alpha$ is either a singlet or an octet. 
If this step is performed for every incoming quark, the total color structure is of the form
\begin{equation}\label{eq:GeneralQCDAmplitudePairedqqbar}
	| \Col \rangle
	=
	\sum{
		\raisebox{-0.45\height}{
			\includegraphics[scale=0.4]{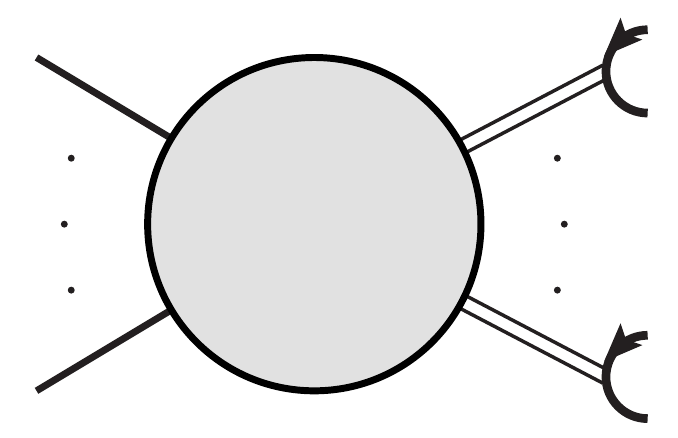}
		}
	}
.
\end{equation}
The order of the gray blob is the same as for the original gray 
blob in \eqref{eq:GeneralQCDAmplitude}, or lower, since 
\eqref{eq:QuarkLineSimplification} does not introduce any loops, 
but contains a singlet piece.
Consider a specific $\qqbar$-pair in $| \Col \rangle$ that is now 
in a common representation, for example $i$ and $j$ in 
\eqref{eq:GeneralQCDAmplitudeQuarkiSingletOctet} being in the 
representation $\alpha$. If the $\qqbar$-pair is also paired into a 
common representation in the basis vector, the quark trace 
gives a factor of $\TR$, $\Nc$ or vanishes.
The resulting color structure is topologically equivalent to a color structure of the same order as \eqref{eq:GeneralQCDAmplitude} with $i$ and $j$ exchanged for a gluon or singlet, 
apart from
possibly
being 
disconnected (due to the first term in \eqref{eq:QuarkLineSimplification}) and containing $d^{abc}$ vertices (from the second term of \eqref{eq:QuarkLineSimplification}). Such color structures are then equivalent to color structures with one less $\qqbar$-pair, and hence the $\qqbar$-pair $ij$ does not require any other manipulation than what has already been shown. 
However, if $i$ and $j$ are not paired in the basis vector, i.e., they do not belong to the same $q\bar{q}$ pair of the basis vector, the color structure is not guaranteed to contain loops of the type in \eqref{eq:Loop1} or \eqref{eq:Loop2}. Hence additional steps are required.

The terms in \eqref{eq:GeneralQCDAmplitudePairedqqbar} when contracted with the basis vectors will be of the form
\begin{equation}\label{eq:VacuumBubbleQuarks_Initial}
	\raisebox{-0.45\height}{
		\includegraphics[scale=0.4]{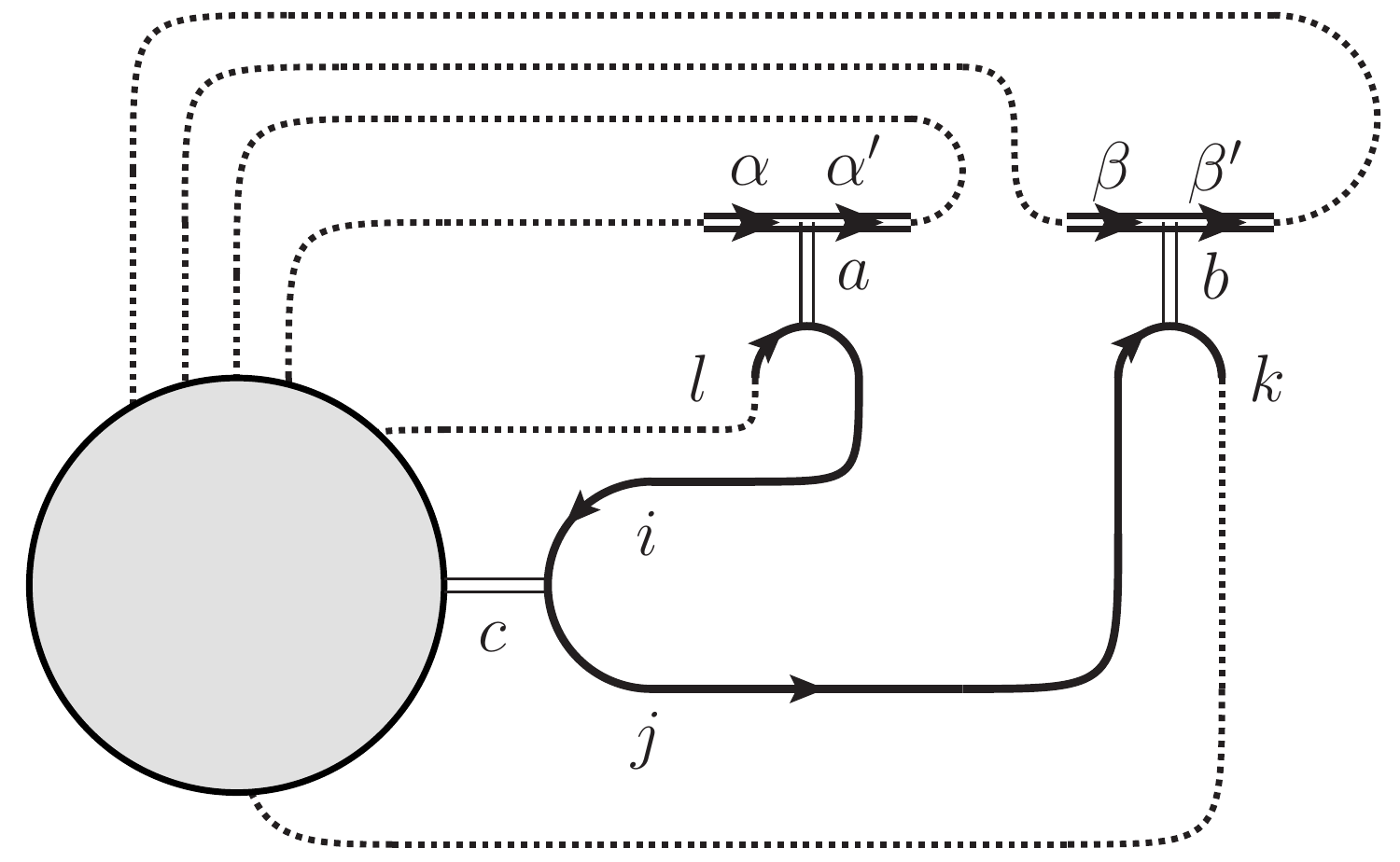}
	}
,
\end{equation}
where the gray blob contains everything else in the vacuum bubble (the rest of the chain of arbitrary representations and the blob in \eqref{eq:GeneralQCDAmplitudePairedqqbar}) and $a$, $b$ and $c$ label the representations of the thin double lines, singlets or octets.  
The goal is to manipulate this color structure into a color structure where the quark-line from $l$ to $k$ is only connected to the basis vector by one representation, a singlet or an octet.
If this is achieved, the quark trace that $l$ and $k$ are part of contains fewer generators, and the procedure can be repeated until the trace is over three or fewer generators.
There are two possibilities for $c$ in \eqref{eq:VacuumBubbleQuarks_Initial}:
\begin{enumerate}
\item If $c$ is a singlet:
  \begin{enumerate}
  \item and at least one of $a$ and $b$ is a singlet, then $l$ and $k$ can be seen as a 
    quark-antiquark pair that is paired in the basis vector. The quark-antiquark pair $lk$ is in a singlet if both $a$ and $b$ are singlets, and otherwise in an octet. 
  \item and both $a$ and $b$ are octets, then \eqref{eq:QuarkLineSimplification} can be used to combine $a$ and $b$ in a singlet or an octet. When $a$ and $b$ are in a singlet they are part of a loop of the form of \eqref{eq:Loop2}, and when they are in an octet they form a loop as in \eqref{eq:Loop1}. Removing these loops will leave a color structure where $l$ and $k$ are in a singlet in the first case, and in an octet in the second case.
  \end{enumerate}
For both (a) and (b), the vacuum bubble is smaller, since the basis vector contains one quark-antiquark pair less;
initially it had the pairs $li$ and $jk$, but they have been replaced by one quark-antiquark pair $lk$.

With $lk$ being in a singlet or an octet and connected to the basis vector, the procedure can be repeated. The color structure can be written on the form of \eqref{eq:VacuumBubbleQuarks_Initial}, with $lk$ taking the place of either $li$ or $jk$, until the new $c$ is an octet or the new $l$ and $k$ are directly connected, in which case the quark trace is over two, or fewer, generators.
\item If $c$ is an octet the color structure is of the form
\begin{equation}\label{eq:VacuumBubbleGluon_Initial}
	\raisebox{-0.45\height}{
		\includegraphics[scale=0.4]{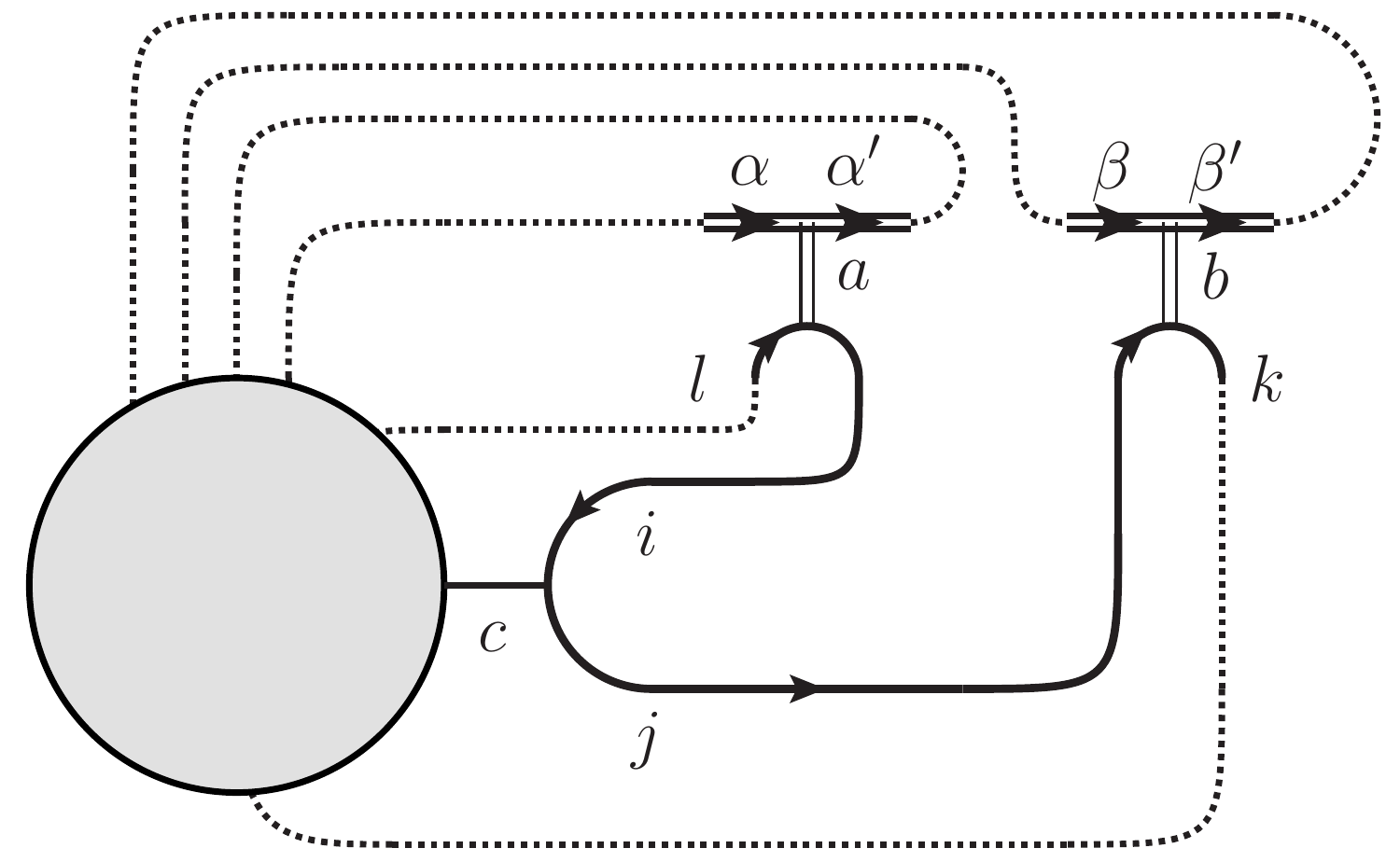}
	}
.
\end{equation}
We wish to manipulate the color structure such that,
instead of $a$, $b$ and $c$, there is only one representation connecting to $lk$
and it connects directly to the basis vector.

Using a completeness relation, \eqref{eq:CRDiagrammatic}, on quark $l$ and antiquark $j$ in \eqref{eq:VacuumBubbleGluon_Initial} gives
\begin{equation}\label{eq:VacuumBubbleGluon_CR}
\frac{d_1}{
\raisebox{-0.45\height}{
	\includegraphics[scale=0.3]{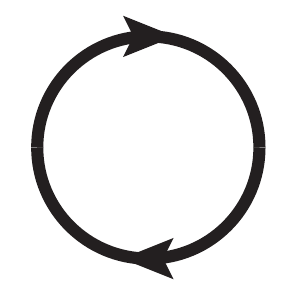}
}
}
	\raisebox{-0.45\height}{
		\includegraphics[scale=0.4]{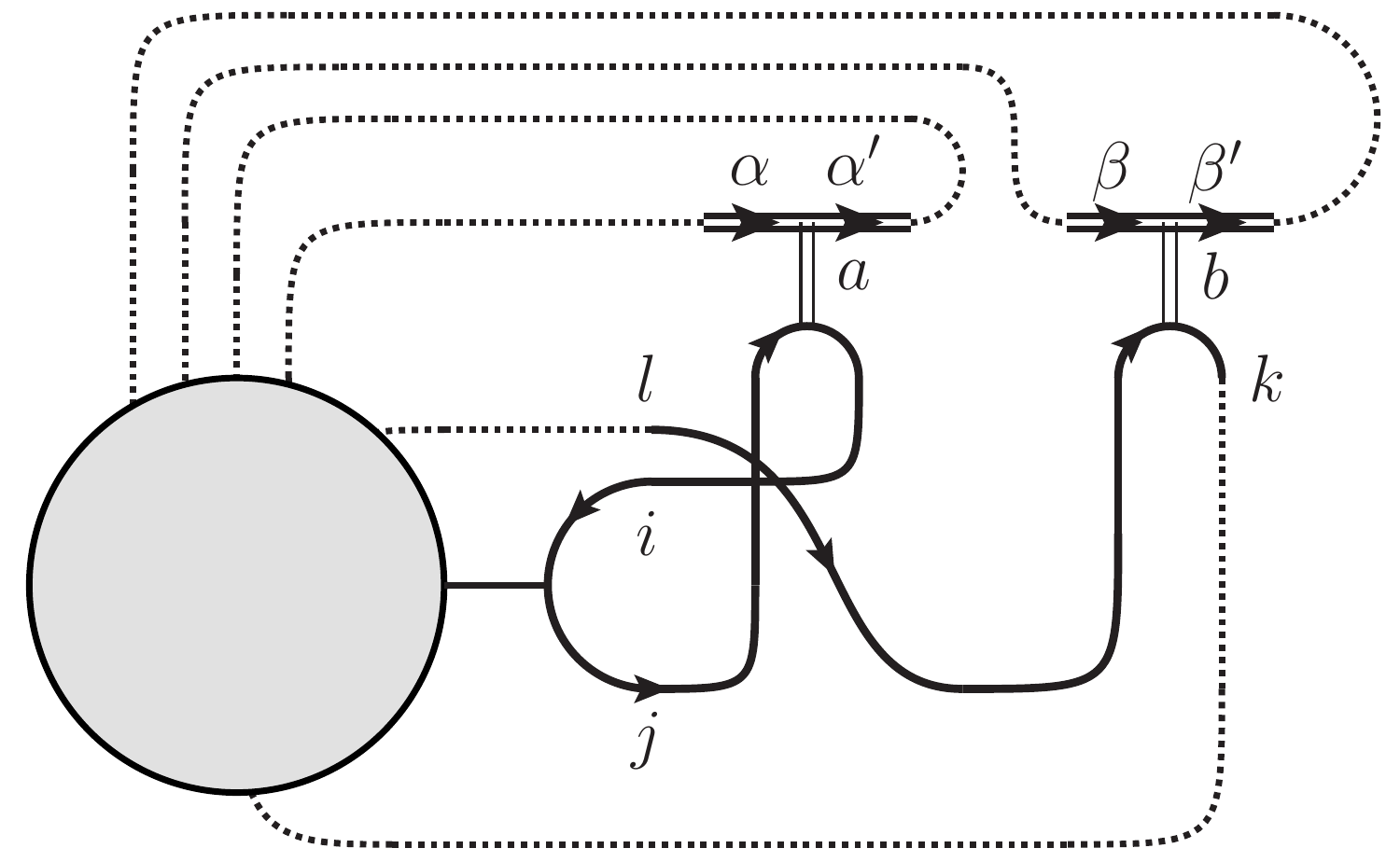}
	}
+
\frac{d_A}{
\raisebox{-0.45\height}{
	\includegraphics[scale=0.3]{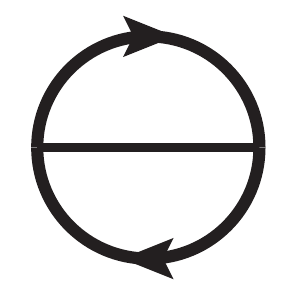}
}
}	
\raisebox{-0.45\height}{
		\includegraphics[scale=0.4]{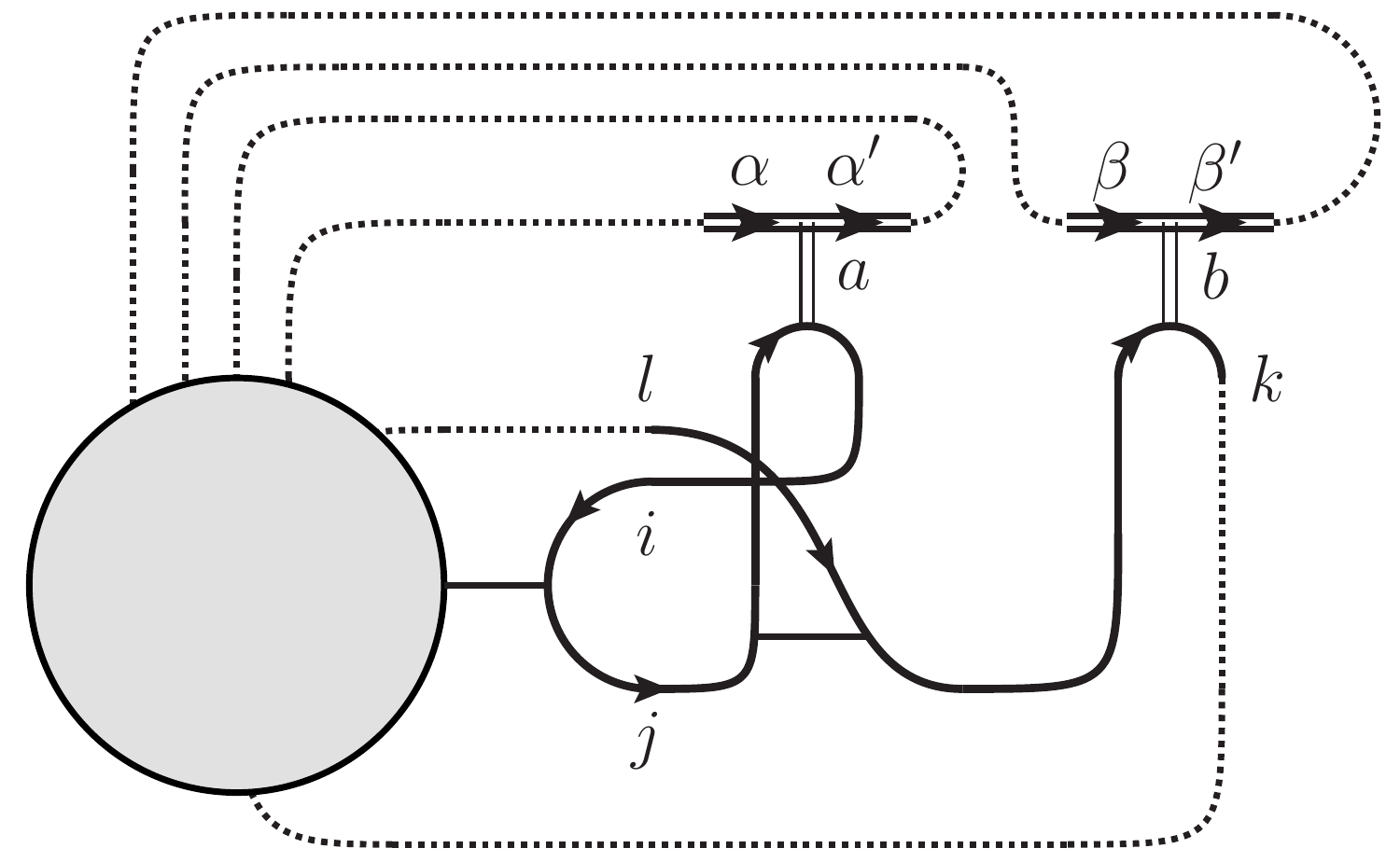}
	}
,
\end{equation}
where the first term vanishes if $a$ is a singlet (since the generators are traceless). For the first term, 
there is only one representation, $b$, connecting to the quark-line $lk$, 
so the procedure can be repeated to remove the rest of the octets/singlets from the trace. 
For the second term of \eqref{eq:VacuumBubbleGluon_CR} another completeness relation can be applied to quark $l$ and antiquark $k$, 
\begin{equation}\label{eq:VacuumBubbleGluon_CR_Term2}
\frac{d_1}{
\raisebox{-0.45\height}{
	\includegraphics[scale=0.3]{Figures/Representations/Quarks/Wigners/Wig3jSinglet}
}
}
	\raisebox{-0.45\height}{
		\includegraphics[scale=0.4]{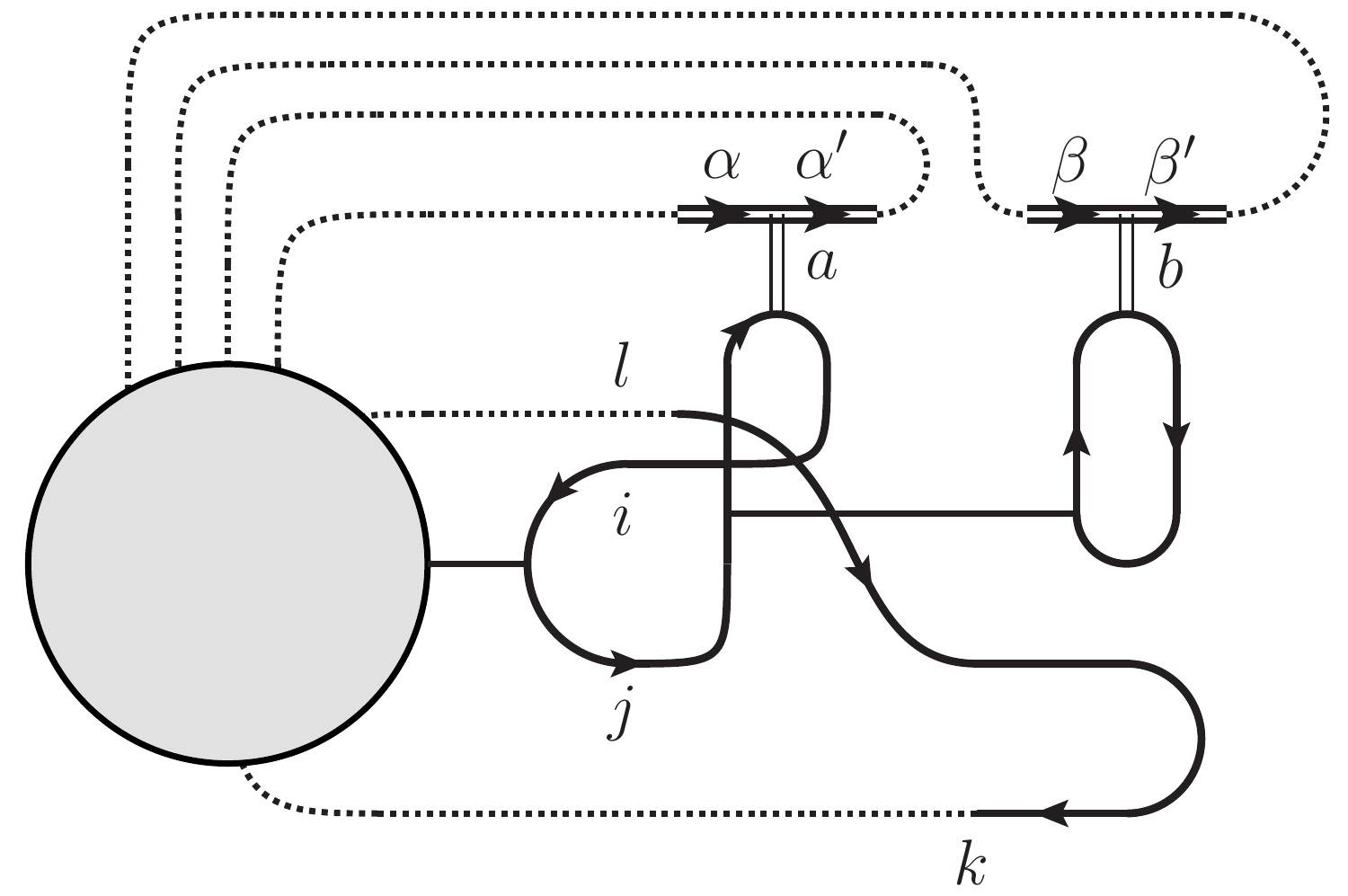}
	}
+
\frac{d_A}{
\raisebox{-0.45\height}{
	\includegraphics[scale=0.3]{Figures/Representations/Quarks/Wigners/Wig3jOctet}
}
}	
\raisebox{-0.45\height}{
		\includegraphics[scale=0.4]{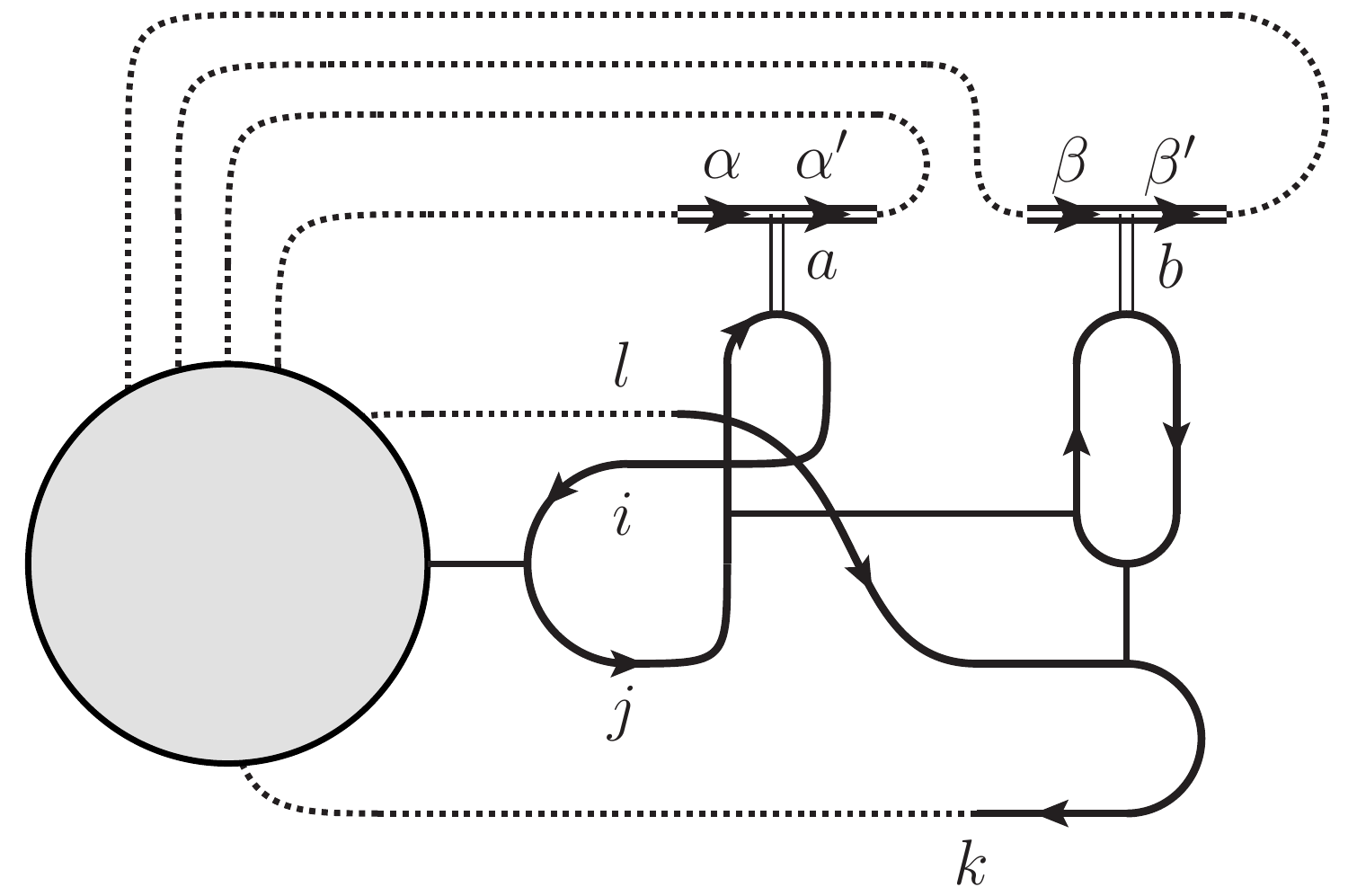}
	}
.
\end{equation}
The first term of \eqref{eq:VacuumBubbleGluon_CR_Term2} vanishes unless $b$ is an octet. As the quark trace with $l$ and $k$ has been disconnected they can be seen as a singlet connected to the chain of arbitrary representations and the procedure can be repeated. For the second term, the different possible combinations of $a$ and $b$ can be divided into two cases.
\begin{enumerate}[(i)]
\item The representations $a$ and $b$ are octets. The traces over three generators can be replaced by $if$ and $d$ vertices and a loop of type \eqref{eq:Loop3} is formed  
\begin{equation}\label{eq:VacuumBubbleGluon_abGluons}
\raisebox{-0.45\height}{
		\includegraphics[scale=0.4]{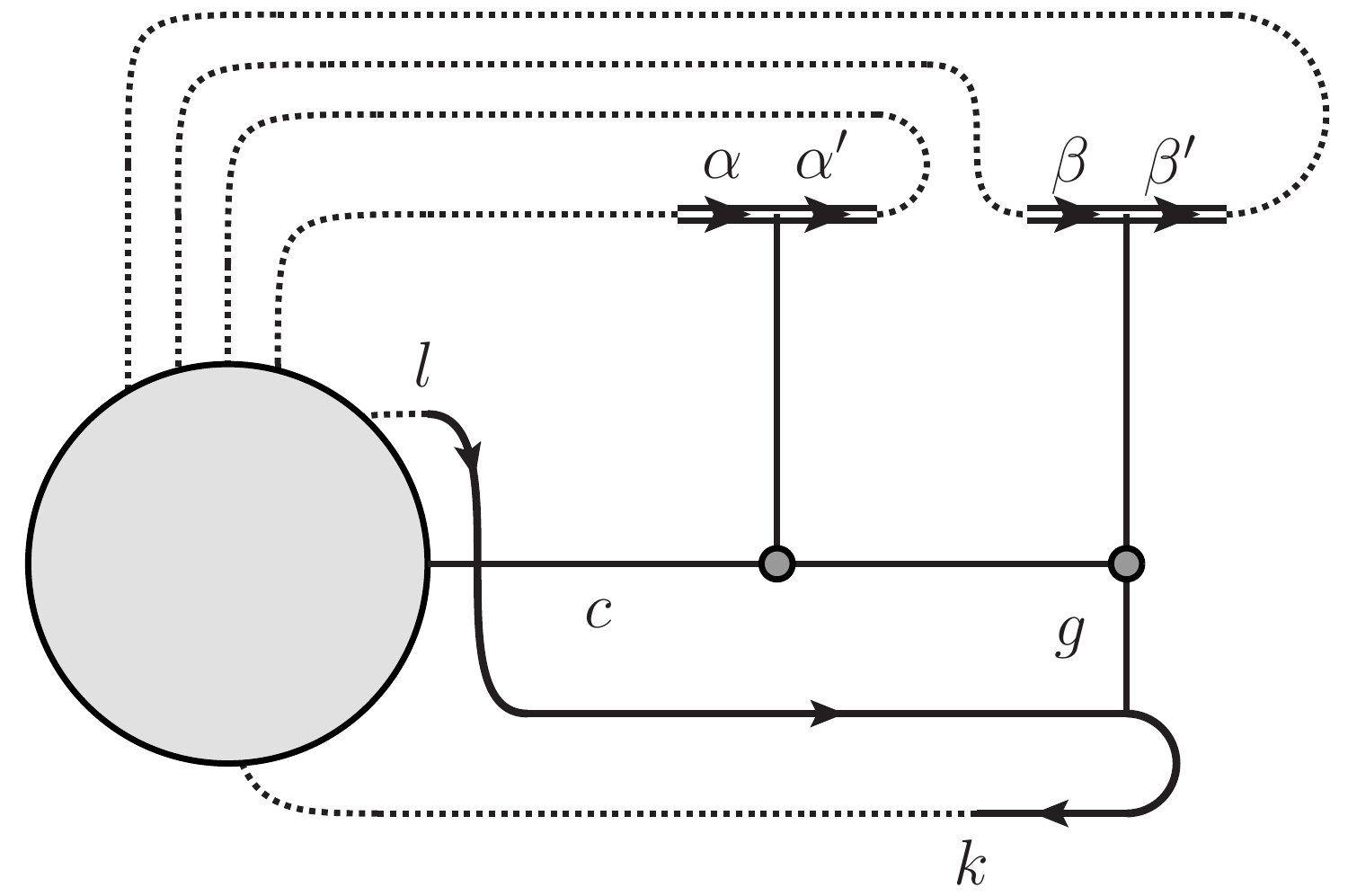}
}
,
\end{equation}
where the two small gray blobs are different combinations of $if$ and $d$, and $g$ labels the gluon connecting to $l$ and $k$. 
Contracting the loop of the type in \eqref{eq:Loop3} leaves a color structure where
$g$ connects the basis vector
to $l$ and $k$, and $c$ connects the big gray blob to the basis vector.
The color structure is now topologically equivalent to a color structure with one less $\qqbar$-pairs, and hence the procedure can be repeated.

\item For the case when $a$ is a singlet (analogous steps can be performed if $b$ is a singlet) and $b$ is either a singlet or an octet, a completeness relation, \eqref{eq:CRDiagrammatic}, can be applied
\begin{equation}\label{eq:VacuumBubbleGluon_abSinglets}
\raisebox{-0.45\height}{
		\includegraphics[scale=0.4]{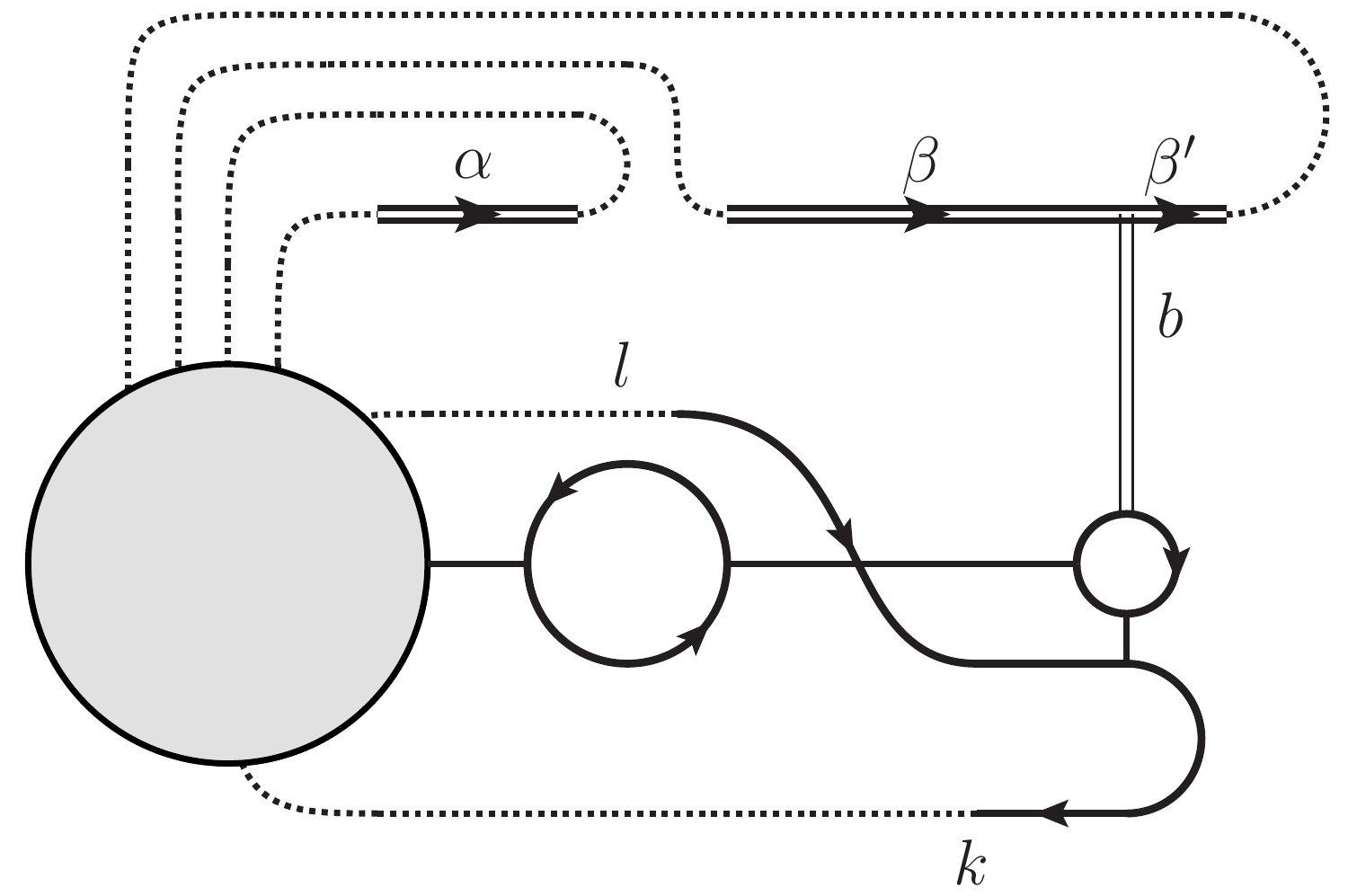}
	}
=
\sum_{\psi}{
\frac{d_\psi}{\includegraphics[scale=0.4]{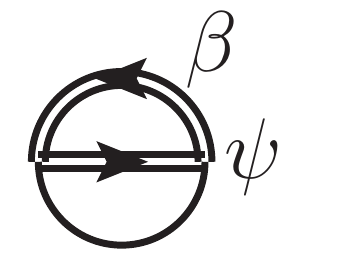}
\hspace{-2.5mm}
}
\raisebox{-0.45\height}{
		\includegraphics[scale=0.4]{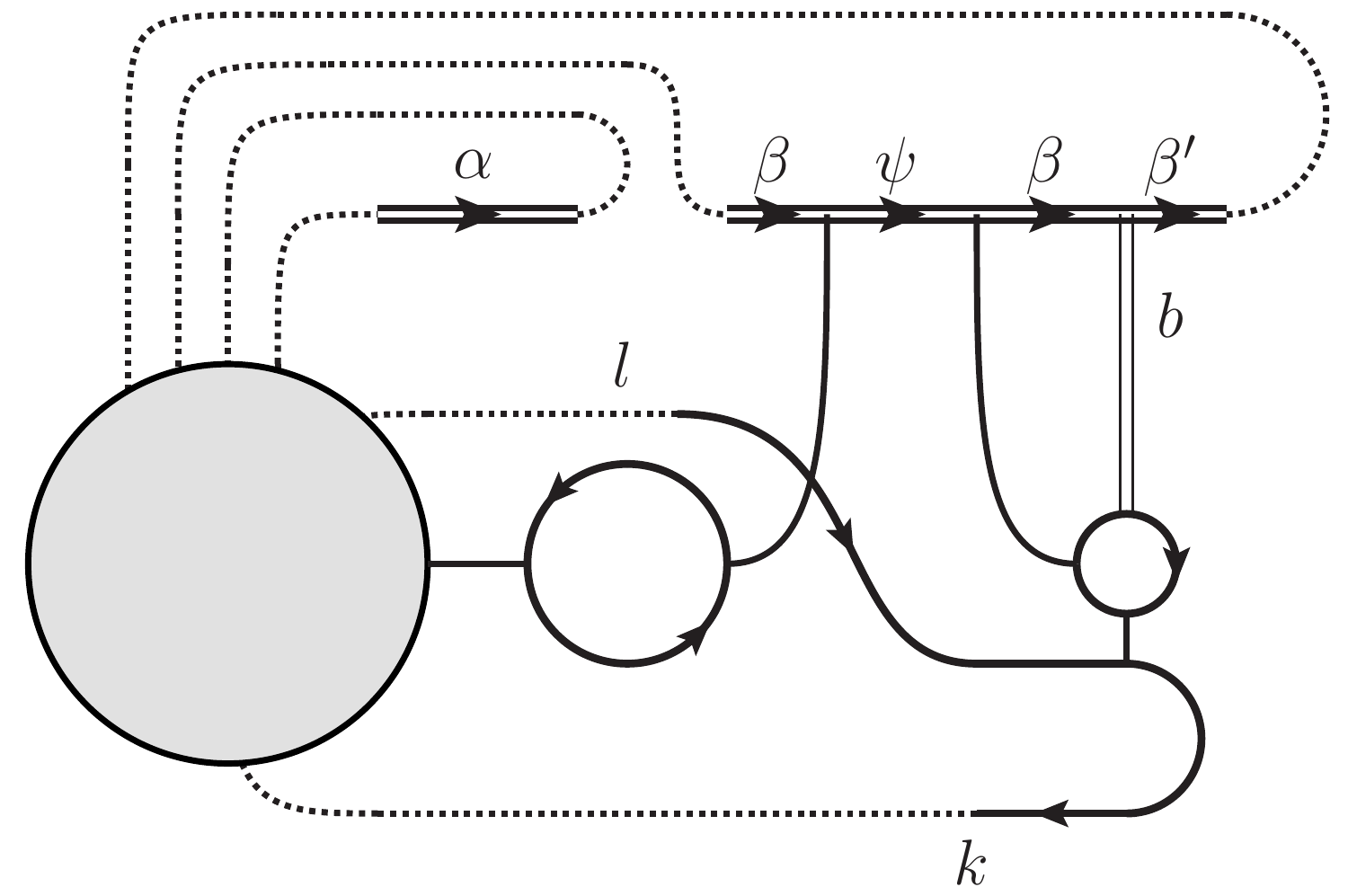}
	}
}
.
\end{equation}
If $b$ is a singlet, then two gluons have been connected to the chain of arbitrary representations,
but as $a$ and $b$ are both singlets, there are now as many gluons connected to the chain of arbitrary representations as if $a$ and $b$ had been octets instead. This is important for the latter part of this appendix, where constraints are put on the representations in the Wigner $6j$ coefficients. 
 Note that if both $a$ and $b$ are singlets, then the completeness relation can be applied to any representation in the chain of arbitrary representations, not just $\beta$. If $b$ was an octet, the trace over three generators can be written as $if$ and $d$ vertices and the resulting vertex correction can be removed as in \eqref{eq:CRApplicationLoopType1LastCR}.
\end{enumerate}
 In both cases, (i) and (ii), the quark trace is shorter, and the procedure can be repeated. 
\end{enumerate}

All possible cases have been covered above, the quark traces are systematically removed by disconnecting representations from them, making them shorter until they are over three, or fewer, generators.
After the quark traces have been removed, the color structure can be contracted using the types of loops required for a gluon-only amplitude of the same order as the original amplitude with quarks (with the addition of loops of type \eqref{eq:Loop2}).
The procedure in this section of the appendix is valid for arbitrary order in perturbation theory, in particular for LO and NLO. By allowing contraction of loops of the type in \eqref{eq:Loop3} any LO and NLO QCD color structure can be completely contracted.

\subsection{Constraints on representations in Wigner coefficients}
The above reasoning proves that only Wigner coefficients of the form of \eqref{eq:Wigner6j3Representations} and \eqref{eq:Wigner6j4Representations} are required, since those are the only forms occurring in \eqref{eq:CRApplicationLoopType1}, \eqref{eq:CRApplicationLoopType1LastCR}, \eqref{eq:CRApplicationLoopType3} and \eqref{eq:CRApplicationLoopType2LastCR}.
We will now move on to put constraints on the representations 
appearing for LO gluon-only color structures, i.e.,
in \eqref{eq:CRApplicationLoopType1}, \eqref{eq:CRApplicationLoopType1LastCR} and \eqref{eq:CRApplicationLoopType2LastCR}. After this, constraints on the representations required for up to NLO with both gluons and quarks are handled, i.e., \eqref{eq:CRApplicationLoopType3}.

An important property of the representations in tensor products between multiple adjoint representations is their so-called first occurrence \cite{Keppeler:2012ih}. The first occurrence, $n_f$, of a representation $M$ is the lowest integer $i$ such that $M\in{}A^{\otimes{}i}$. In \cite{Keppeler:2012ih} it was shown that $n_f$ of the representations in $M\otimes{}A$ can differ from $n_f$ of $M$ by at most $\pm1$. This is a key part in the following derivation of constraints on the representations in the Wigner coefficients.

We want to show that the above contraction procedure never requires 
$\psi$ in \eqref{eq:CRApplicationLoopType1} to have a first occurrence larger than 
$\lfloor{}n/2\rfloor$ for tree-level color structures. 
To prove this, we note that it is enough to prove it for the basis 
vectors with highest possible first occurrence for all involved 
representations.
The representations in the basis vector are
\begin{equation}\label{eq:InitialArbitraryRepresentations}
\alpha_1,\alpha_2,\dots,\alpha_{\lfloor{}n/2\rfloor{-2}},\alpha_{\lfloor{}n/2\rfloor{-1}},\alpha_{\lfloor{}n/2\rfloor},\dots,\alpha_{n-3}
\end{equation}
and the highest first occurrences they can have are given by
\begin{equation}
\label{eq:InitialArbitraryRepresentationsFirstOccurrence}
\begin{cases}
2,3,\dots,\left\lfloor{}\frac{n}{2}\right\rfloor-1,\left\lfloor{}\frac{n}{2}\right\rfloor,\left\lfloor{}\frac{n}{2}\right\rfloor-1,\dots,2,\;\;\;\;\;\;\;\;\;\;\;\;n\;\text{even}
\\
2,3,\dots,\left\lfloor{}\frac{n}{2}\right\rfloor-1,\left\lfloor{}\frac{n}{2}\right\rfloor,\left\lfloor{}\frac{n}{2}\right\rfloor,\left\lfloor{}\frac{n}{2}\right\rfloor-1,\dots,2\;\;\;\;\;\;n\;\text{odd}.
\end{cases}
\end{equation}
The rightmost color structure of \eqref{eq:CRApplicationLoopType1} will be zero if $n_f(\psi)>n_f(\alpha)+1$ or $n_f(\psi)>n_f(\gamma)+1$, 
since the representations can differ by at most $\pm1$ in first occurrence.
From these limits we see that the first occurrence of $\psi$ can only exceed $\lfloor{}n/2\rfloor$ if $n_f(\alpha)\geq\lfloor{}n/2\rfloor$ and $n_f(\gamma)\geq\lfloor{}n/2\rfloor$. 
Noting that in the chain of representations $\alpha$ and $\gamma$ 
are not next to each other (as they are separated by $\beta$),
one can see from \eqref{eq:InitialArbitraryRepresentationsFirstOccurrence}
that $\alpha$ and $\gamma$ cannot both have first occurrence
$\lfloor{}n/2\rfloor$ since they are not adjacent.
Thus the first occurrence of $\psi$ cannot exceed $\lfloor{}n/2\rfloor$.

Stronger constraints can be put on the coefficients by examining 
the first occurrences of $\alpha$, $\beta$, $\gamma$ and $\psi$ 
of \eqref{eq:CRApplicationLoopType1}. 
For even $n$, we see from \eqref{eq:InitialArbitraryRepresentationsFirstOccurrence} 
and \eqref{eq:CRApplicationLoopType1} that the representation $\psi$ can 
only have first occurrence of $\lfloor{}n/2\rfloor$ if 
$n_f(\alpha)=n_f(\gamma)=\lfloor{}n/2\rfloor-1$. 
The consequence of this is that for even $n$ there can be 
at most two representations in coefficients of the form of 
\eqref{eq:Wigner6j4Representations} with first occurrence 
$\lfloor{}n/2\rfloor$, and they cannot be adjacent in the 
coefficient (in \eqref{eq:CRApplicationLoopType1} they would be 
$\beta$ and $\psi$, thus separated by $\alpha$ and $\gamma$). 
If $n$ is odd, then two of $\alpha$, $\beta$ and $\gamma$ can have first occurrence $\lfloor{}n/2\rfloor$, and $\psi$ may also have first occurrence $\lfloor{}n/2\rfloor$, the Wigner coefficient can then contain at most three out of four representations with first occurrence $\lfloor{}n/2\rfloor$. 

For the Wigner coefficients of the form of \eqref{eq:Wigner6j3Representations} in
\eqref{eq:CRApplicationLoopType1LastCR} the first occurrences are the same as for three adjacent representations in the chain of arbitrary representations, \eqref{eq:InitialArbitraryRepresentations}, i.e. for even $n$ there can be at most one representation with $\lfloor{}n/2\rfloor$ and for odd $n$ there can be at most two. For the Wigner $6j$ coefficient occurring in \eqref{eq:CRApplicationLoopType2LastCR}, the same argument applies, no new representation has been added so it cannot contribute to an increase in first occurrences of representations.
These constraints were used in calculating the number of required Wigner 
coefficients for LO gluon-only processes, shown in \tabref{tab:N6j}.

It is now of interest to see how the constraints on the representations 
in the required Wigner coefficients change by including 
NLO diagrams and quarks.
As the only change compared to the previous discussion is 
\eqref{eq:CRApplicationLoopType3}, it is only the constraints on the 
Wigner $6j$ coefficients with three general representations, 
\eqref{eq:Wigner6j3Representations}, that are altered. 
The case of interest is when at least two of the three 
representations have a first occurrence of $\lfloor{}n/2\rfloor$ 
(the case with one representation with $n_f=\lfloor{}n/2\rfloor$ 
is already required for LO gluon color structures),
this case occurs when $\psi$ has a first occurrence 
of $\lfloor{}n/2\rfloor$.
For even $n$, the 
highest first occurrences
appear when 
$n_f(\alpha)=n_f(\gamma)=\lfloor{}n/2\rfloor-1$ and 
$n_f(\beta)=\lfloor{}n/2\rfloor$, 
for which we see from the right hand side of \eqref{eq:CRApplicationLoopType3} 
that coefficients with two representations with first occurrence 
$\lfloor{}n/2\rfloor$ may appear ($\psi$ and $\beta$).
Similarly, for odd $n$, the interesting case is when
$n_f(\alpha)=n_f(\beta)=\lfloor{}n/2\rfloor$ or
$n_f(\beta)=n_f(\gamma)=\lfloor{}n/2\rfloor$
for which all three representations can have 
first occurrence of $\lfloor{}n/2\rfloor$. As is shown in \tabref{tab:N6j}, 
these altered constraints lead to a very small difference in the 
required number of coefficients between tree-level gluon color structures and NLO color structures with quarks and gluons.

\section{Construction history independence of $6j$ coefficients}
\label{sec:Uniqueness6j}

To prove that the $6j$ coefficients which only contain
unique vertices are independent of the 
construction history, we note that the various copies
of a given vector space, such as the $V^{27,35}$- and the
$V^{10,35}$-versions of the vector space corresponding to 
the $35$-plets for three gluons, $A^{\otimes 3}$, are isomorphic.
With unique vertices we mean vertices between representations 
$\alpha, \beta$ and $\gamma$, s.t. the tensor product
$\alpha \otimes \beta$ only contains one instance of the
representation $\gamma$.

Since the vector spaces $V^{27,35}$ and $V^{10,35}$ are isomorphic, 
we can find a transformation $\Uni$ from $A^{\otimes 3}$ to $A^{\otimes 3}$, 
which maps the vector space $V^{27,35}$ to $V^{10,35}$ in a unitary way,
but which maps every other irreducible subspace in $A^{\otimes 3}$ to 0. 
Seen as a matrix (in a suitable basis), $\Uni$ thus has an 
off-diagonal 35-dimensional
matrix block whereas all other elements equal zero.
The projection operators $\Proj^{10,35}$ and $\Proj^{27,35}$ are then 
related as
$\Proj^{10,35}=\Uni \Proj^{27,35} \Uni^{-1}$, in birdtrack notation,
\begin{equation}
  \parbox{3cm}{\epsfig{file=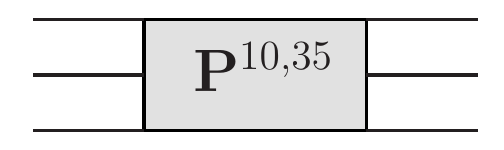,width=3cm}}=
  \parbox{7cm}{\epsfig{file=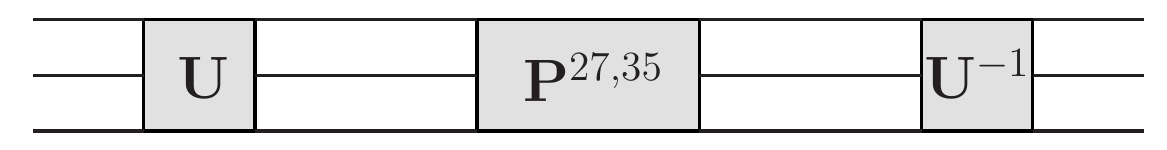,width=7cm}} .
\end{equation}
For the proof of the existence of construction history independent
$6j$ coefficients, the existence of $\Uni$ is enough, but for
explicit construction, we remark that $\Uni$ can be
obtained by noting that any invariant transformation $\mathbf{T}$ between
the irreducible vector spaces $V^{10,35}$ and $V^{27,35}$ will be proportional 
to the unit matrix by Schur's lemma. Since any transformation expressed
in birdtracks is invariant, we can use any non-vanishing $\mathbf{T}$.
Normalizing $\mathbf{T}$ and inserting it in the only non-vanishing
block in $\Uni$ gives us $\Uni$.
 
Using $\Uni$ we can rewrite the $6j$ coefficient 

\begin{equation}
  \parbox{6.5cm}{\epsfig{file=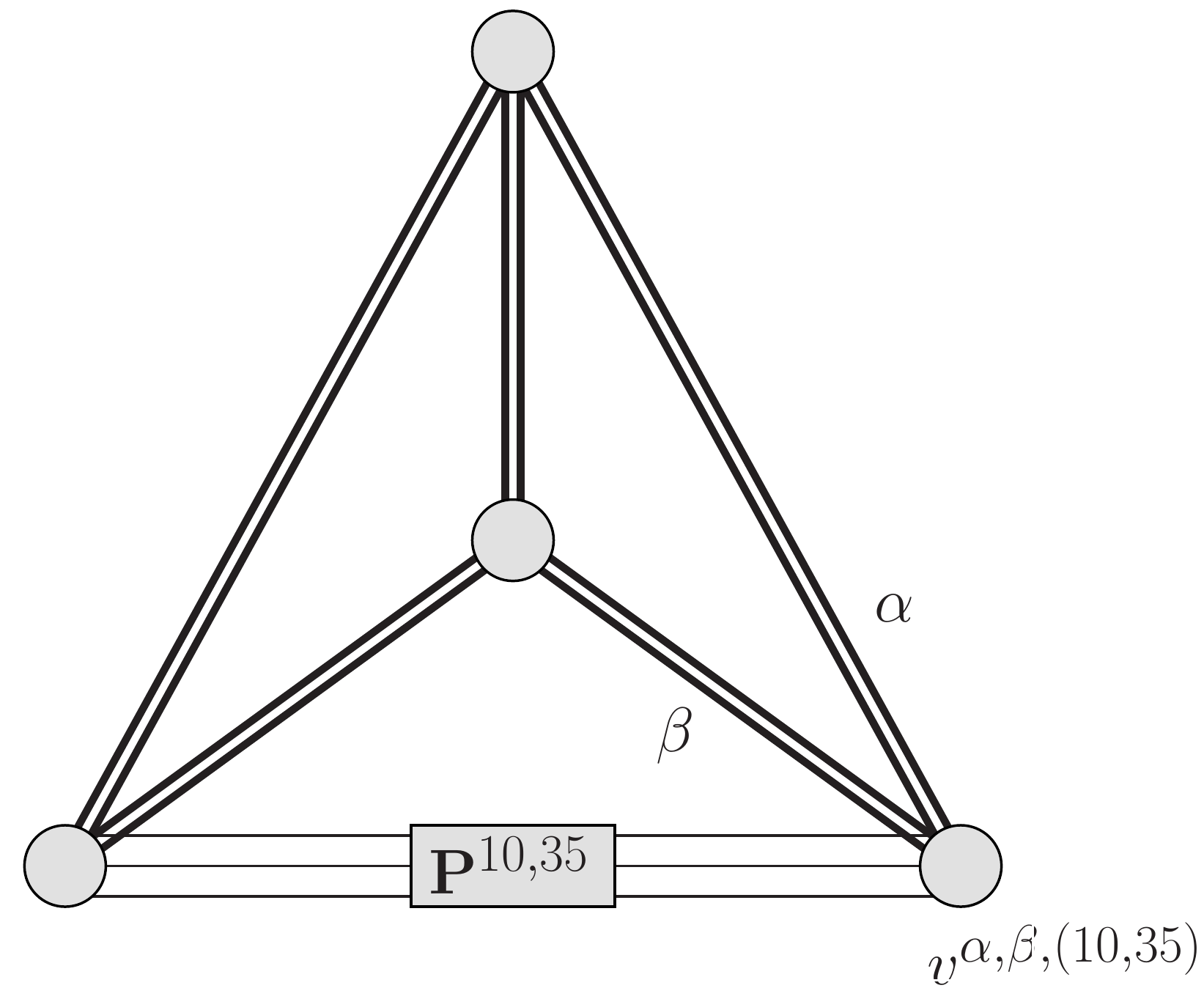,width=6.5cm}}
  =   
  \parbox{6.5cm}{\epsfig{file=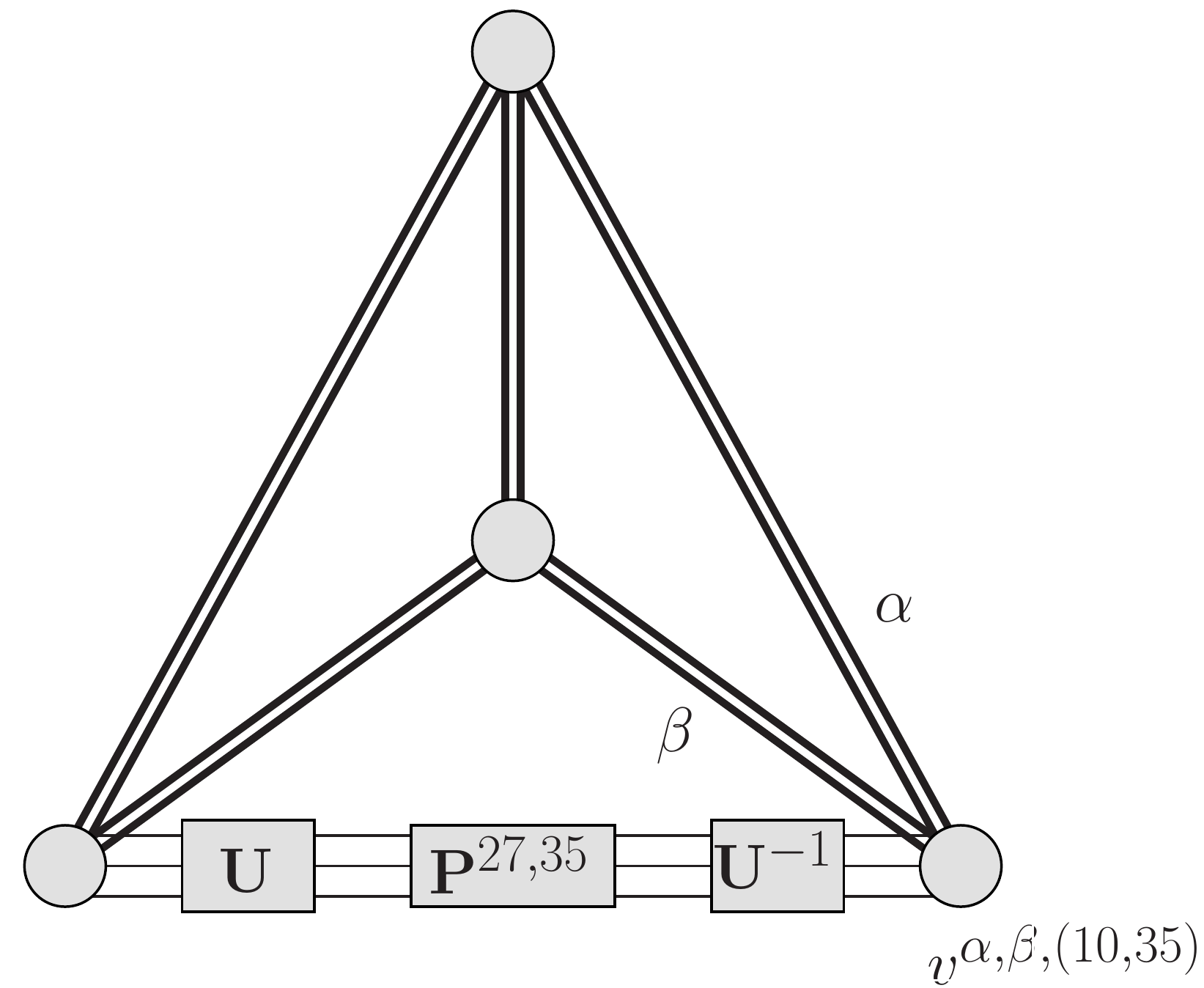,width=6.5cm}}, 
\end{equation}
from which we see that if the vertex $v^{\alpha,\beta,(27,35)}$ is defined
s.t. it equals the encircled region in
\begin{equation}
\parbox{3.5cm}{\epsfig{file=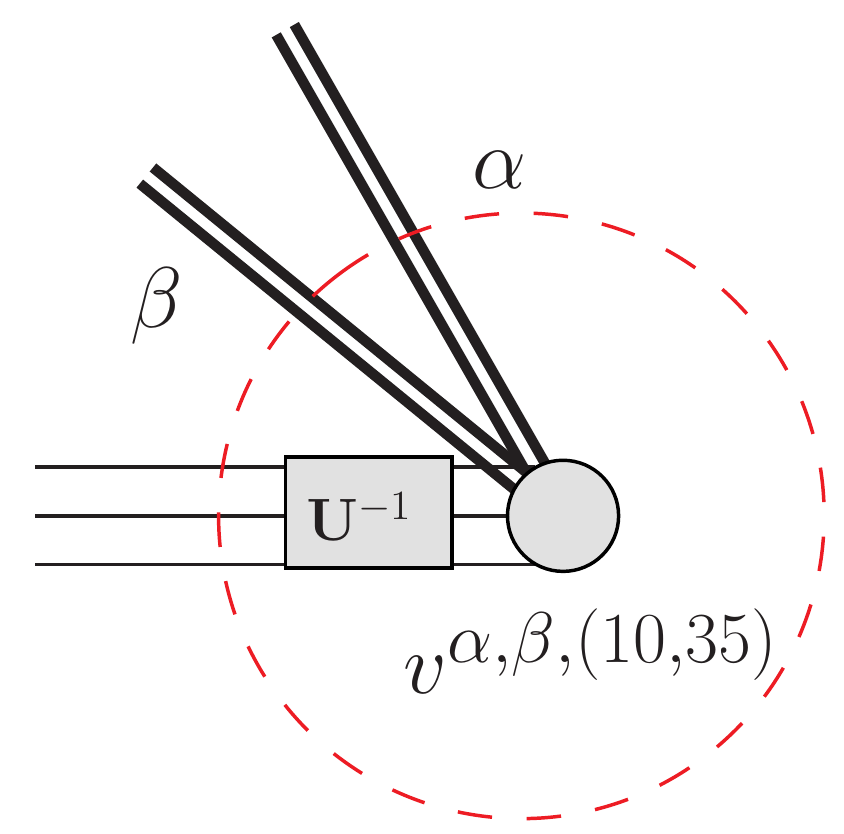,width=3.5cm}}\; =v^{\alpha, \beta, (27,35)} , 
\label{eq:vertex_equality}
\end{equation}
and similarly for the vertex to the left in the $6j$ coefficient, the equality 
is satisfied. For unique and normalized vertices, the $v^{\alpha,\beta,(27,35)}$-vertex 
has by necessity already been defined this way (up to a sign), and we conclude 
that the $6j$ coefficients must agree.
By the same argument all $6j$ coefficients involving unique vertices 
which only differ by the construction history must agree (modulo signs). 
For vertices appearing in several forms, or to fix the sign ambiguity,
we note that if we define the various vertices s.t. \eqref{eq:vertex_equality} 
holds, the $6j$ coefficients agree.

\section{Proof of symmetries of $6j$ coefficients}
\label{sec:Symmetries6js}
The rotation symmetry of the first type of coefficient in \eqref{eq:RotationalSymmetry} is obvious. For the second type of coefficient, the Wigner coefficient can be put into a form where the symmetry is easier to see. Moving the central vertex upwards and to the right, rotating and finally moving the upper right vertex to the middle gives the symmetry,
\begin{equation}\label{eq:RotationalSymmetry4RepsProof}
\raisebox{-0.45\height}{
	\includegraphics[scale=0.4]{Figures/Wigner6js/Wig6j4Reps}
}
\hspace{-3mm}
=
\raisebox{-0.385\height}{
	\includegraphics[scale=0.4]{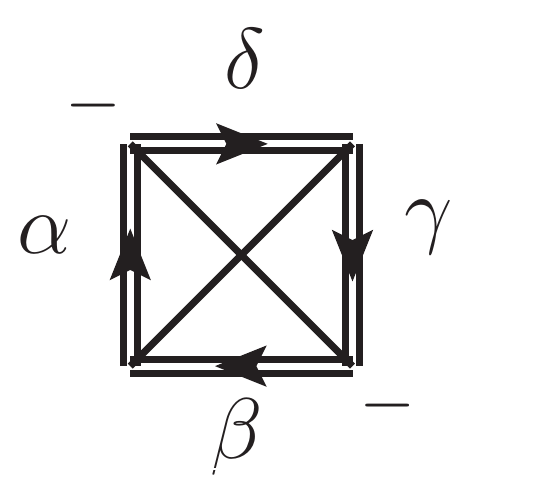}
}
\hspace{-3mm}
=
\raisebox{-0.385\height}{
	\includegraphics[scale=0.4]{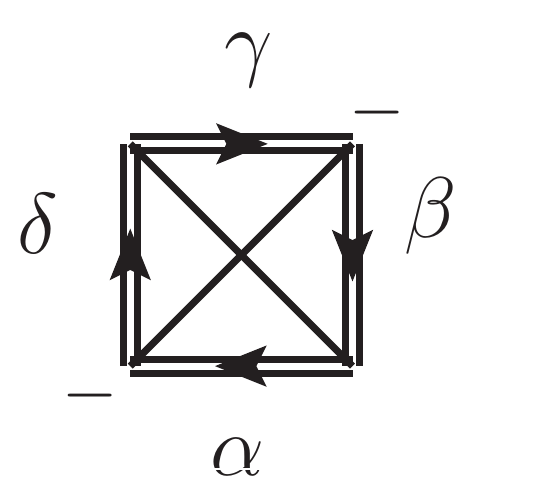}
}
\hspace{-3mm}
=
\raisebox{-0.35\height}{
	\includegraphics[scale=0.4]{Figures/Wigner6js/Wig6j4RepsRotated}
}.
\end{equation}

To prove the conjugation symmetry, \eqref{eq:ConjugationSymmetry}, 
we recall that the color sum in the Wigner coefficients only consists 
of factors with a different number of closed quark-lines, i.e., 
they can be expanded in the fundamental representation, giving a 
polynomial in $N_c$ (up to vertex normalizations). 
This means that the Wigner coefficients are real numbers, 
and as such do not change under conjugation. 

The final symmetry, \eqref{eq:MirrorSymmetry}, is shown by simply 
swapping places of the lower left and lower right vertices.

\bibliographystyle{JHEP}  
\bibliography{Refs} 

\end{document}

%% file: Tables/ExplicitWigTable.tex
\begin{table}[tbp]
\centering
\begin{tabular}{ |l|l|l| }
\hline
\raisebox{-0.45\height}{\includegraphics[scale=0.4]{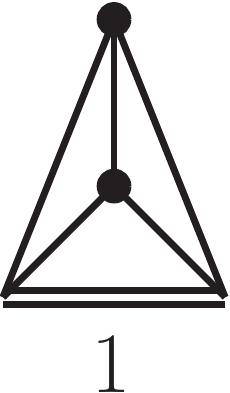}}\hspace{0.5mm}=$
\frac{1}{N_c^2-1}$
&
\raisebox{-0.4\height}{\includegraphics[scale=0.4]{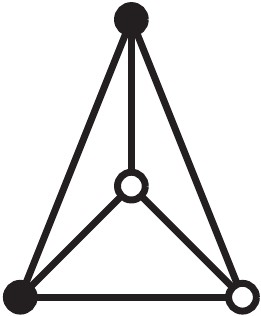}}\hspace{0.5mm}=$
\frac{1}{2 \left(N_c^2-1\right)}$
&
\raisebox{-0.45\height}{\includegraphics[scale=0.4]{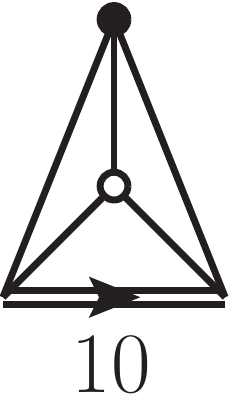}}\hspace{0.5mm}=$
-\frac{1}{\sqrt{N_c^2-4} \left(N_c^2-1\right)}$
\\[6mm]
\hline
\raisebox{-0.4\height}{\includegraphics[scale=0.4]{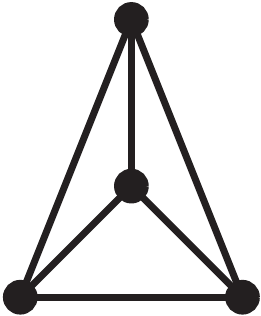}}\hspace{0.5mm}=$
\frac{1}{2 \left(N_c^2-1\right)}$
&
\raisebox{-0.45\height}{\includegraphics[scale=0.4]{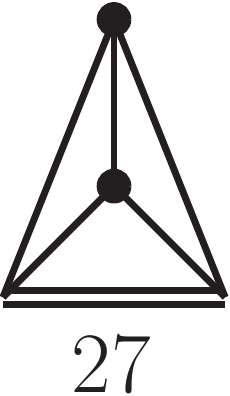}}\hspace{0.5mm}=$
-\frac{1}{N_c \left(N_c^2-1\right)}$
&
\raisebox{-0.45\height}{\includegraphics[scale=0.4]{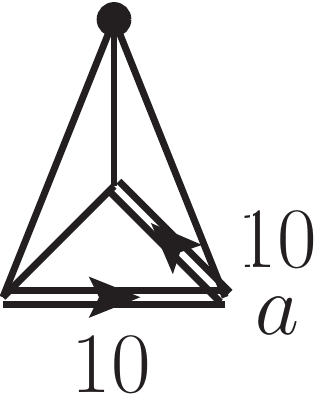}}\hspace{0.5mm}=$
\frac{\sqrt{2}}{\sqrt{N_c^2-4} \left(N_c^2-1\right)}$
\\[6mm]
\hline
\raisebox{-0.45\height}{\includegraphics[scale=0.4]{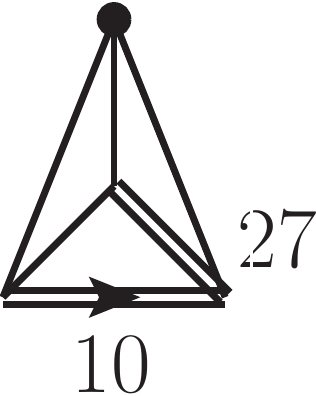}}\hspace{0.5mm}=$
-\frac{1}{\left(N_c-1\right) N_c \sqrt{\left(N_c+1\right) \left(N_c+2\right)}}$
&
\raisebox{-0.45\height}{\includegraphics[scale=0.4]{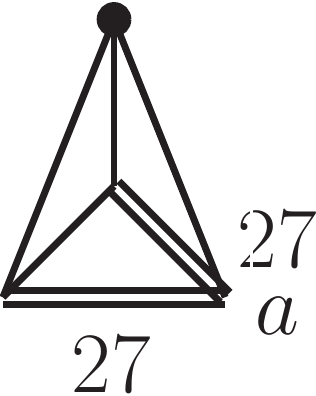}}\hspace{0.5mm}=$
\frac{\sqrt{2}}{\left(N_c-1\right) N_c \sqrt{N_c \left(N_c+3\right)}}$
&
\raisebox{-0.45\height}{\includegraphics[scale=0.4]{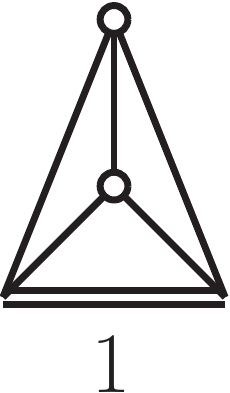}}\hspace{0.5mm}=$
\frac{1}{N_c^2-1}$
\\[6mm]
\hline
\raisebox{-0.4\height}{\includegraphics[scale=0.4]{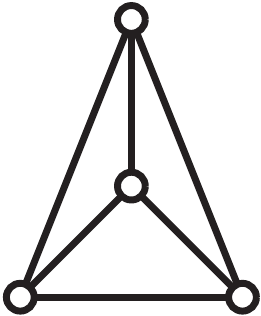}}\hspace{0.5mm}=$
\frac{N_c^2-12}{2 \left(N_c^2-4\right) \left(N_c^2-1\right)}$
&
\raisebox{-0.45\height}{\includegraphics[scale=0.4]{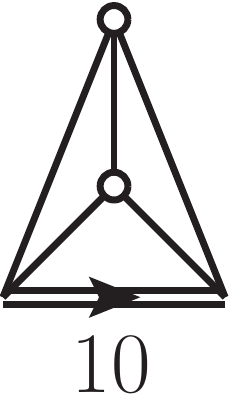}}\hspace{0.5mm}=$
-\frac{2}{\left(N_c^2-4\right) \left(N_c^2-1\right)}$
&
\raisebox{-0.45\height}{\includegraphics[scale=0.4]{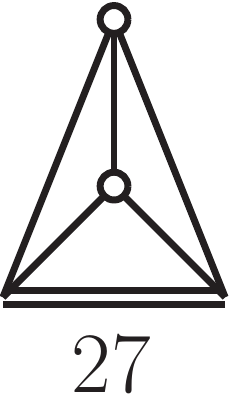}}\hspace{0.5mm}=$
\frac{1}{\left(N_c+2\right) \left(N_c^2-1\right)}$
\\[6mm]
\hline
\raisebox{-0.45\height}{\includegraphics[scale=0.4]{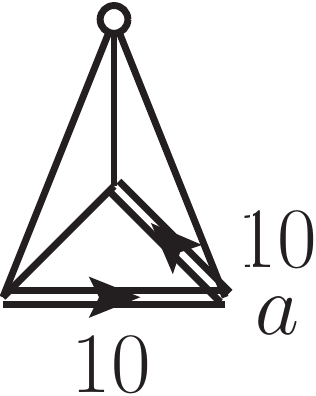}}\hspace{0.5mm}=$
\frac{\sqrt{2}}{\left(N_c^2-4\right) \left(N_c^2-1\right)}$
&
\raisebox{-0.45\height}{\includegraphics[scale=0.4]{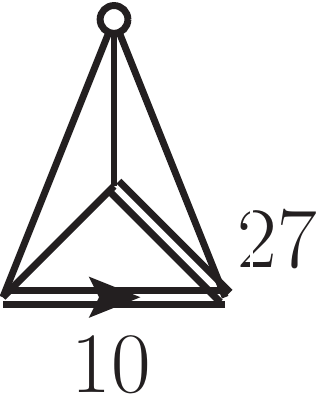}}\hspace{0.5mm}=$
\frac{1}{\left(N_c-1\right) \sqrt{\left(N_c-2\right) \left(N_c+1\right)} \left(N_c+2\right)}$
&
\raisebox{-0.45\height}{\includegraphics[scale=0.4]{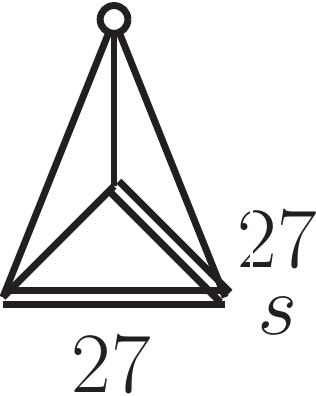}}\hspace{0.5mm}=$
\frac{\sqrt{2} \sqrt{N_c+4}}{\left(N_c-1\right) N_c \left(N_c+2\right) \sqrt{N_c+3}}$
\\[6mm]
\hline
\raisebox{-0.45\height}{\includegraphics[scale=0.4]{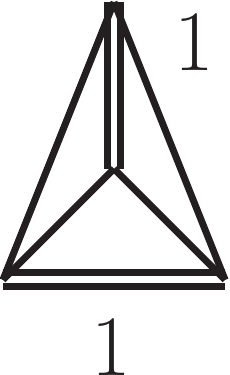}}\hspace{0.5mm}=$
\frac{1}{N_c^2-1}$
&
\raisebox{-0.45\height}{\includegraphics[scale=0.4]{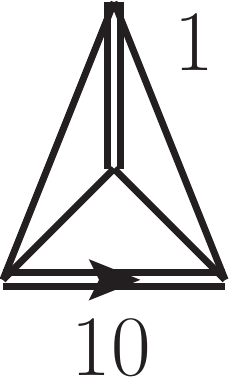}}\hspace{0.5mm}=$
\frac{1}{N_c^2-1}$
&
\raisebox{-0.45\height}{\includegraphics[scale=0.4]{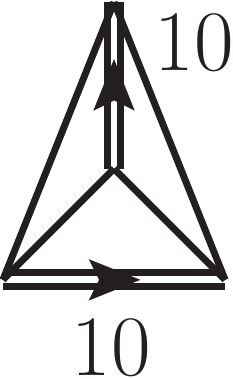}}\hspace{0.5mm}=$
\frac{1}{\left(N_c^2-4\right) \left(N_c^2-1\right)}$
\\[6mm]
\hline
\raisebox{-0.45\height}{\includegraphics[scale=0.4]{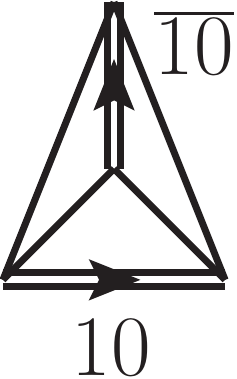}}\hspace{0.5mm}=$
\frac{1}{\left(N_c^2-4\right) \left(N_c^2-1\right)}$
&
\raisebox{-0.45\height}{\includegraphics[scale=0.4]{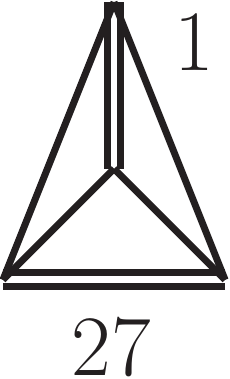}}\hspace{0.5mm}=$
\frac{1}{N_c^2-1}$
&
\raisebox{-0.45\height}{\includegraphics[scale=0.4]{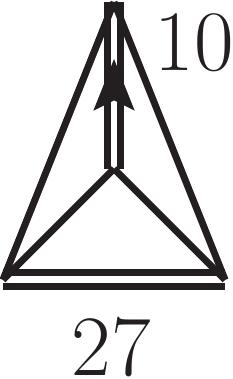}}\hspace{0.5mm}=$
-\frac{1}{N_c \left(N_c+2\right) \left(N_c^2-1\right)}$
\\[6mm]
\hline
\raisebox{-0.45\height}{\includegraphics[scale=0.4]{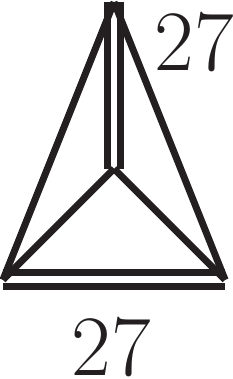}}\hspace{0.5mm}=$
\frac{N_c^2+N_c+2}{N_c^2 \left(N_c+2\right) \left(N_c+3\right) \left(N_c^2-1\right)}$
&
&
\\[6mm]
\hline
\end{tabular}
\caption{\label{tab:4GluonWignerCoefficients} Non-vanishing Wigner $6j$ 
coefficients required for up to NLO QCD color structures with up to four 
external gluons plus 
$\qqbar$-pairs. Note that all vertices (in particular the antisymmetric 
triple-gluon vertices) have been normalized such that the corresponding 
$3j$ coefficient is one, c.f. \eqref{eq:3jNormalization1} and 
\eqref{eq:3j888Normalization1}. 
}
\end{table}